%% file: tesi.tex
 \documentclass[11pt,twoside,final]{huthesis}

\usepackage{epsfig,bm,epsf,float,lscape,deluxetable,latexsym,rotating}

\begin{document}

\input{mathdefs} 


\dsp

\include{titolo}



\include{teoria}   

\include{spin}

\include{urc}

\include{nfw}

\include{angmom}

\include{simulazioni}

\include{conclusioni}


{
\ssp 

}

\appendix
\include{appendix}


\end{document}

%% file: mathdefs.tex
%

\newcommand{\deriv}[2]{\frac{d#1}{d#2}}
\newcommand{\derivc}[3]{\left. \frac{d#1}{d#2}\right|_{#3}}
\newcommand{\pd}[2]{\frac{\partial #1}{\partial #2}}
\newcommand{\pdc}[3]{\left. \frac{\partial #1}{\partial #2}\right|_{#3}}

\newcommand{\bra}[1]{\left\langle #1\right|}
\newcommand{\ket}[1]{\left|#1\right\rangle}
\newcommand{\braket}[2]{\left\langle #1 \left|#2\right.\right\rangle}
\newcommand{\braOket}[3]{\left\langle #1\left|#2\right|#3\right\rangle}

\def\cal#1{\mathcal{#1}}

\def\avg#1{\left< #1 \right>}
\def\abs#1{\left| #1 \right|}
\def\recip#1{\frac{1}{#1}}
\def\vhat#1{\hat{{\bf #1}}}
\def\smallfrac#1#2{{\textstyle\frac{#1}{#2}}}
\def\smallrecip#1{\smallfrac{1}{#1}}

\def\spshalf{{1\over{2}}}
\def\Orabi{\Omega_{\rm rabi}}
\def\btt#1{{\tt$\backslash$#1}}


\def\msun{M_{\odot}}


\def\simgt
{\hbox{\raise0.5ex\hbox{$\; >\lower1.06ex\hbox{$\kern-1.07em{\sim}$}\;$}}}
\def\simlt
{\hbox{\raise0.5ex\hbox{$\; <\lower1.06ex\hbox{$\kern-1.07em{\sim}$}\;$}} }

\def\whs{\thinspace ${\rm W \: Hz^{-1} \: sr^{-1}}$}
\def\lpow{\thinspace ${\rm log_{10}P_{1.4GHz}\: }$}
\def\Re{$R_e\:$}
\def\Me{$M_e(<R_e)\:$}
\def\Mv{$M_{\rm vir}\:$}
\def\Ms{$M_{\rm sph}\:$}
\def\Mbh{$M_{\rm BH}\:$}
\def\Msun{$M_{\odot}\:$}
\def\vv{$V_{\rm vir}\:$}
\def\sv{${\rm \sigma/V_{vir}}\:$}

\def\zv{$z_{\rm vir}\:$}
\def\FJ{Faber-Jackson }
\def\RR{$R_{1.4}^* \:$}
\def\chandra{{\it Chandra\/}}

\newenvironment{trat}
{\begin{list}
  {$\quad -$}
  {\itemsep = 0.25ex\parsep=5pt\topsep = 5mm\leftmargin=7mm}}
{\end{list} }
\def\schrod{Schroedinger's Equation}
\def\helm{Helmholtz Equation}

\def\be{\begin{equation}}
\def\ee{\end{equation}}
\def\bea{\begin{eqnarray}}
\def\eea{\end{eqnarray}}
\def\bean{\begin{mathletters}\begin{eqnarray}}
\def\eean{\end{eqnarray}\end{mathletters}}

\newcommand{\tbox}[1]{\mbox{\tiny #1}}
\newcommand{\half}{\mbox{\small $\frac{1}{2}$}}
\newcommand{\pit}{\mbox{\small $\frac{\pi}{2}$}}
\newcommand{\sfrac}[1]{\mbox{\small $\frac{1}{#1}$}}
\newcommand{\mbf}[1]{{\mathbf #1}}
\def\text{\tbox}

\newcommand{\mV}{{\mathsf{V}}}
\newcommand{\mL}{{\mathsf{L}}}
\newcommand{\mA}{{\mathsf{A}}}
\newcommand{\lB}{\lambda_{\tbox{B}}}  
\newcommand{\ofr}{{(\mbf{r})}}       
\def\ofkr{(k;\mbf{r})}			
\def\ofks{(k;\mbf{s})}			
\newcommand{\ofs}{{(\mbf{s})}}       
\def\xt{\mbf{x}^{\tbox T}}		

\def\ce{\tilde{C}_{\tbox E}}		
\def\cew{\tilde{C}_{\tbox E}(\omega)}		
\def\ceqmw{\tilde{C}^{\tbox{qm}}_{\tbox E}(\omega)}	
\def\cewqm{\tilde{C}^{\tbox{qm}}_{\tbox E}}	
\def\ceqm{C^{\tbox{qm}}_{\tbox E}}	
\def\cw{\tilde{C}(\omega)}		
\def\cfw{\tilde{C}_{\cal F}(\omega)}		

\def\tcl{\tau_{\tbox{cl}}}		
\def\tcol{\tau_{\tbox{col}}}		
\def\terg{t_{\tbox{erg}}}		
\def\tbl{\tau_{\tbox{bl}}}		
\def\theis{t_{\tbox{H}}}		

\def\area{\mathsf{A}_D}			
\def\ve{\nu_{\tbox{E}}}			
\def\vewna{\nu_E^{\tbox{WNA}}}		

\def\dxcqm{\delta x^{\tbox{qm}}_{\tbox c}}	

\newcommand{\rop}{\hat{\mbf{r}}}	
\newcommand{\pop}{\hat{\mbf{p}}}

\newcommand{\sint}{\oint \! d\mbf{s} \,} 
\def\gint{\oint_\Gamma \!\! d\mbf{s} \,} 
\newcommand{\lint}{\oint \! ds \,}	
\def\infint{\int_{-\infty}^{\infty} \!\!}	
\def\dn{\partial_n}				
\def\aswapb{a^*\!{\leftrightarrow}b}		
\def\eps{\varepsilon}				

\def\dhdxt{\partial {\cal H} / \partial x}
\def\dhdx{\pd{\cal H}{x}}
\def\dhdxnm{\left( \pd{\cal H}{x} \right)_{\!nm}}
\def\dhdxnmsq{\left| \left( \pd{\cal H}{x} \right)_{\!nm} \right| ^2}

\def\bcs{\stackrel{\tbox{BCs}}{\longrightarrow}}	

\def\wx{\omega_x}
\def\wy{\omega_y}
\newcommand{\ofro}{({\bf r_0})}
\def\Eb{E_{\rm blue,rms}}
\def\Er{E_{\rm red,rms}}
\def\Es2{E_{0,{\rm sat}}^2}
\def\sb{s_{\rm blue}}
\def\sr{s_{\rm red}}

\def\ie{{\it i.e.\ }}
\def\eg{{\it e.g.\ }}
\newcommand{\etal}{{\it et al.\ }}
\newcommand{\ibid}{{\it ibid.\ }}

\def\gap{\hspace{0.2in}}

%

\newcounter{eqletter}
\def\mathletters{%
\setcounter{eqletter}{0}%
\addtocounter{equation}{1}
\edef\curreqno{\arabic{equation}}
\edef\@currentlabel{\theequation}
\def\theequation{%
\addtocounter{eqletter}{1}\thechapter.\curreqno\alph{eqletter}%
}%
}
\def\endmathletters{\setcounter{equation}{\curreqno}}


\newcommand{\bk}{{\bf k}}
\def\kf{k_{\text F}}
\newcommand{\br}{{\bf r}}
\newcommand{\TL}{{\text{(L)}}}
\newcommand{\TR}{{\text{(R)}}}
\newcommand{\TLR}{{\text{L,R}}}
\newcommand{\VSD}{V_{\text{SD}}}
\newcommand{\GT}{\Gamma_{\text{T}}}
\newcommand{\DEL}{\mbox{\boldmath $\nabla$}}
\def\lf{\lambda_{\text F}}
\def\st{\sigma_{\text T}}
\def\stlr{\sigma_{\text T}^{\text{L$\rightarrow$R}}}
\def\strl{\sigma_{\text T}^{\text{R$\rightarrow$L}}}
\def\aeff{a_{\text{eff}}}
\def\aaeff{A_{\text{eff}}}
\def\gat{G_{\text{atom}}}
\def\arcsec{\hbox{$^{\prime\prime}$}}
\def\sun{\hbox{$\odot$}}
\def\la{\mathrel{\mathchoice {\vcenter{\offinterlineskip\halign{\hfil
$\displaystyle##$\hfil\cr<\cr\sim\cr}}}
{\vcenter{\offinterlineskip\halign{\hfil$\textstyle##$\hfil\cr
<\cr\sim\cr}}}
{\vcenter{\offinterlineskip\halign{\hfil$\scriptstyle##$\hfil\cr
<\cr\sim\cr}}}
{\vcenter{\offinterlineskip\halign{\hfil$\scriptscriptstyle##$\hfil\cr
<\cr\sim\cr}}}}}
\def\ga{\mathrel{\mathchoice {\vcenter{\offinterlineskip\halign{\hfil
$\displaystyle##$\hfil\cr>\cr\sim\cr}}}
{\vcenter{\offinterlineskip\halign{\hfil$\textstyle##$\hfil\cr
>\cr\sim\cr}}}
{\vcenter{\offinterlineskip\halign{\hfil$\scriptstyle##$\hfil\cr
>\cr\sim\cr}}}
{\vcenter{\offinterlineskip\halign{\hfil$\scriptscriptstyle##$\hfil\cr
>\cr\sim\cr}}}}}
\def\degr{\hbox{$^\circ$}}
\def\arcmin{\hbox{$^\prime$}}
\def\arcsec{\hbox{$^{\prime\prime}$}}
\def\utw{\smash{\rlap{\lower5pt\hbox{$\sim$}}}}
\def\udtw{\smash{\rlap{\lower6pt\hbox{$\approx$}}}}
\def\fd{\hbox{$.\!\!^{\rm d}$}}
\def\fh{\hbox{$.\!\!^{\rm h}$}}
\def\fm{\hbox{$.\!\!^{\rm m}$}}
\def\fs{\hbox{$.\!\!^{\rm s}$}}
\def\fdg{\hbox{$.\!\!^\circ$}}
\def\farcm{\hbox{$.\mkern-4mu^\prime$}}
\def\farcs{\hbox{$.\!\!^{\prime\prime}$}}
\def\fp{\hbox{$.\!\!^{\scriptscriptstyle\rm p}$}}
\def\cor{\mathrel{\mathchoice {\hbox{$\widehat=$}}{\hbox{$\widehat=$}}
{\hbox{$\scriptstyle\hat=$}}
{\hbox{$\scriptscriptstyle\hat=$}}}}
\def\sol{\mathrel{\mathchoice {\vcenter{\offinterlineskip\halign{\hfil
$\displaystyle##$\hfil\cr\sim\cr<\cr}}}
{\vcenter{\offinterlineskip\halign{\hfil$\textstyle##$\hfil\cr\sim\cr
<\cr}}}
{\vcenter{\offinterlineskip\halign{\hfil$\scriptstyle##$\hfil\cr\sim\cr
<\cr}}}
{\vcenter{\offinterlineskip\halign{\hfil$\scriptscriptstyle##$\hfil\cr
\sim\cr<\cr}}}}}
\def\sog{\mathrel{\mathchoice {\vcenter{\offinterlineskip\halign{\hfil
$\displaystyle##$\hfil\cr\sim\cr>\cr}}}
{\vcenter{\offinterlineskip\halign{\hfil$\textstyle##$\hfil\cr\sim\cr
>\cr}}}
{\vcenter{\offinterlineskip\halign{\hfil$\scriptstyle##$\hfil\cr
\sim\cr>\cr}}}
{\vcenter{\offinterlineskip\halign{\hfil$\scriptscriptstyle##$\hfil\cr
\sim\cr>\cr}}}}}
\def\lse{\mathrel{\mathchoice {\vcenter{\offinterlineskip\halign{\hfil
$\displaystyle##$\hfil\cr<\cr\simeq\cr}}}
{\vcenter{\offinterlineskip\halign{\hfil$\textstyle##$\hfil\cr
<\cr\simeq\cr}}}
{\vcenter{\offinterlineskip\halign{\hfil$\scriptstyle##$\hfil\cr
<\cr\simeq\cr}}}
{\vcenter{\offinterlineskip\halign{\hfil$\scriptscriptstyle##$\hfil\cr
<\cr\simeq\cr}}}}}
\def\gse{\mathrel{\mathchoice {\vcenter{\offinterlineskip\halign{\hfil
$\displaystyle##$\hfil\cr>\cr\simeq\cr}}}
{\vcenter{\offinterlineskip\halign{\hfil$\textstyle##$\hfil\cr
>\cr\simeq\cr}}}
{\vcenter{\offinterlineskip\halign{\hfil$\scriptstyle##$\hfil\cr
>\cr\simeq\cr}}}
{\vcenter{\offinterlineskip\halign{\hfil$\scriptscriptstyle##$\hfil\cr
>\cr\simeq\cr}}}}}
\def\grole{\mathrel{\mathchoice {\vcenter{\offinterlineskip\halign{\hfil
$\displaystyle##$\hfil\cr>\cr\noalign{\vskip-1.5pt}<\cr}}}
{\vcenter{\offinterlineskip\halign{\hfil$\textstyle##$\hfil\cr
>\cr\noalign{\vskip-1.5pt}<\cr}}}
{\vcenter{\offinterlineskip\halign{\hfil$\scriptstyle##$\hfil\cr
>\cr\noalign{\vskip-1pt}<\cr}}}
{\vcenter{\offinterlineskip\halign{\hfil$\scriptscriptstyle##$\hfil\cr
>\cr\noalign{\vskip-0.5pt}<\cr}}}}}
\def\leogr{\mathrel{\mathchoice {\vcenter{\offinterlineskip\halign{\hfil
$\displaystyle##$\hfil\cr<\cr\noalign{\vskip-1.5pt}>\cr}}}
{\vcenter{\offinterlineskip\halign{\hfil$\textstyle##$\hfil\cr
<\cr\noalign{\vskip-1.5pt}>\cr}}}
{\vcenter{\offinterlineskip\halign{\hfil$\scriptstyle##$\hfil\cr
<\cr\noalign{\vskip-1pt}>\cr}}}
{\vcenter{\offinterlineskip\halign{\hfil$\scriptscriptstyle##$\hfil\cr
<\cr\noalign{\vskip-0.5pt}>\cr}}}}}
\def\loa{\mathrel{\mathchoice {\vcenter{\offinterlineskip\halign{\hfil
$\displaystyle##$\hfil\cr<\cr\approx\cr}}}
{\vcenter{\offinterlineskip\halign{\hfil$\textstyle##$\hfil\cr
<\cr\approx\cr}}}
{\vcenter{\offinterlineskip\halign{\hfil$\scriptstyle##$\hfil\cr
<\cr\approx\cr}}}
{\vcenter{\offinterlineskip\halign{\hfil$\scriptscriptstyle##$\hfil\cr
<\cr\approx\cr}}}}}
\def\goa{\mathrel{\mathchoice {\vcenter{\offinterlineskip\halign{\hfil
$\displaystyle##$\hfil\cr>\cr\approx\cr}}}
{\vcenter{\offinterlineskip\halign{\hfil$\textstyle##$\hfil\cr
>\cr\approx\cr}}}
{\vcenter{\offinterlineskip\halign{\hfil$\scriptstyle##$\hfil\cr
>\cr\approx\cr}}}
{\vcenter{\offinterlineskip\halign{\hfil$\scriptscriptstyle##$\hfil\cr
>\cr\approx\cr}}}}}
\def\diameter{{\ifmmode\mathchoice
{\ooalign{\hfil\hbox{$\displaystyle/$}\hfil\crcr
{\hbox{$\displaystyle\mathchar"20D$}}}}
{\ooalign{\hfil\hbox{$\textstyle/$}\hfil\crcr
{\hbox{$\textstyle\mathchar"20D$}}}}
{\ooalign{\hfil\hbox{$\scriptstyle/$}\hfil\crcr
{\hbox{$\scriptstyle\mathchar"20D$}}}}
{\ooalign{\hfil\hbox{$\scriptscriptstyle/$}\hfil\crcr
{\hbox{$\scriptscriptstyle\mathchar"20D$}}}}
\else{\ooalign{\hfil/\hfil\crcr\mathhexbox20D}}%
\fi}}

\def\getsto{\mathrel{\mathchoice {\vcenter{\offinterlineskip
\halign{\hfil
$\displaystyle##$\hfil\cr\gets\cr\to\cr}}}
{\vcenter{\offinterlineskip\halign{\hfil$\textstyle##$\hfil\cr\gets
\cr\to\cr}}}
{\vcenter{\offinterlineskip\halign{\hfil$\scriptstyle##$\hfil\cr\gets
\cr\to\cr}}}
{\vcenter{\offinterlineskip\halign{\hfil$\scriptscriptstyle##$\hfil\cr
\gets\cr\to\cr}}}}}
\def\lid{\mathrel{\mathchoice {\vcenter{\offinterlineskip\halign{\hfil
$\displaystyle##$\hfil\cr<\cr\noalign{\vskip1.2pt}=\cr}}}
{\vcenter{\offinterlineskip\halign{\hfil$\textstyle##$\hfil\cr<\cr
\noalign{\vskip1.2pt}=\cr}}}
{\vcenter{\offinterlineskip\halign{\hfil$\scriptstyle##$\hfil\cr<\cr
\noalign{\vskip1pt}=\cr}}}
{\vcenter{\offinterlineskip\halign{\hfil$\scriptscriptstyle##$\hfil\cr
<\cr
\noalign{\vskip0.9pt}=\cr}}}}}
\def\gid{\mathrel{\mathchoice {\vcenter{\offinterlineskip\halign{\hfil
$\displaystyle##$\hfil\cr>\cr\noalign{\vskip1.2pt}=\cr}}}
{\vcenter{\offinterlineskip\halign{\hfil$\textstyle##$\hfil\cr>\cr
\noalign{\vskip1.2pt}=\cr}}}
{\vcenter{\offinterlineskip\halign{\hfil$\scriptstyle##$\hfil\cr>\cr
\noalign{\vskip1pt}=\cr}}}
{\vcenter{\offinterlineskip\halign{\hfil$\scriptscriptstyle##$\hfil\cr
>\cr
\noalign{\vskip0.9pt}=\cr}}}}}
\def\bbbr{{\rm I\!R}} 
\def\bbbm{{\rm I\!M}}
\def\bbbn{{\rm I\!N}} 
\def\bbbf{{\rm I\!F}}
\def\bbbh{{\rm I\!H}}
\def\bbbk{{\rm I\!K}}
\def\bbbp{{\rm I\!P}}
\def\bbbone{{\mathchoice {\rm 1\mskip-4mu l} {\rm 1\mskip-4mu l}
{\rm 1\mskip-4.5mu l} {\rm 1\mskip-5mu l}}}
\def\bbbc{{\mathchoice {\setbox0=\hbox{$\displaystyle\rm C$}\hbox{\hbox
to0pt{\kern0.4\wd0\vrule height0.9\ht0\hss}\box0}}
{\setbox0=\hbox{$\textstyle\rm C$}\hbox{\hbox
to0pt{\kern0.4\wd0\vrule height0.9\ht0\hss}\box0}}
{\setbox0=\hbox{$\scriptstyle\rm C$}\hbox{\hbox
to0pt{\kern0.4\wd0\vrule height0.9\ht0\hss}\box0}}
{\setbox0=\hbox{$\scriptscriptstyle\rm C$}\hbox{\hbox
to0pt{\kern0.4\wd0\vrule height0.9\ht0\hss}\box0}}}}
\def\bbbq{{\mathchoice {\setbox0=\hbox{$\displaystyle\rm
Q$}\hbox{\raise
0.05\ht0\hbox to0pt{\kern0.4\wd0\vrule height0.9\ht0\hss}\box0}}
{\setbox0=\hbox{$\textstyle\rm Q$}\hbox{\raise
0.05\ht0\hbox to0pt{\kern0.4\wd0\vrule height0.9\ht0\hss}\box0}}
{\setbox0=\hbox{$\scriptstyle\rm Q$}\hbox{\raise
0.05\ht0\hbox to0pt{\kern0.4\wd0\vrule height0.8\ht0\hss}\box0}}
{\setbox0=\hbox{$\scriptscriptstyle\rm Q$}\hbox{\raise
0.05\ht0\hbox to0pt{\kern0.4\wd0\vrule height0.8\ht0\hss}\box0}}}}
\def\bbbt{{\mathchoice {\setbox0=\hbox{$\displaystyle\rm
T$}\hbox{\hbox to0pt{\kern0.25\wd0\vrule height0.95\ht0\hss}\box0}}
{\setbox0=\hbox{$\textstyle\rm T$}\hbox{\hbox
to0pt{\kern0.25\wd0\vrule height0.95\ht0\hss}\box0}}
{\setbox0=\hbox{$\scriptstyle\rm T$}\hbox{\hbox
to0pt{\kern0.25\wd0\vrule height0.95\ht0\hss}\box0}}
{\setbox0=\hbox{$\scriptscriptstyle\rm T$}\hbox{\hbox
to0pt{\kern0.25\wd0\vrule height0.95\ht0\hss}\box0}}}}
\def\bbbs{{\mathchoice
{\setbox0=\hbox{$\displaystyle\rm S$}\hbox{\raise0.5\ht0\hbox
to0pt{\kern0.38\wd0\vrule height0.45\ht0\hss}\hbox
to0pt{\kern0.52\wd0\vrule height0.5\ht0\hss}\box0}}
{\setbox0=\hbox{$\textstyle \rm S$}\hbox{\raise0.5\ht0\hbox
to0pt{\kern0.38\wd0\vrule height0.45\ht0\hss}\hbox
to0pt{\kern0.52\wd0\vrule height0.5\ht0\hss}\box0}}
{\setbox0=\hbox{$\scriptstyle \rm S$}\hbox{\raise0.5\ht0\hbox
to0pt{\kern0.38\wd0\vrule height0.45\ht0\hss}\raise0.05\ht0\hbox
to0pt{\kern0.52\wd0\vrule height0.45\ht0\hss}\box0}}
{\setbox0=\hbox{$\scriptscriptstyle\rm S$}\hbox{\raise0.5\ht0\hbox
to0pt{\kern0.38\wd0\vrule height0.45\ht0\hss}\raise0.05\ht0\hbox
to0pt{\kern0.52\wd0\vrule height0.45\ht0\hss}\box0}}}}
\def\bbbz{{\mathchoice {\hbox{$\sf\textstyle Z\kern-0.4em Z$}}
{\hbox{$\sf\textstyle Z\kern-0.4em Z$}}
{\hbox{$\sf\scriptstyle Z\kern-0.3em Z$}}
{\hbox{$\sf\scriptscriptstyle Z\kern-0.2em Z$}}}}
\def\ts{\thinspace}

\newcommand{\LB}{Landauer-B\"{u}ttiker}

%
%


\newcommand{\afebase}{${\rm [\alpha/Fe]_b}$}
\newcommand{\zbase}{${\rm Z_b}$}
\newcommand{\afenew}{${\rm [\alpha/Fe]_n}$}
\newcommand{\znew}{${\rm Z_n}$}
\newcommand{\fa}{${\rm f_{\alpha}}$}
\newcommand{\ffe}{${\rm f_{Fe}}$}

%% file: titolo.tex


\title{THE ROLE OF GALAXY FORMATION IN \ \ \ \ \ \ \ \ \ \ \ \ \ \ \ \ \ \ THE STRUCTURE AND DYNAMICS OF \ \ \ \ \
\ \ \ \ \ \ \ \ \ \ \ \ \ \ \ \ \ \ \ \ \ \ \ \ \ \ \ DARK
MATTER HALOS}
\author{Chiara Tonini}
\degreemonth{Trieste, 24$^{\rm th}$ October} 
\degreeyear{2007}
\degree{Doctor Philosophiae in Astrophysics}
\field{Physics}
\department{Physics}
\advisora{Prof. Paolo Salucci} 
\advisorb{Dr. Andrea Lapi}
\advisorc{Prof. Carlos S. Frenk}

\maketitle
\copyrightpage

\newpage
\addcontentsline{toc}{section}{Table of Contents}
\tableofcontents



\begin{citations}

\bigskip 
\bigskip 
\bigskip 

\noindent
The original material this Thesis is made of has been published in the following
papers:

\noindent
Chapter 2:

\begin{quote}
        ``Measuring the spin of spiral galaxies''

        \textbf{Tonini, C.}, Lapi, A., Shankar, F. \& Salucci, P. \ 2006, ApJL, 638,
        L13
\end{quote}
\noindent
Chapter 3: 
\begin{quote}
        ``The universal rotation curve of spiral galaxies II. The dark matter
	distribution out to the virial radius''

	Salucci, P., Lapi, A., \textbf{Tonini, C.}, Gentile, G., Yegorova, I. \& Klein, U. 2007, MNRAS, 378, 41
\end{quote}
\noindent
Chapter 4: 
\begin{quote}
        ``$\Lambda$CDM halo density profiles: where do actual halos converge
        to NFW ones?''

        Gentile G., \textbf{Tonini, C.} \& Salucci P. 2007, A\&A, 467, 925
\end{quote}
\noindent
Chapter 5: 
\begin{quote}
        ``Angular momentum transfer in dark matter halos: erasing the cusp''

        \textbf{Tonini, C.}, Lapi, A., Salucci, P. \ 2006, ApJ, 649, 591
\end{quote}
\noindent
Chapter 6: 
\begin{quote}
        ``Infalling of substructures and dark matter halo evolution''

        \textbf{Tonini, C.} \& Frenk, C.S., in preparation
\end{quote}
\noindent
Appendix A:
\begin{quote}        
        ``Mass modelling from rotation curves''

        \textbf{Tonini, C.} \& Salucci, P. \ 2004, in \textit{Baryons in
        Dark Matter Halos}, ed. R. Dettmar et al. (http://pos.sissa.it), 89
\end{quote}

\end{citations}

\begin{acknowledgments}

\bigskip
\bigskip

\it
\noindent
This Thesis is the result of 4 unforgettable years, and I would like to thank
all the people who shared them with me, making them light and joyful and
exciting. We've come a long way together, and looking back, I'm happy with
what I see... What a ride!

\noindent
To my Advisor Paolo Salucci, you guided me with patience and 
trusted me and allowed me to follow my own way, 
even if it was unorthodox... You taught me the job, and it was fun! 
To Andrea Lapi, thank you for being a friend and a great collaborator. I learned so much
from you, working with you has been 
exciting... yeah, we got there before everybody else!

\noindent
To Carlos Frenk, you gave me a fantastic chance and taught me a lot, in many ways. 

\noindent
Finally, to Frank vdB, you are the funniest and most improbable role model ever!

\noindent
A special place in my heart is for the Tigers, Simona, Irina and Melita, 
your friendship made this time and place special for me, it felt like home.
Simo, we've been together from the very beginning to the very end, I have 
no words to say how much you mean to me. 

\noindent
Irins, thank you for sharing this years with 
me... You know what I'm talking about! 

\noindent
Meliz, meeting you and talking to you
was a joy and a relief... 

\noindent
To Tana delle Tigri, thanks... 
no place at Sissa will ever be like that again.

\noindent
Ale! Fede! Where would I be without you? You've 
been with me always, with the sun and the rain, you've seen the best and the
worst of me... Thank you for being around. 

\noindent
To Fabio, you saved my life. Thank you.

\noindent
To Alberto and Sara, revolution was fun! Keep on rockin' guys, 
see you at another stage!

\noindent
Carlo and Enrica, I miss you always, everytime we meet is special, I'm ready
for the next crazy adventure...

\noindent
Thanks to my brothers-in-Radio Fred and Luca, it's been awesome!! And thanks 
to all of Phasetransition, those were days of inspiration.

\noindent
To my friends forever, my crepe-sister Fede, Puppi, Giada, 
Guido, Alex, Schiab, Skizzo, Michele,  
Marghi, Paolino, Lidia, Reni,
Tom, Novel, Antonio, Paolone, Beppe \& Mara, Dunja, Manu, Max, Beppe M., Laura, 
Luca Tornatore, the Durham guys, Kristen Juan Valeria and Nick, Mauro, Elena, 
Adriano (we share a big big love), I love you all, so much.

\noindent
To Chris and Eddie, thank you for the sunshine. 

\noindent
A special mention to PJ, thank you for the soundtrack.

\newpage

\noindent
To all the friends I won't be seeing around so often any more... Even if we'll be apart, 
scattered across oceans and lands, in weird places full of very weird people, 
all because of that crazy
day when we chose this funny kind of life... You're all in me, 
I know we'll always be together.

\noindent
See you soon, another time, another place.

\noindent
Chiara

\end{acknowledgments}

\dedication

\begin{quote}
\hsp
\em
\raggedleft

To my Family

with Love

\end{quote}

\newpage

\begin{abstract}

\bigskip
\bigskip
\bigskip

\noindent
Galaxy formation is a complex matter. It involves physics on hugely different
scales, forms of matter still unknown, and processes poorly understood. 
The current theory of the formation of structures in the Universe 
predicts the assembly of dark matter halos through hierarchical clustering of
small objects into larger and larger ones under the effect of gravity. 
In the potential wells of halos the baryons collapse to 
form galaxies, and the delicate balance between gravity, dissipative
processes, star formation and feedback shapes the variety of systems we see
today. At the galactic scales, some fundamental theoretical predictions 
fails to show up in the observations, regarding the structure and dynamics of
the dark matter halo.  

My Thesis addresses this problem by taking an evolutionary approach.  
I analysed in detail the many and different observational evidences of a discrepancy
between the predicted halo equilibrium state and the one inferred from the
measurable observables of disk galaxies, as well as of the scaling relations existing
between the angular momentum, geometry and mass distribution of the luminous
and dark components, and realised that they all seem to point towards the same
conclusion: the baryons hosted
inside the halo, by collapsing and assembling to form the galaxy, perturbed
the halo equilibrium structure and made it evolve into new configurations.

From the theoretical point of view, the behaviour of dark matter halos as
collisionless systems of particles makes their equilibrium structure and mass
distribution extremely sensitive to perturbations of their inner dynamics. 
The galaxy formation occurring inside the halos is a tremendous event, and the
dynamical coupling between the baryons and the dark matter during the
protogalaxy collapse represents a perturbation of the halo dynamical structure
large enough to trigger a halo evolution, according to the relative mass and
angular momentum of the two components. 

My conclusion is that the structure and dynamics of dark matter halos, as
well as the origin of the connection between the halo and galaxy properties, 
are to be understood in in terms of a joint evolution of the baryonic and dark
components, originating at the epoch of the collapse and formation of the
galaxy.

\end{abstract}

\newpage

\startarabicpagination


%% file: teoria.tex
\chapter{Introduction: hierarchical clustering in a CDM Universe}

\noindent
In this Chapter I introduce the theory of structure formation, outlying the
background physics and providing a general framework for the original material of this Thesis.
I start with a brief description of the current cosmological model and the 
growth of perturbations in the primordial Universe, to define the paradigm of
hierarchical clustering, in its latest formulation.
I then analyze the equilibrium structure of dark matter halos as collisionless
systems of particles, describing their phase-space properties and the
physical mechanisms that shape them. In particular, I focus on the
hierarchical clustering prediction of a self-similar halo, with a
characteristic phase-space structure, and describe in
detail its properties and the unsolved problems it presents, both from the
theoretical point of view and when compared to
observations. 
I finally introduce the topic of the scaling relations between the dark matter and
the baryonic component hosted by the halos, that represents the starting point
for the models of galaxy formation and the prelude for this Thesis.

I warn the reader that the material contained in this Chapter gives a very general
outline of the theoretical framework without entering in too much detail, and refer her/him to
the provided references for an exhaustive portrait of the topic.

\newpage

\section{The smooth Universe}  

\noindent
The standard cosmological model relies on the basic observational evidence
that the Universe at large scales is homogeneous and isotropic \cite{padma},
as stated by the \textit{cosmological principle}, and confirmed by
data from the Cosmic Microwave Background, and the distributions 
and correlation functions of galaxies and clusters (2dFGRS, \cite{colless}),
while on smaller and smaller scales it features an increasing complexity. 
To understand the present-day pattern of structures we need to consider the 
evolution of the Universe as a whole, as a smooth expanding background, and 
the gravitational growth of primordial inhomogeneities in its matter
components, that decouple from it and develop characteristic properties.     
 
The Universe is not in a stationary state, it evolves with an 
\textit{expansion factor} $a(t)$ that completely characterizes its dynamics.
The only global evolution compatible with homogeneity and isotropy is 
of the kind $ \dot{\textbf{r}}= \textbf{v}(t)=H(t) \ \textbf{r}$, where 
$\textbf{v}$ is the time-variation of the \textit{proper distance} between two points in
space, defined as 
$\textbf{r}=a(t) \ \textbf{x}$, where $\textbf{x}$ is the
so-called \textit{comoving distance} between the two objects and is constant. 
The rate of expansion $H(t)=\dot{a}(t)/a(t)$ is called the 
\textit{Hubble parameter}, and $v(t)=H(t) \ \textbf{r}(t)$ is referred to as the \textit{Hubble law}.  

The dynamics of the evolving Universe is governed by the Friedmann's
equations; the first one is a a continuity 
equation for the density and pressure of any matter-energy
component present in the Universe:
\be
\frac{d(\rho_i a^3)}{da}=-3 a^2 p_i~.
\label{friedmann1}
\ee
The second equation describes the link between the energy densities and the
geometry of space-time:
\be
\left( \frac{\dot{a}}{a} \right)^2 = \frac{8 \pi G}{3} \sum_i \rho_i-\frac{k}{a^2}
\label{friedmann2}
\ee
After the equation of state $w_i=p_i/\rho_i$ has been specified for a given
species, the first equation yields the evolution of the corresponding energy density:
\be
\rho_i(a)=\rho_i(a_0) \ \left( \frac{a_0}{a} \right)^3 \ exp \left[ -3
  \int_{a_0}^{a} \frac{dx}{x} w_i(x) \right]~.
\ee
Once the $\rho_i$ are known, the second Friedmann's equation
(\ref{friedmann2}) determines the expansion factor: 
\be
\left( \frac{\dot{a}}{a} \right)^2 = H_0^2 \sum_i \Omega_i \left(
\frac{a_0}{a} \right)^{3(1+w_i)} -\frac{k}{a^2}~,
\label{second}
\ee
where $H_0=100h \ km \ s^{-1} \ Mpc^{-1}$ is the Hubble parameter today, and
$\Omega_i \equiv \rho_i(a_0)/\rho_c$ is the energy density in terms of the
\textit{critical density} $\rho_c \equiv 3 H_0^2/8\pi G$ that yields a flat
($k=0$) Universe: 
this corresponds to $\Omega \equiv \sum_i \Omega_i =1$. As it turns out (see the
most recent data from WMAP-3, \cite{wmap3}), 
our Universe yields a value of $\Omega$ compatible with flatness
($\Omega\simeq 1.02 \pm 0.02$).
By evalutating Eq.~(\ref{second}) at the present epoch one
obtains $k/a_0^2=H_0^2(\Omega-1)$, which allows to set the conventional values
$a_0=1$ and $\dot{a}_0=H_0$.  

The evolution of the Universe is thus determined by its
species' content at any given time; 
the simplest solutions to the Friedmann's equations are obtained whenever an
energy density component dominates over the others, as often happens at
different epochs; so, in general, 
$a(t) \propto t^{2/[3(1+w)]}$ for $w>-1$, and $a(t) \propto e^{\alpha t}$
with $\alpha=const$ for $w=-1$.  
Moreover, in the current models, the components present in the Universe are each
characterized by a constant $w_i$, so that $\rho_i(a) \propto a^{-3(1+w_i)} \propto
(1+z)^{3(1+w_i)}$. 

The energy density of the Universe is dominated today by an exotic component called  
\textbf{dark energy}, discovered by observing the recession velocity of 
high-redshift Ia supernovae; it accounts for $\Omega_{DE} \sim 0.73$, with 
a negative equation of state, that causes accelerated expansion at present
epochs for any value $w<-1/3$. The nature of this component is unknown; 
according to the equation of state, each model makes different
predictions on the future evolution of the Universe; current models include
a \textit{cosmological constant} $\Lambda$ 
with $w=-1$, a \textit{phantom energy} with $w<-1$, or
a more general \textit{quintessence} with $-1 <w < -1/3$.

The remaining components of the energy density are dominated by matter in
non-relativistic form, of an unknown nature, that we call \textbf{dark matter}; 
its presence was hypotized already in the '30s by Zwicky (\cite{zwicky},\cite{zwicky2}),
and has been confirmed ever since from observations of the dynamics of
galaxies and clusters, and of gravitational lensing. 
In the current scenario of structure formation, it is thought to
consist of particles which interact only through gravity;
candidates that satisfy the requirement and are compatible with the standard
cosmology are neutrinos, axions, the lightest supersymmetric particle,
``jupiters'' and black holes of mass $< 100 M_{\odot}$. These fall under the
common definition of \textbf{cold} dark matter; at early times these particles
are non-relativistic, with mean velocities that are small relative to the 
mean expansion of the Universe (this requirement excludes light neutrinos with
masses $< 30 eV$ \cite{white1}. 
For both the dark matter ($\Omega_{DM} \simeq
0.23$) and the baryons ($\Omega_B \simeq 0.04$), the equation of state is  
$w=0$, and the density is diluted like $\rho \propto a^{-3} \propto (1+z)^3$.

The ultra-relativistic matter, mainly in the form of \textbf{neutrinos}
($\Omega_\nu < 0.015$), and
the \textbf{radiation} ($\Omega_R \simeq 5 \ 10^{-5}$) 
are characterized by $w=1/3$, with a dilution of $\rho \propto a^{-4} \propto
(1+z)^4$. As a consequence, in an expanding Universe the energy density of
radiation decreases more quickly than the volume expansion.

The relevant component in the present discussion is the dark matter (DM); 
in the standard cosmogony, the cosmic structures we see today had their origin
in the early Universe from quantum-generated DM energy density perturbations,
grown by gravitational instability (\cite{padma},\cite{peacock}). 
In the next Section, I will briefly review the theory of the linear growth of the 
primordal perturbations in the case of non-relativistic dark matter, and the
subsequent non-linear evolution leading to hierarchical
clustering. \footnote{A complete description of these processes has to be done
in the General Relativity formalism, even for non-relativistic matter, to
account for the initial stages when perturbations are super-horizon.
However, for the purpose of
introducing the formation of dark matter halos, it is sufficient to follow the
evolution of the DM perturbations after they enter the horizon, when the
Newtonian approximation can be safely used.}

\subsection{Ripples in the pond}

\noindent
While the Universe as a whole can be described as homogeneous and isotropic at
large scales, on smaller scales it becomes progressively more inhomogeneous
and clumpy. 
If the primordial Universe had indeed been totally uniform, it would have
never developed any of the structures seen today; on the contrary, even the slightest 
inhomogeneity would have been dramatically amplified by gravitational instability.
Consider Eq.~(\ref{friedmann2}), after
defining $\rho=(\rho_0 a_0^3)/a^3$ and differentiating with respect to $t$:
\be
\ddot{a}=-\frac{4\pi G \rho_0}{3a^2}=-\left( \frac{2}{9t_0^2} \right) \frac{1}{a^2}
\label{friedmanndiff}
\ee
If we perturb $a(t)$ slightly and have $a(t)+\delta a(t)$, such that the
corresponding fractional density perturbation relative to the smooth
background $\bar{\rho}$ is 
$\delta \equiv (\rho-\bar{\rho})/\bar{\rho}=-3(\delta a/a)$, we find that
$\delta a$ satisfies the equation 
\be
\frac{d^2}{dt^2} \delta a = \left( \frac{4}{9t_0^2} \right) \frac{\delta
  a}{a^3}=\frac{4}{9} \frac{\delta a}{t^2}~.
\label{deltaa}
\ee
The growing solution to this equation is $\delta a \propto t^{4/3} \propto
a^2$. Hence the density perturbation, parameterized by its  
\textit{density contrast} $\delta$, grows as $\delta \propto a$.

\section{The hierarchical clustering}

\noindent
The current paradigm for structure formation finds its roots in the 
pioneering work by Peebles 
(\cite{peeblesdicke},\cite{peebles70},\cite{peebles73},\cite{peebles74},\cite{peebles80}),
who established the \textbf{hierarchical clustering} or \textbf{isothermal} 
theory. In this scenario, structure builds up through the aggregation of
nonlinear objects into larger and larger units. The original formulation of
the model predates the Cold Dark Matter (CDM) version; it describes the
clustering of baryonic structures, but encounters a
number of difficulties and is soon abandoned. In the CDM model 
(\cite{peebles82},\cite{blumenthal2},\cite{davis},\cite{bardeen}) the build-up of
structures is governed by the dark dissipationless component, that evolves under gravity from
an initially gaussian distribution of primordial perturbations; small
fluctuations first, and then larger and larger ones, become nonlinear and
collapse when self-gravity dominates their dynamics, to form virialised, gravitationally
bound systems. As larger perturbations collapse, the smaller objects embedded
in them cluster to form more complex patterns. 
In the meanwhile, the dark matter provides the potential wells within which the
gas cools and forms galaxies under dissipative collapse \cite{padma}.

The success of the CDM model stems from a variety of sources. First of all,
most of the mass in the Universe does appear to be in the DM form; second, if
this matter has interacted only through gravity since early times, it is
possible to reconcile the very small observed amplitude of fluctuations in the
Cosmic Microwave Background with the massive nonlinear structures in the
present Universe. Moreover, the large-scale distribution of galaxies is
consistent with the patterns resulting from gravitational amplification of
gaussian density fluctuations. This is a simple and natural condition in CDM
models, where the galaxies are indeed expected to trace the DM
distribution on large scales \cite{white3}.  

In a radiation-dominated Universe, there are two independent perturbation modes of the coupled
radiation-gas mixture for which the density contrast is non-decreasing in
time. The isothermal mode describes fluctuations in the photon-to-baryon rate,
while the radiation temperature is almost uniform. The other mode, called
adiabatic, describes fluctuations in the radiation temperature, with a
constant photon-to-baryon rate (and was championed by Zel'dovich and
collaborators \cite{white1}). 

If the models are refined to accomodate dark matter, the two modes lead to very
different scenarios. The adiabatic mode is the one that grows faster, and it
leads to the formation of bigger structures compared to the isothermal; in
this scenario, cluster-size objects form first, and galaxies are originated
from fragmentation of such structures. This is inconsistent with the
observational evidence that galaxies are older than larger-scale objects
\cite{white2}. Moreover, the natural dark matter candidate for such a scenario
is hot, in the form of neutrinos. Numerical simulations showed that this kind
of dark matter is unable to reproduce the observed distribution of
galaxies and clusters \cite{white1}. In the isothermal scenario objects of
galactic scale form by aggregation and merging, while large-scale structures
are essentially random and have little influence on galaxy properties. 

The CDM model, although taking its first steps from the isothermal picture, features
a power spectrum of the fluctuations significantly redder than white noise
(the power density at galactic scales is well below that on cluster scales);
as a result, collapse on galactic scales occurs more recently in the CDM than
in the old isothermal picture. Compared with the old scenario, 
the CDM cosmogony is much less clearly hierarchical, 
with the interesting consequence that 
the galaxy formation inside a dark matter halo is not really an
independent event with respect to the formation of the halo itself, and of bigger
and smaller structures. Protogalactic collapse is neither the falling together
of a single smooth perturbation nor the merging of a set of well equilibrated
precursor objects, but lies somewhere between the two. In addition, while in
general galaxies form before the larger structures in which they are embedded
(like clusters), the temporal separation of the two processes is not enough
for them to be independent. As a result, \textbf{biases} arise in the galaxy
population, meaning that the properties of galaxies strongly depend on the
large-scale environment surrounding them \cite{white1}.

The most convenient way to describe the growth of structures is to use the
Fourier transform $\delta_{\textbf{k}}(t)$ of the density contrast, treated as a
realisation of a random process. The \textbf{power spectrum} of the fluctuations   
at a given wavenumber $k$ is $P(k,t) \equiv \langle |\delta_{\textbf{k}}(t)|^2 \rangle$,
defined as a statistical quantity averaged over an ensemble of possibilities;
the isotropy of the Universe implies that the power spectrum depends only on
the magnitude $|\textbf{k}|$ of the wavenumber \cite{padma}.

At early times, the DM density fluctuation field is represented as a
superposition of plane waves $\delta(x)= \int \delta_k e^{i(\vec{k}\cdot \vec{x})} d^3k$;
the Fourier modes are independent, with phases randomly distributed on
($0,2\pi$]. The distribution of $\delta(x)$ at any arbitrary
set of positions ($x_1, x_2, ...$) is a multivariate gaussian.
Such fields
are known as \textit{gaussian random fields} and are predicted by a wide class of
theories for the origin of structure in the Universe \cite{white1}.  
In this picture, $P(k,t)$ gives the complete statistical description of the initial
conditions and the subsequent evolution of the density field, down to nonlinearity.

The power per logarithmic band in $k$ is given by 
\be
\Delta_k^2(t)=\frac{k^3 |\delta_k(t)|^2}{2\pi^2}=\frac{k^3 P(k,t)}{2\pi^2}~.
\ee
Observations show the power spectrum to be
fairly smooth \cite{padma}, and hence it can be approximated by a power law in
$k$ locally at any given time.
Of interest for the $\Lambda$CDM model, the CMB data at very large scales
are well fitted with a simple power-law like
\be
P(k)=P(K_0)\left( \frac{k}{k_0} \right)^{n_s-1}
\ee 
with spectral index $n_s \approx 0.95$, indicating an almost
scale-free power spectrum. However, inflationary models favour a running
spectral index, $n_s=n_s(k)=d \textrm{ln} P / d \textrm{ln} k $, so that
\be
P(k)=P(K_0)\left( \frac{k}{k_0} \right)^{n_s(k_0)+ln(k/k_0)d n_s/d ln (k)}~;
\ee 
the debate is still open about the best-fit to the newest data (\cite{wmap1},\cite{wmap3}).
Notice that the requirement of hierarchical clustering, that small objects form first, is
ensured if $P(k,t)$ is a decreasing function of mass, or correspondingly,
an increasing function of the spatial wavenumber $k$.

The amplitude of the fluctuations at a given scale $M$ is related to the mean
square fluctuation $\sigma$, defined as
\be
\sigma^2_M=\left( \frac{\delta M}{M} \right)^2 = \frac{1}{2\pi^2}
\int_0^{k_{max}=1/R} k^2 P(k) dk~,
\ee
where the mass scale $M$ is associated to the linear scale $R$ by $M=(4\pi/3)\bar{\rho}(t_0)R^3$.
Notice that 
\be
\sigma^2_M=\left( \frac{\delta M}{M} \right)^2 = \frac{1}{2\pi^2}
\int_0^{k_{max}=1/R} k^3 P(k) d\textrm{ln}k=\int_0^{k_{max}} \Delta_k^2 d\textrm{ln}k ~,
\ee
so that the power at scale $k$, $\Delta_k^2(t)$, can be viewed as the contribution to the mean square
fluctuation in the logarithmic band $k$, or at the scale $R \approx k^{-1}$:
\be
\Delta_k^2(t)=\left( \frac{\delta \rho}{\rho} \right)^2_{R\simeq k^{-1}} =
\left( \frac{\delta M}{M} \right)^2_{R\simeq k^{-1}} \cong \sigma^2(R,t)~.
\ee
Given the power spectrum defined above, the mean square fluctuation is related
to the linear scale $R$ by  
\be
\sigma^2_R \propto k^{3+n_s-1} \propto R^{-(3+n_s-1)}~.
\ee
The normalization of the power spectrum is usually obtained through the value
of $\sigma_8\equiv \sigma(8 \ Mpc \ h^{-1})$; in this work, I take $\sigma_8 =0.8$,
an average value between the first- and third-year WMAP data (\cite{wmap1},\cite{wmap3}).

At any given time, the power spectrum $P(k,t)$ completely characterizes the
pattern of density fluctuations.
The gravitational potential generated by a perturbation $\delta$ in a region of size
$R$ is $\phi \propto \delta M/R \propto \bar{\rho} \delta R^2$. In an
expanding Universe $\bar{\rho} \propto a^{-3}$ and $R \propto a$, while the
perturbation grows as $\delta \propto a$, making $\phi$ constant in time. This
implies that the perturbations that were present in the Universe at the time
when radiation decoupled from matter would have left their imprint on the
radiation field. Photons climbing out of a potential well lose energy and are redshifted 
by an amount $\Delta \nu / \nu \approx \phi/c^2$, hence we expect to see a
temperature fluctuation in the microwave radiation corresponding to 
$\Delta T/T \approx \Delta \nu / \nu \approx \phi/c^2$.
The largest potential wells would have left their imprint on the CMB at the
time of decoupling; galaxy clusters constitute the deepest wells in the
Universe, with escape velocities of the order $v_{esc} \approx \sqrt{GM/R}
\approx 10^3 km/s$, that leads to $\Delta T/T \approx (v_{esc}/c)^2 \approx 10^{-5}$.
The proof that the theory was sound was indeed found in the microwave
background radiation, when such temperature fluctuations were
observed (COBE \cite{cobe}, \cite{padma}).

When the self-gravity of the growing perturbation dominates over the
expansion, the density contrast reaches some critical value 
$\delta (R,t)\rightarrow \delta_c \approx 1$ 
and the evolution goes nonlinear; matter
collapses to form a bound structure on the scale $R=R_{nl}$, that  
at any time $t$ obeys the relation
\be 
R_{nl}(t) \propto a(t)^{2/(n+3)} = R_{nl}(t_0)(1+z)^{-2/(n+3)}~;
\ee
in other words, structure of mass $M \propto R_{nl}^3$ form at a redshift
$z$ where
\be
M_{nl}(z)=M_{nl}(t_0)(1+z)^{-6/(n+3)}~.
\ee 
Once formed and virialised, such gravitationally bound structures remain frozen at a
mean density defined as 
\be
\rho_{halo} \simeq \Omega \rho_c \Delta_{vir} (1+z)^3~,
\ee
where $\Delta_{vir}$ is the density contrast at the redshift $z$ of formation,
and $\rho_{back}(z)=\rho_c \Omega (1+z)^3$ is the background density of the
Universe at $z$, in terms of the critical density 
$\rho_c=2.8\times 10^{11}\,h^2\,M_{\odot}\, \mathrm{Mpc}^{-3}$. 
Given a halo of mass $M_{vir}$, the virial radius is determined as 
\be
R_{vir} = [3\, M_{vir}\,\Omega_M^z/  4\pi\,\rho_c\,\Omega_M\, (1+z)^3\,\Delta_{vir}]^{1/3}~;
\ee
at any given redshift $z$, the density parameter is 
$\Omega_M^{z}=\Omega_M\,(1+z)^3/[(1-\Omega_M)+\Omega_M\,(1+z)^3]$ in terms of
today's value $\Omega_M$, and the density contrast is
$\Delta_{vir}=18\,\pi^2+82\,(\Omega_M^z-1)-39\,(\Omega_M^z-1)^2$.
The cosmological parameters adopted throughout this work are  
consistent with WMAP-3 \cite{wmap3}; in particular, I consider
a flat ($k=0$) Universe with Hubble parameter
$h=H_0/100~\mathrm{km~s}^{-1}~\mathrm{Mpc}^{-1}=0.71$, matter density 
$\Omega_M\approx 0.27$ and $\Delta_{vir}\approx 100$ at $z=0$. 

After defining the circular velocity associated with the mass $M_{vir}$ as 
\be
v_c^2 \equiv \frac{GM_{vir}}{r} \equiv \frac{4\pi G}{3}\rho_{halo}r^2~, 
\ee
the redshift of formation of the halo can be written as
\be
(1+z) \simeq 5.8 \left( \frac{200}{\Omega \Delta_{vir}} \right)^{1/3} \frac{(v_c/200 \
  km/s)^{2/3}}{(r/h^{-1}\ Mpc)^{2/3}}~.
\ee 
Notice the manifestation of the hierarchical clustering; smaller scales are
the first to go nonlinear and virialise, forming objects of higher mean density;
while larger perturbations grow and collapse, the small fluctuations on top of
them cluster and merge, building up the hierarchy.

\section{The equilibrium structure of dark matter halos}

\noindent
The virialised dark matter objects originated from the gravitational growth of
perturbations and the subsequent evolution through hierarchical clustering are
called \textbf{halos}. In the state-of-the-art version of the theory, 
they are collisionless systems of particles; the particles move under the influence of
the \textit{mean} potential generated by the whole mass distribution, while
the two-body encounters are negligible. 

A system made of a large number of particles moving under the influence of a 
smooth potential is fully described in phase-space, at any time $t$, by 
specifying the number of particles $f(\textbf{x},\textbf{v},t)d^3 \textbf{x}
d^3\textbf{v}$ having positions in the
volume $d^3 \textbf{x}$ centered on $\textbf{x}$ and velocities in the volume
$d^3\textbf{v}$ centered on $\textbf{v}$. The quantity
$f(\textbf{x},\textbf{v},t)$ is a probability density, called
\textbf{phase-space distribution function} (DF) and is always $f \ge 0$ for physical
systems \cite{BT}.
Given the particles' initial coordinates and velocities, their phase-space
trajectories are determined by Newton's law; in other words, 
if $f(\textbf{x},\textbf{v},t_0)$ is known at some time $t_0$, the
information it contains is sufficient to evaluate $f(\textbf{x},\textbf{v},t)$
at any later time $t$. The DF thus fully specifies the evolution of the system.

Consider a system of particles moving along their orbits under a gravitational
potential \textbf{$\Phi$},
 giving rise to a flow of points in phase space; if we write the phase-space coordinates as 

\be
(\textbf{x},\textbf{v}) \equiv \textbf{w} \equiv (w_1,...,w_6)~,
\label{coordinates}
\ee
then the velocity of the flow is
\be
\dot{\textbf{w}} = (\dot{\textbf{x}},\dot{\textbf{v}}) = (\textbf{v}, -\textbf{$\nabla$}\textbf{$\Phi$})~.
\label{flow}
\ee
The 6-dimensional vector $\dot{\textbf{w}}$ is a generalised velocity that
bears the same relationship to $\textbf{w}$ as the 3-dimensional velocity
$\textbf{u}=\dot{\textbf{x}}$ does to $\textbf{x}$.

The flow described by Eq.~(\ref{flow}) conserves the particles; in the absence
of collisions, particles do not jump from one point of phase-space to another,
but follow smooth trajectories. Hence, $f(\textbf{w},t)$ satisfies a
continuity equation, like the matter density of an ordinary fluid \cite{BT}:
\be
\frac{\partial f}{\partial t} + \sum_{\alpha=1}^6 \frac{\partial (f
  \dot{w}_\alpha)}{\partial w_\alpha} = 0~.
\label{continuity}
\ee 
If integrated over volume, this equation shows that there is a balance between
the rate at which particles enter and exit a given phase-space volume. 

It is easy to see that the flow described by $\dot{\textbf{w}}$ has the special property that
\be
\sum_{\alpha=1}^6 \frac{\partial \dot{w}_\alpha}{\partial w_\alpha} =
\sum_{i=1}^3 \left( \frac{\partial v_i}{\partial x_i} + \frac{\partial
  \dot{v}_i}{\partial v} \right) = \sum_{i=1}^3 -\frac{\partial}{\partial v_i}
\left( \frac{\partial \textbf{$\Phi$}}{ \partial x_i} \right) = 0~;
\label{zeroflow}
\ee
notice that $(\partial v_i / \partial x_i) = 0$ between independent
coordinates in phase space, and the last equality comes from the fact that
\textbf{$\nabla \Phi$} does not depend on the velocities.
Combining (\ref{continuity}) and (\ref{zeroflow}) leads to the 
\textbf{collisionless Boltzmann equation}:
\be
\frac{\partial f}{\partial t} + \sum_{\alpha=1}^6 \dot{w}_\alpha 
    \frac{\partial f}{\partial w_\alpha} = \frac{\partial f}{\partial t} 
    + \sum_{i=1}^3 \left(v_i \frac{\partial f}{\partial x_i} - 
    \frac{\partial \Phi}{\partial x_i} \frac{\partial f}{\partial v_i}\right) = 0~,
\label{B1}
\ee
equivalently written as
\be
\frac{\partial f}{\partial t} + \textbf{v} \cdot \textbf{$\nabla$} f -\textbf{$\nabla \Phi$} \cdot
\frac{\partial f}{\partial \textbf{v}} = 0~,
\label{Bvector}
\ee
or simply
\be
\frac{\mathrm{d} f}{\mathrm{d} t} = 0~.
\ee
This is the fundamental equation governing the dynamics of a collisionless system, and states that
the flow of points in phase space is incompressible; the DF 
around the phase point of any given particle moving through phase space is constant.
This does not hold if encounters are not negligible \cite{BT}.

The DF is conserved in the motion through phase-space, therefore it represents a
steady-state solution of the Boltzmann equation. This allows to introduce the
concept of \textbf{integral of motion} in the potential \textbf{$\Phi$}(\textbf{x}).
By definition, a function of the phase-space coordinates
$\textbf{I}(\textbf{x},\textbf{v})$ is an integral if and only if
\be
\frac{d}{dt} I([\textbf{x}(t),\textbf{v}(t)]=0
\label{intmot}
\ee
along all orbits. With the equations of motion this becomes
\be
\frac{dI}{dt}=\nabla I \cdot \frac{d \textbf{x}}{dt} + \frac{\partial
  I}{\partial \textbf{v}} \cdot \frac{d \textbf{v}}{dt}=0~, \ \ \ \textrm{or}
\ \ \ \textbf{v} \cdot \nabla I - \nabla \Phi \cdot \frac{\partial I}{\partial
\textbf{v}}=0~.
\ee
But this is exactly Equation (\ref{Bvector}), and this leads to the
formulation of the \textbf{Jeans theorem}:
\noindent
\textit{any steady-state solution of the collisionless Boltzmann equation
  depends on the phase-space coordinates only through integrals of motion in
  the potential $\Phi$, and any function of the integrals yields a steady
  state solution of the collisionless Boltzmann equation}. 
A proof of the theorem can be found in \cite{BT}.
For systems of interest in the present work, the potential is regular (see
  the same reference); in this case, the \textbf{Strong Jeans theorem} holds:
\noindent
\textit{the DF of a steady-state system in which almost all orbits are regular
with incommensurable frequencies may be presumed to be a function only of
three independent isolating integrals}.

The Jeans theorem assures that the DF of a system can always be expressed in terms
of the integrals of motions; as will be clear in Chapter 5, this is
particularly useful because, by this property, the DF determines the symmetry of the system.
Consider a steady-state spherical system; a simple extension of the strong Jeans theorem 
\cite{lybell} states that the integrals of motion 
that describe it are the energy $E$, the modulus of the total angular momentum
$L^2$ and its $z$-component $L_z$ \cite{BT}.
If the system is nonrotating, the DF will
be a constant in $L_z$, so that the explicit dependence on $L_z$ will
disappear; if the system is also symmetric in the components of the velocity
dispersion, the explicit dependence on $L^2$ will also disappear.    
In this sense, the DF gives information about the symmetries governing the
system's evolution. Any perturbation acting from outside, that
causes a symmetry breaking in the velocity dispersion tensor, alters the
system's evolution and transforms the DF describing its equilibrium.

The quantity $f$ is unfortunately not accessible to direct measurement in 
real systems. In fact, one can count the particles in any given volume 
$d^3 \textbf{x} \ d^3 \textbf{v}$,
but as the latter shrinks and contains only a few particles, 
the result will radically depend on
the particular volume chosen, with wild fluctuations that invalidate the reliability of the result.  
Only by treating $f$ as a probability density, it correctly describes the
properties of the system; in analogy with the wave function in a quantum dynamics
system, it is not measurable in itself, but the physical information is
contained in the expectation values of some phase-space functions
$Q(\textbf{x},\textbf{v})$ obtained through $f$. 
As an example, given a dark matter halo characterised by some $f(r,v)$, the
spatial density profile (in real space) is given by:
\be
\rho (r) = \int_0^{\infty} f(r,v) d^3v~, 
\label{densityfromf}
\ee 
and the mean velocity of the particles in the given volume is:
\be
\overline{v}_i \equiv \frac{1}{\rho} \int f v_i d^3v~; 
\ee
in general, all the macroscopic observables of the system will be average quantities
over the DF:
\be
\langle O \rangle = \frac{ \int O \ f(r,v) d^3v}{\int f(r,v) d^3v}~.
\ee
If one wants to estimate the DF in some point (\textbf{x},\textbf{v}), 
the best approximation to the actual one is the average of $f$ over a small
volume centered on the point. This quantity is called the \textbf{coarse-grained}
distribution function $\bar{f}$, and is complicated and tricky to use, because
it does not satisfy the Boltzmann equation. 
However, it can be measured in numerical simulations, and as I will explain
later on, it can be related to the entropy of the system. 

In general, the Boltzmann equation itself is of little practical
use since, given the high number of variables of $f$, the complete solution is 
difficult to obtain. However, valuable information on the system is 
contained in the equation's moments.
By integrating Eq. (\ref{B1}) over the velocity, one obtains a continuity equation,
\be
\frac{\partial \rho}{\partial t}+ \frac{\partial (\rho \overline{v}_i)}{\partial
  x_i}=0~,
\label{cont}
\ee
after considering that $x_i,v_i,t$ are independent variables, that
$f(x,v)=0$ for sufficiently large velocities, and after applying the divergence
theorem. 
By multiplying Eq. (\ref{B1}) by $v_j$ and integrating over velocity again,
one obtains
\be
\frac{\partial (\rho \overline{v}_j)}{\partial t} + \frac{\partial (\rho \overline{v_i
    v_j})}{\partial x_i} +\rho \frac{\partial \Phi}{\partial x_j} =0~,
\label{cont2}
\ee
where 
\be 
\overline{v_i v_j} \equiv \frac{1}{\rho} \int v_i v_j f d^3v~.
\label{doppiamedia}
\ee
The mean value of $v_i v_j$ can be decomposed in a contribution from the
streaming motions $\overline{v}_i \overline{v}_j$ and one from the velocity dispersion:
\be
\sigma^2_{ij} \equiv \overline{(v_i-\overline{v}_i)(v_j-\overline{v}_j)} = \overline{v_i v_j} -
\overline{v}_i \overline{v}_j~;
\label{sigma}
\ee
by subtracting $v_j$ times Eq. (\ref{cont}) from Eq. (\ref{cont2}), and making
use of Eq. (\ref{sigma}), we obtain the analog of Euler's equation of the flow
of a fluid, for our system of collisionless particles:
\be
\rho \frac{\partial \overline{v}_j}{\partial t} + \rho \overline{v}_i \frac{\partial
  \overline{v}_j}{\partial x_i} = -\rho \frac{\partial \Phi}{\partial x_j}
-\frac{\partial (\rho \sigma^2_{ij})}{\partial x_i}~.
\label{euler}
\ee
The last term of the r.h.s. acts like a pressure force
$-\nabla p$, where $-\rho \sigma^2_{ij}$ is a stress tensor that describes a generic
anisotropic pressure. The set of equations (\ref{cont}, \ref{cont2},
\ref{euler}) are known as the \textbf{Jeans equations}.
The valuability of these equations (and in particular of Eq.~\ref{euler}) is
in their relating observationally accessible quantities that satisfy
the Boltzmann equation, like the system's streaming
velocity, velocity dispersion, density profile, 
without actually solving the Boltzmann equation
itself. As an example, given a system with known density profile, the Jeans'
equations yield the velocity dispersion profile compatible with equilibrium.
Notice that this process in general allows for more than one solution, 
since we lack the analogous of
the equation of state of a fluid, that would link the components of
the tensor $\textbf{$\sigma$}^2$ to the density; in practice, we overcome the
difficulty by making some assumption about the symmetry of
$\textbf{$\sigma$}^2$, which is a valid semplification if the assumption itself
is physically sound \cite{BT}.   

A particularly useful application of the Jeans's equations in the present work
is the study of spherically symmetric systems in steady states; after defining the 
spherical coordinates 
(\textbf{x},\textbf{v}) $\equiv$ ($r, \theta, \phi, v_r, v_\theta, v_\phi$), 
we have that 
$\overline{v}_r=\overline{v}_\theta=\overline{v}_\phi=0$; in this case,
$\overline{v^2_i}=\sigma^2_i$, and  
\be
\frac{d(\rho \sigma^2_r)}{dr}+\frac{\rho}{r} \left[ 2\sigma^2_r
  -\left(\sigma^2_\theta +  \sigma^2_\phi \right) \right] = -\rho \frac{d\Phi}{dr}~.
\label{jsfera}
\ee
Consider a dark matter halo, such that both the density and the velocity
structure are invariant under rotations about the center of mass; hence, the
halo itself does not rotate, and 
\be
\sigma^2_\theta = \sigma^2_\phi~.
\ee
It follows that the velocity ellipsoids are spheroids with their symmetry axes
pointing to the center of the system. The \textbf{anisotropy parameter} is
defined as
\be
\beta \equiv 1-\frac{\sigma^2_\theta}{\sigma^2_r}
\label{betadef}
\ee
and describes the degree of anisotropy of the velocity distribution at each
point. With these assumtpions, Eq.~(\ref{jsfera}) becomes
\be
\frac{1}{\rho} \frac{d(\rho \sigma^2_r)}{dr} + 2 \frac{\beta \sigma^2_r}{r}
= -\frac{d\Phi}{dr}~.
\label{jsfera2}
\ee
If we can measure the quantities $\rho$, $\beta$ and $\sigma^2_r$ as 
functions of radius, then Eq.~(\ref{jsfera2}) allows to determine the mass
profile and the circular velocity of the system, through
\be
v^2_c=\frac{GM(r)}{r} = -\sigma^2_r \left( \frac{d \ ln \rho}{ d \ ln \ r
}+\frac{d \ ln \sigma^2_r}{ d \ ln \ r } +2 \beta \right)~.
\ee

\subsection{The NFW: entropy stratification and self-similarity}

\noindent
A real theoretical understanding of the equilibrium structure of dark matter
halos, or of the processes that 
lead to it, is still lacking in the general
picture of structure formation \cite{white2}. 
However, in the Seventies numerical simulations were developed and used 
to understand the mechanisms of gravitational clustering, and the
evolving quality of the codes, together with the increasing resolution and
computational power, made them the preferred tool to study the formation of
Cold Dark Matter (CDM) halos.
The success of numerical simulations in reproducing the observed
dynamical properties of galaxies and larger systems depends on the scale
investigated, and there is no agreement about the actual shape of DM halos and
the mass distribution of substructures,
due to inconsistencies between the results of simulations and observations
(this topic will be extensively addressed in the next Chapters); however, simulations 
indeed reproduce well the mechanism of hierarchical clustering, and the latter
enjoys a much broader consensus in being the actual process responsible for structure
formation. \footnote{There is still, however, much debate regarding this
  issue, and models describing other scenarios, such as the monolithic
  collapse of structures, are not ruled out yet.} 
Having given simulations the credit they deserve, it is worth to investigate
some features of the simulated halos that may seem quite surprising. 

\begin{figure}
\epsfig{file=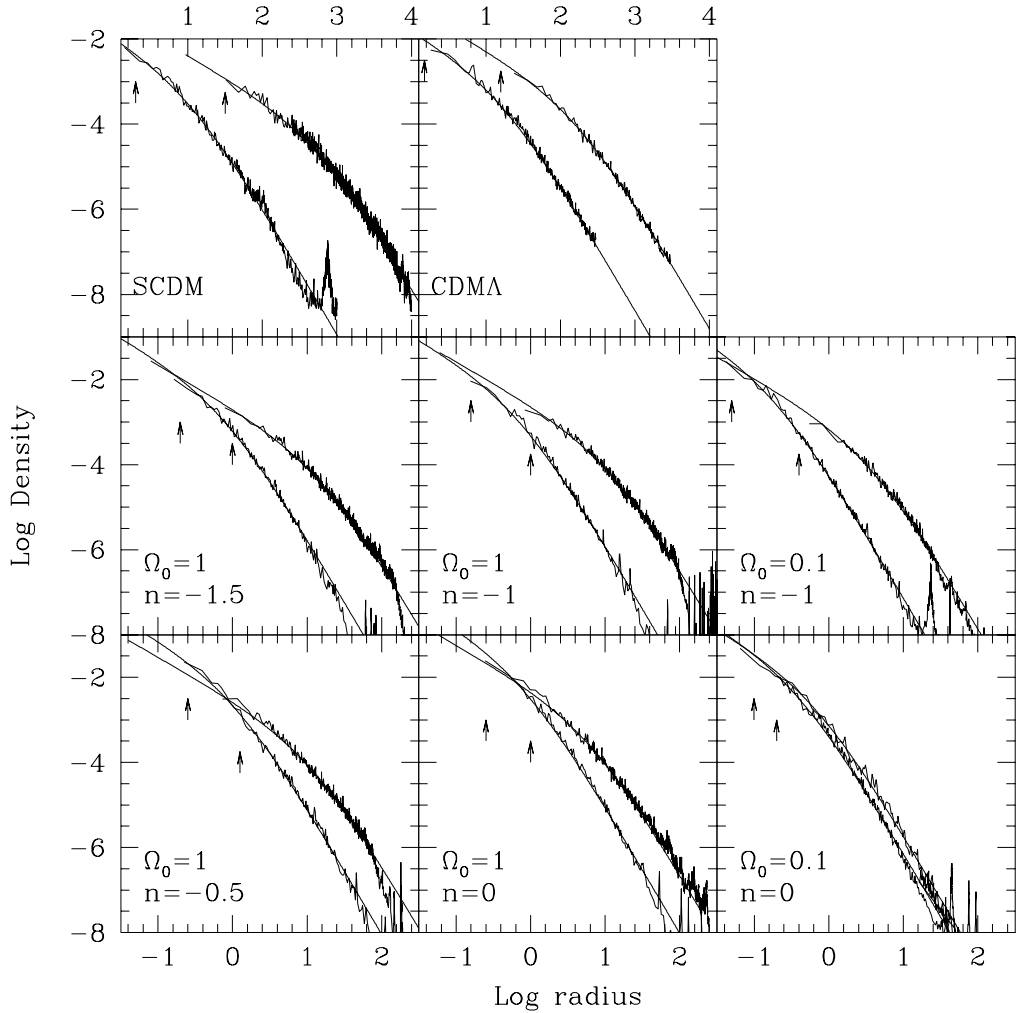,scale=0.7}
\caption{Density profiles of simulated halos in different cosmologies
  \cite{nfw97}. In each panel, the lower-mass halo is represented by the
  leftmost curve; the \textit{solid smooth curve} is the NFW fit. \textit{Left
  panels}: Standard CDM model ($\Lambda=0$). \textit{Right panels}:
  $\Lambda$CDM model. In each panel the varying cosmological parameters are
  specified. Radii are in kiloparsecs (\textit{scale at top}); 
  the arrows indicate the softenign length in each simulation.}
\label{fignfw}
\end{figure}

The most evident property of halos born through hierarchical clustering
is the self-similarity: no matter the
mass scale, they all belong to a one-parameter family of curves, known as the
Navarro, Frenk \& White (NFW, \cite{nfw96},\cite{nfw97}) profile
\be
\rho(x) = {M_{vir}\over 4\pi R_{vir}^3}\, {c^2\, g(c)\over x\, (1+cx)^2}~;
\label{eq:rho_nfw}
\ee
here $M_{vir}$ and $R_{vir}$ are the virial mass and radius, 
$x$ is the normalized radial coordinate $x \equiv r/R_{vir}$, 
$c$ is called the \textit{concentration parameter}, and $g(c)\equiv
[\ln{(1+c)}-c/(1+c)]^{-1}$. 
The NFW is the spherically averaged fit to the density profiles obtained in simulations,
and it is interesting to see that it holds for any Einstein-de Sitter Universe
\cite{cole}, rather than being characteristic of the particular cosmology of the
$\Lambda$CDM model. Fig.~(\ref{fignfw}) is taken from \cite{nfw97} and shows the NFW fits to halos
of different masses in different cosmologies (Standard CDM and $\Lambda$CDM,
with different density parameters and spectral indexes), highlighting the NFW
flexibility in reproducing simulated systems in different universes. 

Although in principle the NFW is a two-parameters family of curves (namely
$M_{vir}$ and $c$), from statistical analysis of the simulated halos it turns
out that there is a mild anticorrelation between the concentration and the halo
mass, that was originally parameterized by \cite{nfw97} (see also
\cite{bullock},\cite{momao}); 
I obtained the relation
\be
c=9.5 \left( \frac{M_{vir}}{10^{12}M_{\odot}} \right)^{-0.13}
\label{bullock}
\ee
at $z=0$, after re-evaluating it with the cosmological parameters used in this work
(see for comparison \cite{gnedin} and \cite{dutton06}). 
As it turns out, the concentration $c$ increases with the
redshift of formation while decreasing with the halo mass, thus fulfilling the
hierarchical clustering requirements. 

Regarding the velocity distribution structure, although it can vary from
particular simulation to simulation, there is general agreement now that the
NFW $\sigma$ profile is isotropic in the centre of the halo, and becomes
radially anisotropic moving outwards, mirroring the hierarchical mass
accretion (\cite{cole},\cite{thomas},\cite{huss}).

The NFW result has been confirmed by a number of subsequent studies
(see for instance \cite{cole},\cite{fuku},\cite{metal98},\cite{huss},\cite{jingsuto}), 
although there is some disagreement 
regarding the innermost value of the logarithmic slope  $\gamma$. NFW argued that a
fitting formula where $\gamma=(1+3y)/(1+y)$ (where $y=r/r_s$ is the radial coordinate in
units of a suitably defined scale-radius $r_s$) provides a very good fit to
the density profiles of simulated halos over two decades in radius. 
Some authors (see \cite{metal98},\cite{ghigna},\cite{fuku2}) have argued that
$\gamma$ converges to a value of $\sim -1.5$ near the center,
rather than $-1$ as expected from the NFW fit. Others \cite{andrej} initially
obtained much shallower inner slopes ($\gamma \sim -0.7$) in their numerical
simulations, but have now revised their conclusions; these authors now argue
that CDM halos have steeply divergent density profiles but, depending on
evolutionary details, the slope of a galaxy-sized halo at the innermost resolved
radius may vary between $-1.0$ and $-1.5$. 

From the theoretical point of view, a number of plausible arguments have been
advanced in order to try and explain the innermost behaviour of dark matter
density profiles from collisionless dynamics' principles. These efforts, however, tend
to give non-unique results and have so far been unable to explain the remarkable
similarity in the structure of dark matter halos of widely different mass formed
in a variety of cosmogonies (\cite{evans2},\cite{syer},\cite{nuss},\cite{lokas}).
As claimed in \cite{white1}, the self-similarity would put a constraint on the
slope of the power spectrum of primordial fluctuations, pinning it down
betwenn $-3 < n < -1$. However, as will also be discussed in this Thesis, the
self-similarity of halos can well be a numerical artifact or a lack of
relevant physics in the inputs of simulations.

\begin{figure}
\centering\epsfig{file=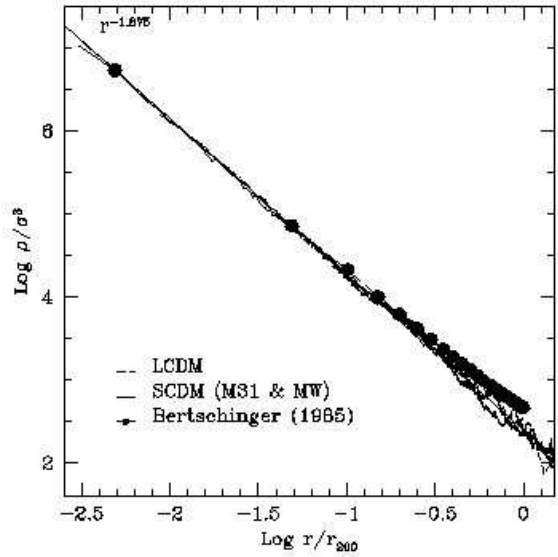,scale=0.6}
\caption{The phase-space density $\rho / \sigma^3$ as a function of radius, as
  shown in \cite{taylor}, with the Bertschiner's solution compared
  to the NFW .}
\label{tay1}
\end{figure}

The self-similarity of dark matter halos can be investigated in phase-space,
where it shows some more interesting facts.
By examining the coarse-grained phase-space structure of CDM halos, Taylor \&
Navarro \cite{taylor} argued of a pattern followed by all CDM-NFW halos regardless of virial
mass; the \textit{phase-space density}, a quantity defined in terms of the
spatial density and the velocity dispersion profiles,
can be expressed as a power law as a function of radius: $\rho /\sigma^3
\approx r^{- \alpha}$, with $\alpha \approx 1.875$; see Fig.~(\ref{tay1}).

Moreover, this slope coincides with that of the self-similar solution derived
by Bertschinger \cite{bert} for the equilibrium phase-space structure of an object
forming under spherical secondary infall of a gas of adiabatic index $\gamma=5/3$ onto a
point-mass seed in an unperturbed Einstein-de Sitter universe. Interestingly
enough, the spatial density profile of the Bertschinger's
system is a power law of constant slope $r^{2 \alpha -6}$, quite different
from the NFW. Notice that for the gas, the quantity $\rho^{5/2}/P^{3/2}$, where $P$ is the
pressure, is equivalent to the phase-space density, and is a measure of the
local entropy of the system: the two are inversely
proportional. Thus, for a DM halo, the phase-space density can be interpreted
as an entropy measure, and the power-law dependence on radius of $\rho
/\sigma^3$ describes an equilibrium configuration characterized by an entropy
profile with a minimum in the center of the halo, and increasing outwards.
It may well be that this entropy stratification is a fundamental property that
underlies the self-similarity of CDM halos \cite{taylor}.

\begin{figure}
\centering\epsfig{file=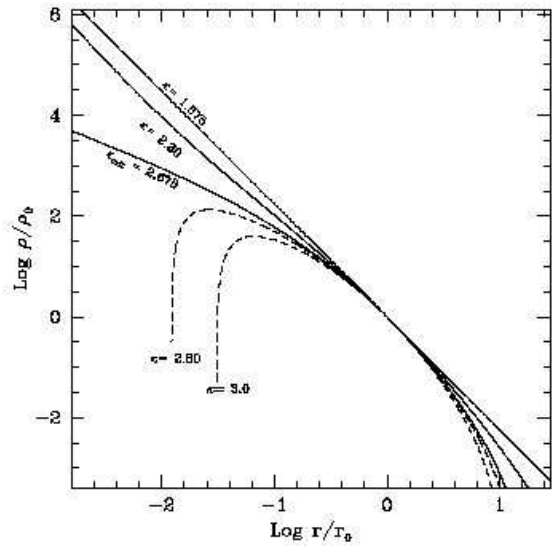,scale=0.6}
\caption{Density profiles corresponding to the Bertschinger's solution, for
  different values of the parameter $\alpha$; the NFW corresponds to the
  critical solution, beyond which the profiles are unrealistic \cite{taylor}.}
\label{tay4}
\end{figure}

\begin{figure}
\centering\epsfig{file=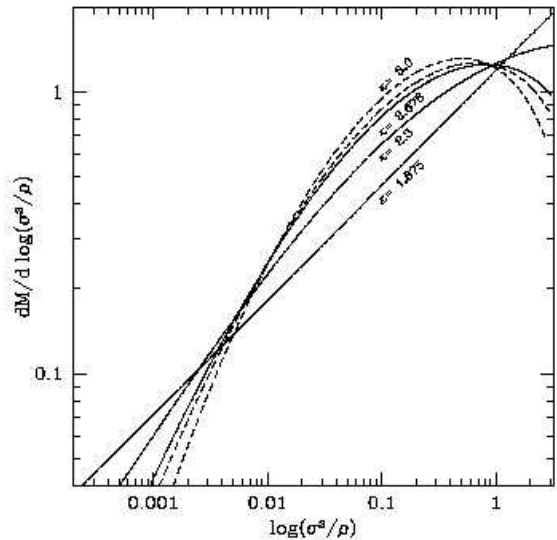,scale=0.6}
\caption{The phase-space distribution functions for the curves of
  Fig.~(\ref{tay4}), showing that the NFW is the maximally mixed
  configuration, corresponding to the most peaked DF \cite{taylor}.}
\label{tay3}
\end{figure}

There is more. Assuming that the phase-space density profile of a
collisionless isotropic system is a power law of slope $\alpha=1.875$, 
the Jeans' equations are
satisfied by a one-parameter family of density profiles; the parameter $k$ can be
chosen as the ratio between the velocity dispersion and the circular velocity
at the radius where the latter peaks. The Bertschinger's solution can be
recovered for $k=\alpha=1.875$, and the spatial density profile is also a
power law. Very interestingly, the NFW profile belongs to this same family of
curves, with $k \simeq 2.678$; this correspond to a critical value, above
which there is no sensible, non-vanishing, monothonic profile in the center of
the halo; see Fig.~(\ref{tay4}).     
Also, this is the critical value that correspond to the maximally mixed, most
uniform phase-space distribution function, as shown if Fig.~(\ref{tay3}). 
In other words, given the entropy
stratification constraint, the NFW is the highest-entropy profile. 
The theoretical explanation of these features is unknown \cite{taylor}.

\subsection{Phase-space mixing and violent relaxation}

\noindent
One of the questions that are still unanswered is,
how does a system choose its equilibrium state?
Two scenarios are possible in principle (see for comparison \cite{BT}): 
(i) the final configuration of the halo is a reflection of the particular
initial conditions of hierarchical clustering, \textit{i.e.} the spectrum of
fluctuations; 
 (ii) the final state is favoured by some fundamental physical
principle, that erases the memory of initial conditions. 
Smooth theoretical halos belong to the second class of objects; 
simulated halos belong to an hybrid class between the two.
In fact, despite the analogy of (ii) with the description of the NFW as the maximally
mixed configuration compatible with entropy stratification,  
the NFW itself is just the spherically averaged 
fit to the mass distribution of each simulated halo or subhalo, with the
averaging procedure smoothing out the substructures,
while the mass function actually conserves the hierarchy all the way through.

An elegant way to treat the equilibrium configuration of smooth halos would be
to use the \textbf{maximum entropy} principle, exploiting the phase-space
distribution function as a probability density; the equilibrium configuration 
would then be the one that maximises the quantity 
\be
S \equiv -\int_{phase \ space} f \ ln f d^3\textbf{x} d^3\textbf{v}~.
\label{entropy}
\ee
Unfortunately, for self-gravitating systems of point masses with a finite total
mass and energy, the entropy does not have a maximum, but can be increased
with no limits simply by increasing the system's degree of central
concentration \cite{BT}.




An important consequence of this fact is that dark matter halos are not by
themselves in long-term thermodynamical equilibrium, but constantly evolve into
states of higher concentration and entropy. Althoug this is not the place to
investigate such issues, it could be that the cuspy structure of the NFW stems directly
from the absence of a maximum entropy state.

To understand the evolution of a system of point particles like a dark matter
halo, with no underlying principle leading to equilibrium, 
two processes have to be taken into account. 

\begin{figure}
\epsfig{file=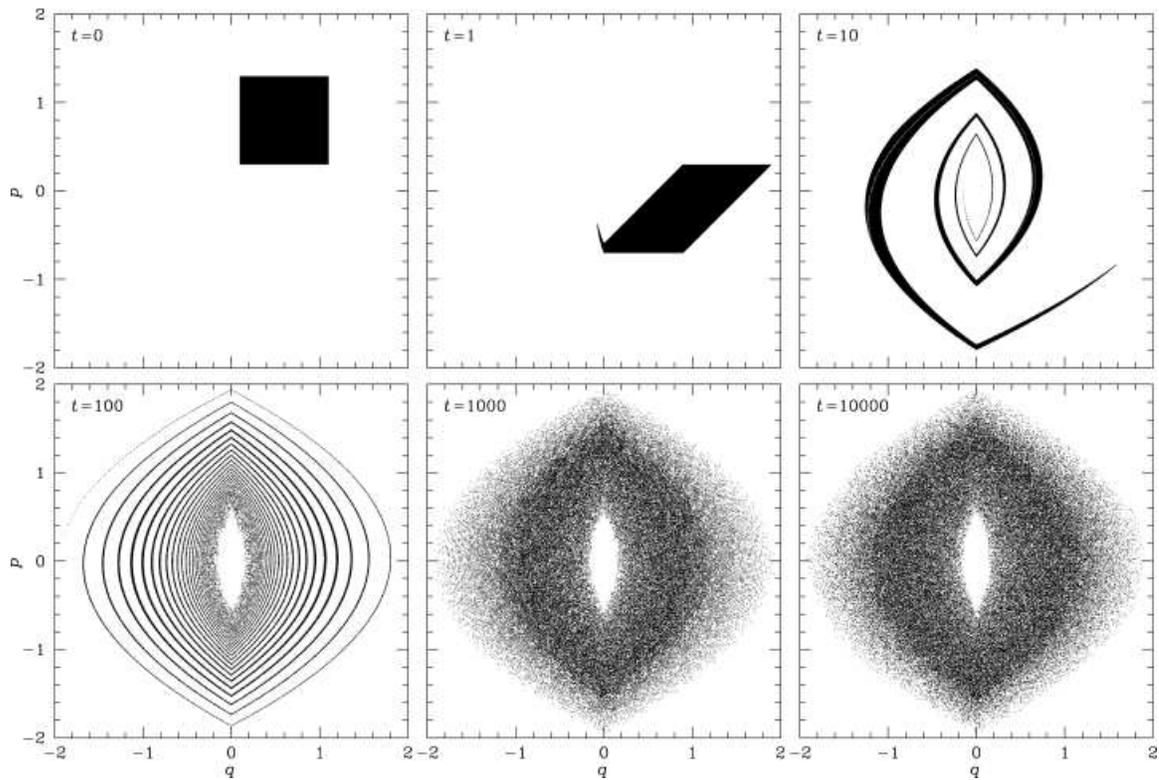,scale=0.6}
\caption{Demonstration of phase mixing; the evolution of a system of
  phase-space points under $1D$ gravity. The DF is constant at all times ($1$
  or $0$), but at late times, the coarse-grained system features a smooth
  distribution and a lower density \cite{dehnen}.}
\label{azz}
\end{figure}

The first is called \textbf{phase mixing}, and it is crucial in understanding
the behaviour of halos in numerical simulations, as will be clear in the following.
Consider the system of $N$ points shown in Fig.~(\ref{azz}), that at some
initial time $t_0$ occupies a volume of phase-space with constant $f=1$. The
system evolves in the Hamiltonian $H=p^2/2+|q|$ of a point mass in $1D$
gravity; as time passes,
volumes of $f=1$ (black) get stretched out and folded together with volumes 
of $f=0$ (white). The evolution is governed by the Boltzmann's equation, thus the
distribution function $f$ is conserved;
the density of phase points in the spiral pattern into which the
system evolves is the same as in the original configuration. 
However, the true $f$ is not accessible to observations; an observer can only
measure the average DF in a small volume around each point, \textit {i.e.} 
the \textit{coarse-grained} distribution function $\bar{f}$. 
At sufficiently small times, the observer can resolve small volumes around a
point such that there are no white regions inside, and 
the measured $\bar{f}$ is actually the true $f$;  but as time goes by,
around $t \sim 100$ any distinction is barred by the finite resolution of any
observation, and the observer can only measure an average of black regions blended with
white ones. This makes her/him see a $\bar{f}$ decreasing in time
around each point. 
In addition, the true $f$ shows a very complex and continuosly evolving pattern, 
while equilibrium is reached in the coarse-grained sense, \textit{i.e.} after a while
$\partial_t \bar{f}=0$ \cite{dehnen}.

The meaning of the collisionless Boltzmann's equation can be analyzed from this
point of view; while stating that $f$ is constant, it also insures that 
$\bar{f}$ is \textit{not increasing} along any trajectory in phase-space
accessible by the system. The density of points in phase-space is diluted 
while the incompressible fluid represented by $f$ is mixed with ``air'', in
the form of stripes void of phase points. 

The entropy defined by Eq.~(\ref{entropy}) is constant, since $f$ is
constant. However, by replacing $f$ with $\bar{f}$ one obtains an entropy
$\bar{S}$ that increases anytime $\bar{f}$ decreases along an orbit due to
phase mixing; this sounds familiar, resembling the increase in entropy of a
thermodynamical system when it moves towards thermal equilibrium \cite{BT}.
In this sense, the evolution of the coarse-grained DF towards a minimum phase-space
density brings the system towards more relaxed, more probable states. Notice
however, that unless we can define a minimum phase-space density allowed for
the system, we cannot find a maximum entropy state, not even in the
coarse-grained sense; this is the case of the
NFW, where the phase-space density is an unbroken power law with no
minimum. For such systems, we can only argue that they cannot evolve from a state
$A$ to a state $B$ if $\bar{S}(B) < \bar{S}(A)$.




In the scenario of hierarchical clustering, where structures are formed
through accretion of smaller units, another process is fundamental in
shaping the final state of dark matter halos, and is called 
\textbf{violent relaxation} \cite{lybell}. 
Rapid changes in the overall gravitational potential, during the collapse
and the whole merger history of dark matter halos, provide relaxation
on relatively short timescales (compared to the collisional case, for
instance); this process governs the nonlinear
evolution of halos into a final state characterised
by structural regularities, with a universal density profile that does
not depend on mass, on cosmological parameters or on the initial fluctuation 
spectrum (after smoothing out subtructures) \cite{white2}. 

During violent relaxation, the volume of phase-space accessible by the system
increases \cite{BT}.
This can be understood in the following way: consider a system of
point masses uniformly distributed in a sphere, with a given distribution of initial velocities
(like a DM halo), and consider a
particle located somewhere in the sphere. Because of
self-gravity, the particle will be pulled towards the centre of the
system, and at the same time the potential well will deepen because of 
the infalling mass (all the other particles behave more or less in the same
way than our test particle); the particle will gain a lot of kinetic energy, and when
it passes near the centre and is on its way out again, the system is
re-expanding, thus the potential well the particle has to climb is shallower 
than before. The particle will reach the potential at which it originally
started with an excess of kinetic energy. Thus, the phase-space region of
velocities and positions it can reach is wider than before \cite{BT}.

Notice the difference between mixing and violent relaxation. The first process
decreases the phase-space density around any point 
while conserving the phase-space volume; the second increases the 
phase-space volume accessible by the system, not necessarily altering the
density of points. However, violent relaxation can increase the phase dilution
by providing additional volume to the system, so it can be considered as
another source of mixing. 
  
While nobody knows whether mixing is well represented in numerical
simulations, and the role it potentially plays, 
violent relaxation is thought to be responsible for the
universal density profiles, as well as the spin and shape distributions, of CDM simulated
halos (as stated in \cite{white2}). In particular, the (smoothed) NFW density profile is the outcome of
such process; its dependence on the halo mass, or on the initial power
spectrum or cosmological context, is entirely upon the value of the concentration $c$,
while the distribution itself is universal \cite{white2}.  
Moreover, the NFW fitting formula is a
good representation of the mass distribution of halos originated through
dissipationless hierarchical clustering in \textit{any} Einstein-de Sitter
universe \cite{cole}, and the systematic dependences of $c$, 
on $P(k)$ and $\Omega$ can always be
understood in terms of different formation times in different cosmologies
\cite{white2}.
Gravitational potential fluctuations during the collisions and mergings which
characterize hierarchical clustering in simulations are evidently strong enough to cause
convergent evolution \cite{white2}. 

\section{Observations \textit{vs} simulations: the cusp issue and the problem
  of substructures}

\noindent
When the predictions of the CDM model are tested against observations, some
unpleasant facts about the halo equilibrium structure come into the spotlight. 

The most succesful tool in studying the mass distribution of
galactic dark matter halos is the spirals' rotation curves, that is
particularly useful in investigating the inner density profile. 
Dynamical analysis of these curves shows, with increasing precision, that the
favoured density profile (the best in fitting the observed data) is flat,
corelike, with a central density significanlty lower than the NFW 
(see for instance \cite{salucciburkert},\cite{borr},\cite{deblok1},\cite{weldrake},
\cite{swaters1},\cite{gentile04},\cite{simon},
\cite{gentile05},\cite{gentile07}).
To solve this striking disagreement, several authors (\textit{e.g.}, 
\cite{spergel},\cite{ahn}) recurred to a framework
different from the standard cosmology; 
however, as argued by \cite{hogan} (and see references therein), 
neither self-interacting nor warm DM cannot explain
the observed features of the inner halos. 
Alternatively \cite{taylor}
argue that the discrepancy arises from the imperfection of the numerical
approach in simulating 
regions where the overdensities exceed $\sim 10^6$ and where the particles may have
completed thousands of orbits during a Hubble time.

In addition, the DM density profiles inferred from observations do not show
any self-similarity; both the amplitude and shape of the profile depend strongly on
the halo mass, as will be extensively discussed in Chapter 3.

The evolution of numerical simulations in 30 years has underlined another very
important issue about hierarchical clustering, and another significant offset
between theory and observations.
By measuring the two-point correlation function of galaxies, we can infer a
well defined power-law down to subgalactic scales, where the distribution is
highly nonlinear.
In the original formulation of the theory, Peebles \cite{peebles78} argued that the
galaxy distribution and the mass distribution inside galaxies form an unbroken,
scale-invariant or fractal-like hierarchy, thus reflecting the dynamical
stability of this arrangement. 
In the same year, White \& Rees \cite{whiterees} argued the contrary; a virialised clump 
of non-dissipative dark matter would not mantain a hierarchical structure, but
would evolve into a monolithic halo with a well defined centre and a smooth
mass distribution. As an example, consider a single cluster halo,
containing many galaxies and not a single ``supergalaxy''; the dark matter
component is smoothed by virialization, while the baryons in single galaxies
are concentrated enough by dissipative processes to avoid ``overmerging''
later on \cite{white1}.  

This is a sore point for numerical simulations. Still today, simulations
confirm the theoretical expectation of a self-similar subhalo population, 
accoding to which galactic systems are simply scaled-down versions of clusters
\cite{donghia3}, with the result that simulated galaxies and groups 
feature an overaundance of substructures, with a mass function at small scales
that is between one and two orders of magnitude higher than inferred from
observations (see Chapter 6, and for instance \cite{kly},\cite{moo},\cite{wilma},\cite{kleyna}). 
In addition to not being observed, such an overabundant subhalo population 
would represent a serious problem for disk galaxies;  
the presence of too much substructure would in fact mess up the observed 
galactic scaling relations (see Chapter 2), and cause dynamical heating and 
disruption of the fragile thin disk. 
In ordet to fit spirals into the hierarchical picture, 
the dark matter halos hosting them must assemble and be well relaxed relatively early, 
meaning that the last major merging events are expected to be
sufficiently far in the past, to allow the disk to form and evolve unperturbed.
In contrast, although galaxies in clusters are embedded in common halos, these halos
can still feature significant substructure; in fact,
they are relatively young, and still far from equilibrium \cite{white1}.
Notice that the difference in scale between these systems mirrors not only 
their different formation times and ages, but also a different mean density 
or concentration $c$.

The problem regarding real halos is twofold: 
on one hand, it may be that the equilibrium configuration of dark matter halos
is always smooth, and that cluster halos simply have not had enough time to
completely relax; on the other hand, there may be a mass-dependence trend for the
survival of substructures, with baryons condensing in the subunits that may
play a significant role (see Chapter 6).

The cusp issue and the substructure problem arising in simulations are linked. 
The explanation of the self-similarity of the NFW may reside in the radial stratification
of the phase-space density, as explained before; if this is the case, 
the shape of the profile and the preservation of the hierarchy
on the whole range of halo masses seem to be
interconnected, in the sense that the NFW is the natural final state of the
hierarchical clustering process. Hence, to deny the NFW 
means to deny the whole picture of structure formation as it is
formulated today. This actually does not seem acceptable, given the number of
successes of the model.

On the other hand, it is a fact that the self-similarity itself is not proven
in real systems, nor is the validity of the density profile. 

One way out is admitting that we are
massively biased when we investigate DM halos observationally, since we
actually see only the luminous component that evolves inside them; we
do not really know how a pure DM halo looks like, or whether the presence of the
galaxy affects its shape. 
In this case, there is a way to reconcile theory and observations, by
modelling the halo reaction to galaxy formation and its subsequent evolution
into different equilibrium structures, triggered by dynamical interactions
between the dark matter and the baryons. 
This is the point of view adopted in this Thesis.

\section{Galaxy formation and scaling relations}

\noindent
Galaxies form from the collapse of baryons inside the potential wells of 
dark matter halos, so their properties are expected to be 
regulated by the structure and formation
history of their hosts, and the environment surrounding them. 
The dependence of the galactic properties on those of the halo has been
observed, in the form of \textit{scaling relations} between the mass
distribution and geometry of the galaxy and the halo structural parameters.
The mass and angular momentum of the material available for galaxy formation,
as well as the rate of interactions between galaxies, are determined by the
halo structure and its evolution in the hierarchical picture (\cite{white3}).    
In addition, the global properties of galaxies depend on how the gas cools into clouds and how
these fall and collapse into the halo potential, and on the characteristics of
star formation and feedback. 

The current theory of galaxy formation is a hybrid between numerical
simulations, that account for the dark component, and semi-analytical models
that take care of the baryonic one. 
The newest codes for SPH simulations are
indeed refined and use the latest semi-analytical prescriptions to reproduce
the observed features of galaxies; the general picture is thought to be
understood (\cite{whitefrenk},\cite{dalcanton},\cite{momaowhite}),
although many observational evidences are still not accomodated in the
scenario (see above). Among these, spiral galaxies present problematic issues, regarding their
dynamics and the geometry of their mass distribution. 
In fact, all the processes leading to galaxy formation affect each other in highly
nonlinear ways, and involve a wide range of scales; simulations are not always
the best tool in investigating such mechanisms, and semi-analytical models
sometimes perform better.
Here I present the current picture of galaxy formation, along with
its shortcomings \footnote{The particular model discussed here is intended 
  to give a general picture of the current theory; newer models by
  other authors may differ in some details.}.  
  
In the current scenario (\cite{dalcanton},\cite{momaowhite}), after
decoupling baryons are trapped into the
potential wells of the growing dark matter perturbations, and initially follow
the same evolution patterns; the two components are initially well mixed,
\textit{i.e.} they share the same phase-space structure. 
They both start to collapse in some overdense regions that will become
the centers of the dark matter halos, and while the density is low their
dynamical behaviour is identical; however, contrary to the baryonic one, the dark matter collapse is
dissipationless, and it halts when the system virialises.  

The baryonic collapse proceeds, leading to densities high enough for
radiative cooling to become effective; at this stage, the process accelerates
and the baryons dynamically decouple from the dark component, fragmenting and condensing
into self-gravitating units. The gravitational potential of the halo regulates
the infall of the clouds towards the denser central regions of the system. 

Tidal torques, originating with the mass accretion of the halo, spin up the
clouds that, depending on the amplitude of the angular momentum $J_z$ and on the details of the
trajectories towards the inner regions, may or may not collapse along a 
preferential direction, forming a disk or a spheroidal component. At the same
time, density in the central regions reaches the density threshold for star
formation to begin; depending on the availability of gas, \textit{i.e.} on the
clouds reaching the center of the halo, star formation regulates the 
amount of feedback and balances the subsequent evolution, by expelling some 
baryon fraction and regulating the infall of material.  

From the results of numerical simulations and a set of semi-analytic
prescriptions, a number of predictions regarding the structure of disk
galaxies have been made (\cite{dalcanton},\cite{momaowhite}). Of particular
interest in this Thesis,
the mass distribution and the geometry of the disk are expressed as functions
of the halo mass and angular momentum.  

The angular momentum $J_z$ of simulated halos is characterized by
the dimensionless \textbf{spin parameter} \cite{peebles69}
\be
\lambda \equiv \frac{J_z \ |E|^{1/2}}{G \ M_{vir}^{5/2}}~,
\ee  
that can be approximated by the ratio between the rotational and circular
velocity $\lambda \approx V_{rot}/V_c$, and expresses the degree of ordinate
motions around the axis of rotation. 
The effective size of the disk forming inside the halo in simulations strongly depends on the 
clouds' angular momentum along the spin axis, which in turn
depends on the halo's $J_z$; the spin parameter
is therefore a useful tool to link the geometrical properties of the disk with
the dynamics of the halo. 

For a NFW halo of virial mass $M_{vir}$ and concentration $c$, the mass profile is
\be
M(x)=M_{vir} g(c) \left( ln(1+cx)-\frac{cx}{1+cx} \right)~,
\ee  
in terms of the radial coordinate $x=r/R_{vir}$. The total energy of the halo
is recovered as 
\be
E=-\frac{G \ M_{vir}^2}{2 \ R_{vir}} \ f_c~,
\ee
with $f_c$ being a shape factor, function of the halo concentration. Given
the circular velocity $V_c^2=GM/r$, the halo total angular momentum is then
\be
J_z=\sqrt{2/f_c} \lambda M_{vir}R_{vir}V_c~.
\ee
The baryons inside the halo are supposed to collapse in a disk of mass 
\be
M_D=m_d \ M_{vir}~, 
\ee
with $m_d$ a universal fraction \cite{momaowhite}. As a consequence, the
total disk angular momentum is also a universal fraction $j_d$ of that of the halo: 
\be
J_D=j_d \ J_z~. 
\ee
Notice that this assumption is highly unrealistic, since the mass-to-light
ratio in spirals is a strong function of mass, with small systems
significantly more dark matter-dominated with respect to the massive ones 
(\cite{salpe},\cite{bell},\cite{shankar}).

The disk profile is
assumed exponential, with scale-length $R_D$, and the specific angular
momentum $j_z$ is supposed to be conserved between the baryonic and dark
component during the collapse.
The resulting disk scale-length as a function of the halo parameters is then:
\be
R_D=\frac{1}{\sqrt{2}} \left( \frac{j_d}{m_d} \right) \lambda R_{vir}
f_c^{-1/2} f_R(c,\lambda,m_d,j_d)
\ee
where 
\be
f_R(c,\lambda,m_d,j_d)=2\left[ \int_0^\infty e^{-u} u^2
  \frac{V_c(R_Du)}{V_{vir}}du \right]^{-1}~;
\ee
for a detailed description, refer to \cite{momaowhite}. From these
prescriptions, the total rotation curve of the system halo $+$ disk galaxy is
obtained simply as $V_{TOT}^2=V_D^2+V_H^2$.

As will be discussed in the next Chapter, observations of disk galaxies show
that the dependences of the disk properties on halo dynamics are highly nonlinear,
and that the simple assumptions of this model are not suitable to yield the
observed disk geometries or masses \cite{chiaraspin}. 

\section{Plan of the Thesis}

\noindent
In Chapter 2 I will introduce a refined model of galaxy formation, that is
based on the observed scaling relations existing between the dark and
luminous components of galaxies, that circumvents the difficulties presented
in the previous Section, and makes predictions about the dynamical properties
of galactic halos from those of the disk galaxies. This allows me to constrain
the halo mass accretion history.

In Chapter 3 I will present an observational counterpart to the NFW, that can
be practically used to infer the global properties of halos (like the virial
mass for instance) from the observed rotation curves, and that introduces 
a breaking in the self-similarity of halos.

In Chapter 4, I will further analyse the properties of the NFW, higlighting
that the discrepancy between the expected mass distribution and the data
coming from real systems is not due to numerical artifacts, but stems from the
lack of a proper modelling of all the physical mechanisms involved in 
galaxy formation.

In Chapter 5, I will finally present a theoretical model of halo evolution
triggered by galaxy formation. First, I will describe the evolution of the
halo equilibrium structure in phase-space, when it is perturbed by 
angular momentum transfer, that modifies the anisotropy profile with the
injection of tangential motions; according to the amplitude of the
perturbation, the halo can acquire cored equilibrium configurations. Second, I
will present a plausible physical mechanism to account for such a halo
evolution, in the form of dynamical friction exerted by the DM on the baryonic
clouds collapsing to form the protogalaxy. 

In Chapter 6, I will describe my work-in-progress about  
the numerical simulation of dynamical friction, for the double purpose of
testing the theoretical model of Chapter 5, and of addressing the issue of
the disruption of substructures in galactic DM halos.   

In Chapter 7 I will finally conclude.

%% file: spin.tex
\chapter{The Spin of Spiral Galaxies}

\noindent
I present a model for computing the angular momentum
and spin parameter distribution function of dark matter 
halos hosting real spiral galaxies, entirely based on the observed scaling relations
between the geometrical and dynamical properties of the galaxies and their
hosts. I then use the spin parameter inferred from the observations as a tool 
to constrain the mass accretion history of dark matter halos.
 
\section{Introduction}

\noindent
The spin of DM halos is thought to be originated during major mergers, 
where large-scale tidal interactions transform the orbital angular momentum of
the colliding objects into a coherent rotation of the final halo \cite{fall}.
Although this mechanism can explain the halo spin, 
the connection to the rotational motions of the galaxies 
hosted inside the halos is not completely understood. 
In fact, even if at early stages the baryons share the same phase-space
structure of the dark component, and in particular have the same specific
angular momentum distribution, they then undergo a number of physical
processes (including radiative cooling and collapse, dynamical friction, 
star formation, heating and shocks, supernova winds and AGN activity)
that dynamically decouple them from the DM to an unknown degree. 
For this reason, any determination of the halo dynamics from the baryons has
to face the question of whether the baryons actually conserve any memory of
it at all. 
On the other hand, it is plausible that the baryons and the host halo continue
to dynamically interact even after the galaxy formation, expecially when one
considers global motions that, like the rotation around the spin axis, 
occur on scales comparable to the halo size and are therefore not largely
affected by mixing, relaxation and local processes generally linked to feedback.  
So, as far as the spin is concerned, it is not far-fetched to expect a certain 
degree of self-regulation in the system galaxy-dark halo, in order to mantain a stable
equilibrium configuration between the two components \cite{chiaraspin}.  

In the simplest scenario, the ratio bewteen the baryonic and the DM angular
momenta is constant. In other words, the angular momentum per unit mass is
conserved, and the only processes that affect its distribution
are purely gravitational.  
This is a plausible picture if the baryons and the DM initially have 
similar phase-space distributions \cite{chiaraspin}. 

The collapse of the luminous component leading to the formation of a disk
galaxy follows a preferential direction, thus setting the disk spin axis.
The geometry of the disk is directly related to the initial dynamical state of the
baryons, in particular the angular momentum distribution; provided a suitable
scaling, this relates to the dynamics of the dark halo. 
In particular, \cite{fall} provided a link between the
disk scale length and the halo angular momentum. 

The tight connection between
halo dynamics and disk geometry is quantified by the \textbf{spin parameter}
$\lambda$ \cite{peebles69}, that proves to be a fundamental tool in the
study of the formation of both the DM halo and the galaxy. 

The physical meaning of the spin parameter is to represent the degree of 
rotation around a given axis in the motions of an object. 
In other words, one can represent the spin parameter as
\be
\lambda \sim \frac{V_{rot}}{V_c}
\ee 
where $V_c=\sqrt{GM(r)/r}$ and $V_{rot}$ is the actual rotational velocity. 
A rotationally supported disk has $V_{rot}/V_c=1$, 
while an object dominated by velocity dispersion has $ V_{rot}/V_c\simeq 0$. 

Mo and collaborators \cite{momaowhite} described a procedure for the computation 
of the scale length of a disk embedded in a NFW 
dark halo of given mass and spin parameter, as yielded by
numerical simulations (see Chapter 1 for a more detailed description).
The model made use of
a set of assumptions on the properties of the baryons, as dependent on the 
host halo: (i) the mass of the galactic disk is a
$universal$ fraction of the halo's; (ii) the total angular momentum of
the disk is also a fixed $universal$ fraction of the halo's; (iii) the disk is
thin and centrifugally supported, with an exponential surface
density profile. The success of the model in reproducing real systems is
limited \cite{chiaraspin}. 

In the present Chapter, I address the same issue from the opposite point of 
view, \textit{i.e.} given the galactic observables, I infer the dark matter
global properties.
In particular, I provide a method for estimating the halo angular
momentum $J_z$ and the spin parameter $\lambda$ of a DM halo hosting 
a spiral galaxy of measured mass and scale length.
Instead of assuming arbitrary scalings between the baryons 
and the DM, the model
relies on a series of empirical relations linking the disk geometry and mass
distribution with the mass distribution and the dynamics of the halo,
through the rotation curve. 

For the sake of simplicity, I'll make use of the specific angular 
momentum conservation assumption, but it will be clear how to
straightforwardly extend the method to more general cases. 

More in particular, in a direct comparison with the Mo et al. method, 
we relax (i), and use instead
an empirical relation that links the disk mass to that of its DM
halo \cite{shankar};
as for (ii), the baryonic angular 
momentum is therefore not a universal fraction of that of the halo, 
but rather depends on the
mass-to-light ratio and on the chosen angular momentum conservation 
law between the components; (iii), the disk is still centrifugally
supported, stable, and distributed according to an exponential
surface density profile, but I'll also take into account the gaseous
(HI+He) component, that turns out to give an important contribution to the
overall angular momentum distribution, expecially in small galaxies.
As for the shape of the DM profile, I'll present the results for a Burkert
halo, in comparison with the NFW \cite{burkert}. 

\section{The halo angular momentum}

\noindent
Statistical studies about the observed properties of spirals, obtained from
rotation curves and photometric measurements, allow for a set of scaling
relations between the mass distribution and geometry of the systems galaxy $+$
dark halo, in a range of masses including most of the population, with the
exception of dwarves. 
For the purpose of computing the halo angular momentum from the spiral
observables, one needs to determine the ratio between the disk and halo mass,
the link between the halo mass and the disk geometry, and the mass
distribution of all the components; the gaseous baryonic
component, lesser important mass-wise, is significant in its contribution to
the angular momentum due to the fact that it is more diffuse.   

\begin{figure}
\centering\epsfig{file=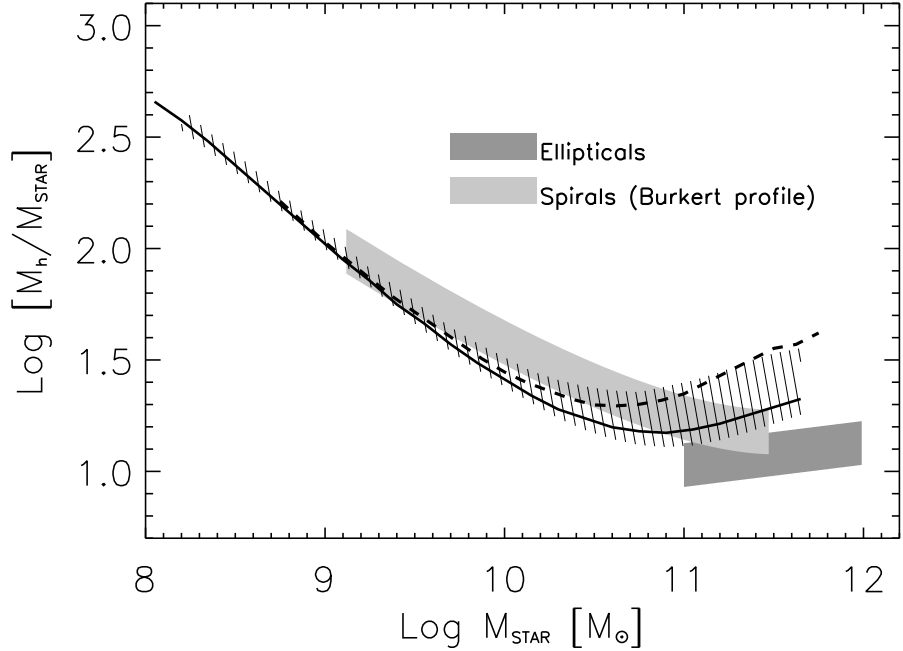,scale=0.7} 
\caption{DM halo-to-stellar mass ratio as a function
of the stellar mass \cite{shankar}. 
The \textit{dark shaded} area
represents the data on giant elliptical galaxies \cite{gerhard};
the \textit{light shaded} area represents data on spiral
galaxies (\cite{pss96},\cite{salucciburkert}).}
\label{shankarfig}
\end{figure}

The total mass of the stellar disk $M_D$ that resides in a halo of
mass $M_{H}$ has been derived by the
statistical comparison of the galactic halo mass function extracted from
N-body simulations with the observed 
stellar mass function \cite{shankar}:
\be
\label{MvirMd}
M_{D} \approx 2.3 \times 10^{10} \ M_{\odot} \ \frac {(M_{H}/3 \
10^{11} \ M_{\odot} )^{3.1} }{1+(M_{H}/3 \ 10^{11} \
M_{\odot})^{2.2}}~;
\ee
it holds for halo masses between $10^{11}$ and about $3 \times
10^{12}\, M_{\odot}$, with an uncertainty around $20\%$, mainly due 
to the mass-to-light ratio used to derive the stellar
mass function from the galaxy luminosity function; it is shown in
Fig.~(\ref{shankarfig}).

This relation replaces the assumption (i) about a universal
mass-to-light ratio for spirals by \cite{momaowhite}, and is physically well
motivated (\cite{salpe},\cite{bell},\cite{shankar}). In fact,
the balance between two processes, namely cooling and feedback,  
leads to a ratio between the mass of the stellar disk and the halo
mass that increases with the latter;
the smaller the halo, the more the efficiency of feedback is enhanced,
while the efficiency of baryonic cooling and 
collapse is diminished by the lesser halo gravitational potential.

The stellar disk is thin, gravitationally supported, 
with an exponential surface density profile described by
\be
\Sigma_D(r) = {M_D\over 2 \pi \ R_D^2}\, e^{-r/R_D}~.
\ee
The characteristic scale-length $R_D$ is the key parameter in defining the
disk geometry, and is estimated from the disk mass, through dynamical 
mass determinations \cite{pss96}: 
\be
\log{{R_D}\over \mathrm{kpc}} = 0.633+0.379\,\log{M_D\over
10^{11}\,M_{\odot}}+0.069\, \left(\log{M_D\over 10^{11}\,
M_{\odot}}\right)^2~.
\label{mdrd}
\ee
This result is consistent with the scale-lengths inferred in previous works
(\cite{dale},\cite{simard},\cite{courteau}).

The halo mass distribution is inferred from the rotation curve;
the majority of the observations of the rotational velocities of spirals yield
a Burkert DM density profile as the best fit \cite{burkert}.
A statistical treatment is given by the Universal Rotation Curve (URC$_0$,\cite{pss96}),
a two-parameters family of curves for a Burkert halo, determined by the halo 
effective core density $\rho_0$ and core radius $R_0$. 
The halo density profile is thus given by:
\be
\rho_H(r)=\frac{\rho_0 \,R_0^3}{(r+R_0)(r^2+R_0^2)}~.
\label{burkert}
\ee
Correspondingly, the cumulative mass profile is given by:
\be
M_{H}(<r)=4 \, M_0 \, \left[\ln\left( 1+\frac{r}{R_0} \right) -
\tan^{-1}\left(\frac{r}{R_0} \right) + \frac{1}{2}\,\ln\left(
1+\frac{r^2}{R_0^2}\right) \right]~, 
\label{burkertmass}
\ee
with $M_0=1.6\,\rho_0\,R_0^3$ being the mass contained inside the
radius $R_0$. 

\begin{figure}
\centering\epsfig{file=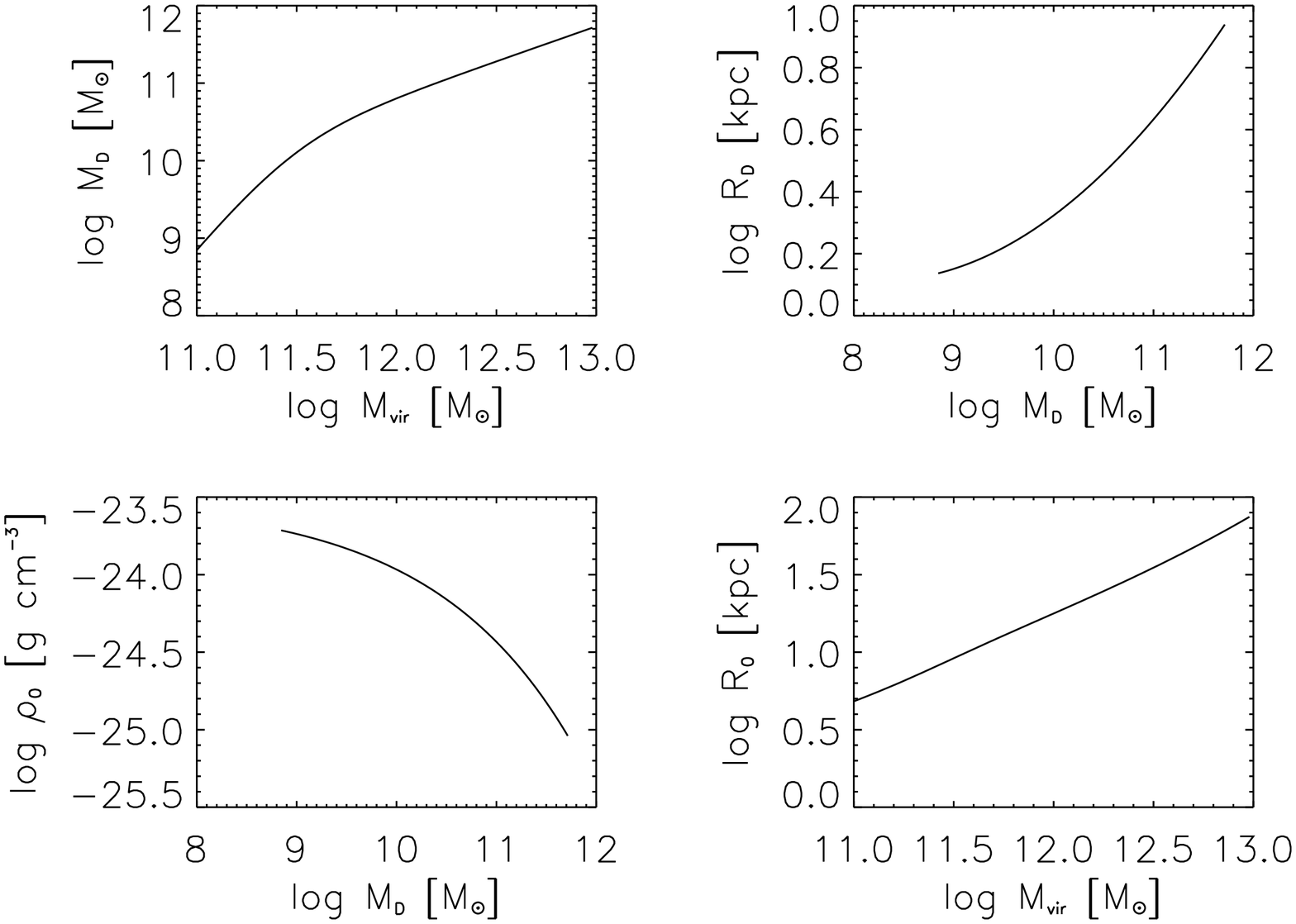,scale=0.4} 
\caption{A summary of the empirical scaling relations adopted in the model;
  \textit{top left}: stellar disk - halo mass; \textit{top right}: disk mass -
  scale-length; \textit{bottom left}: disk mass - halo central density;
  \textit{bottom right}: halo mass - core radius.}
\label{scalings}
\end{figure}

The URC$_0$ itself provides a scaling for $\rho_0$ as a function of
the disk mass:   
\be
\log{\rho_0\over \mathrm{g}~\mathrm{cm}^{-3}} = -23.515-0.964\,
\left({M_D}\over 10^{11}\,M_{\odot}\right)^{0.31}~. 
\label{rhoMd}
\ee
For each given disk mass, the halo is therefore completely determined;
the halo mass $M_H$ is computed through 
Eq.~(\ref{MvirMd}), the virial radius of the halo is set by the cosmology
through
\be
R_{H} = [3\, M_{H}\,
\Omega_M^z/ \, 4\pi\,\rho_c\,\Omega_M\, (1+z)^3\,\Delta_{H}]^{1/3}
\label{Rvir}
\ee
the density $\rho_0$ is obtained through Eq.~(\ref{rhoMd}), and  
the core radius $R_0$ is then computed numerically by
requiring that the mass $M_{H}(< R_{H})$ inside $R_H$ given by the
r.h.s. of Eq.~(\ref{burkertmass}) equals the virial mass $M_{H}$. 
The resulting $R_0$ vs. $M_H$ relation is approximated within a few
percents by 
\be
\log{(R_0/\mathrm{kpc})}\approx
0.66+0.58\,\log{(M_H/10^{11}\,M_{\odot})}~;
\label{r0mh}
\ee
notice that these values of $R_0$ are obtained through a mass modeling
of the whole mass distribution out to $R_H$, rather than  
from the decomposition of the inner rotation curves of single 
galaxies \cite{salucciburkert}.
A summary of the scalings adopted by the model is given in
Fig.~(\ref{scalings}).

Before considering the gas distribution in addition to the stars, I will compute the disk angular
momentum and the Tully-Fisher relation for my mass spectrum, 
in order to compare this model with
the NFW-Mo et al. theory and with observations.

The total circular velocity of the system is
\be
V^2_{c}(r)=V^2_D(r)+V^2_H(r)~. 
\ee
For a thin, centrifugally supported
disk the circular velocity is given by 
\be
V_D^2(r) = \frac{G \, M_D}{2 \, R_D} \, x^2 \, (I_0 \, K_0-I_1 \, K_1)|_{x/2}~;
\label{vdisk}
\ee
here $x=r/R_D$ and the quantity
$B=I_0\,K_0-I_1\,K_1$ is a combination of the modified Bessel
functions that accounts for the disk asphericity \cite{freeman}. The halo
circular velocity is simply 
\be
V_H^2 (r)= G M_{H}(<r) / r~, 
\ee
and it is useful to define the virial velocity $V_{H} \equiv \sqrt{G
  M_{H}/R_{H}}$. 
Given the scaling relations
(\ref{MvirMd}), (\ref{mdrd}), (\ref{rhoMd}) and (\ref{r0mh}) linking
the basic quantities of the system, only the disk mass is needed to
completely determine the shape and amplitude of the
velocity profile. Note
that all the uncertainties on these relations combine to give a $10\%-20\%$
total error on the determination of the velocity profile (see
\cite{toninisalucci} and Appendix A).

\begin{figure}
\centering\epsfig{file=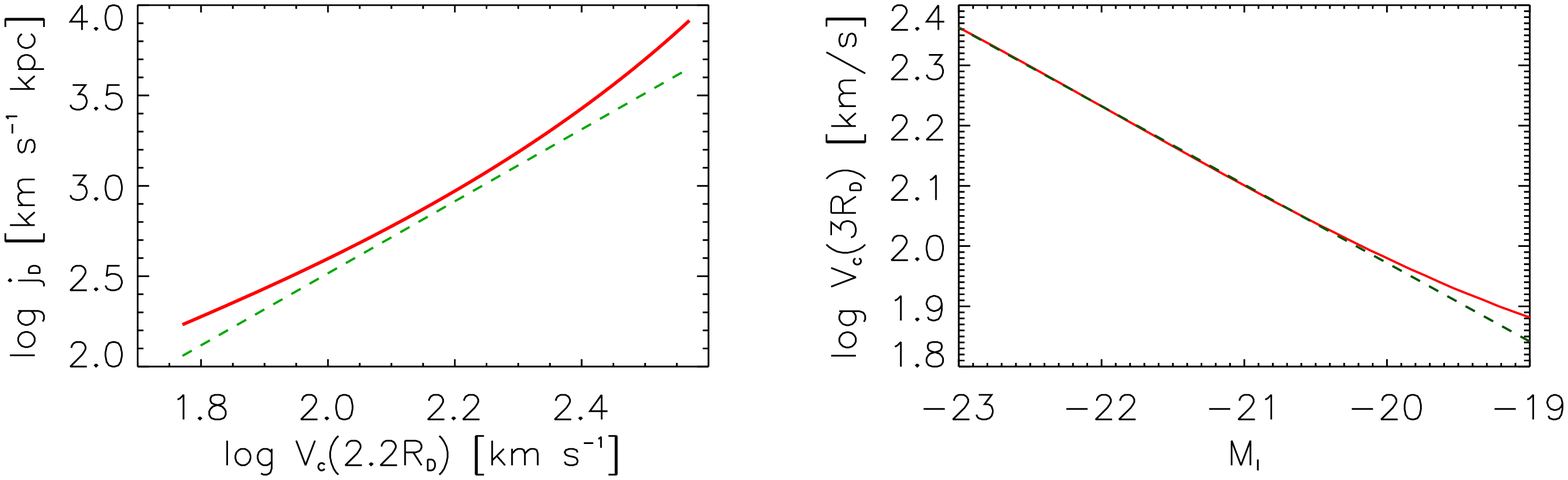,height=6cm, width=13cm} 
\caption{Left panel: the specific angular momentum
of the disk as a function of the rotation velocity at $2.2\,R_D$.
The \textit{solid line} is the result from this model, adopting the
Burkert profile; the \textit{dashed line} is the best-fit relation from
the data collected by \cite{navarrosteinmetz}, see their Figure
3. Right panel: the Tully-Fisher relation. The \textit{solid line} 
represents the result from this model and the \textit{dashed line} 
illustrates the fit to the data by \cite{giovannelli}.}
\label{tully}
\end{figure}

A way of checking the performance of this model is through the computation of
the Tully-Fisher relation. I obtained the B-band luminosity from 
the stellar disk mass, through the relation \cite{shankar}
\be
\log \left( \frac{L_B}{L_{\odot}} \right) \approx 1.33+0.83 \, \log \left(
\frac {M_D}{M_{\odot}} \right)~; 
\ee
I then converted the related magnitude in 
the I-band, through the mean colour $B-I\approx 2$ \cite{fukugita}.
In Fig.~(\ref{tully}, \textit{right}) I compare the resulting TF at $r=3\, R_D$ with the
data by \cite{giovannelli}, finding an excellent agreement.

The angular momentum of the disk is obtained as
\be
J_D=2 \pi \int^{\infty}_0 \Sigma_D(r) \, r \, V_{c}(r)\, r\,dr = M_D
\, R_D \, V_{H} \, f_R~,
\label{Jd}
\ee
with $x=r/R_D$, $f_R = \int^{\infty}_0 x^2\, e^{-x} \, V_{c}(xR_D)/V_{H}~ dx$
acting as a shape factor, and $M_D=2 \pi \,\Sigma_0\, R_D^2$. 
Note that $J_D$ depends linearly on both the mass and on the radial extension of the
disk, while the DM distribution enters the computation through the
integrated velocity profile, encased into the shape factor $f_R$;
the latter slowly varies (by a factor $1.3$ at most) throughout our
range of halo masses. 

\begin{figure}
\centering\epsfig{file=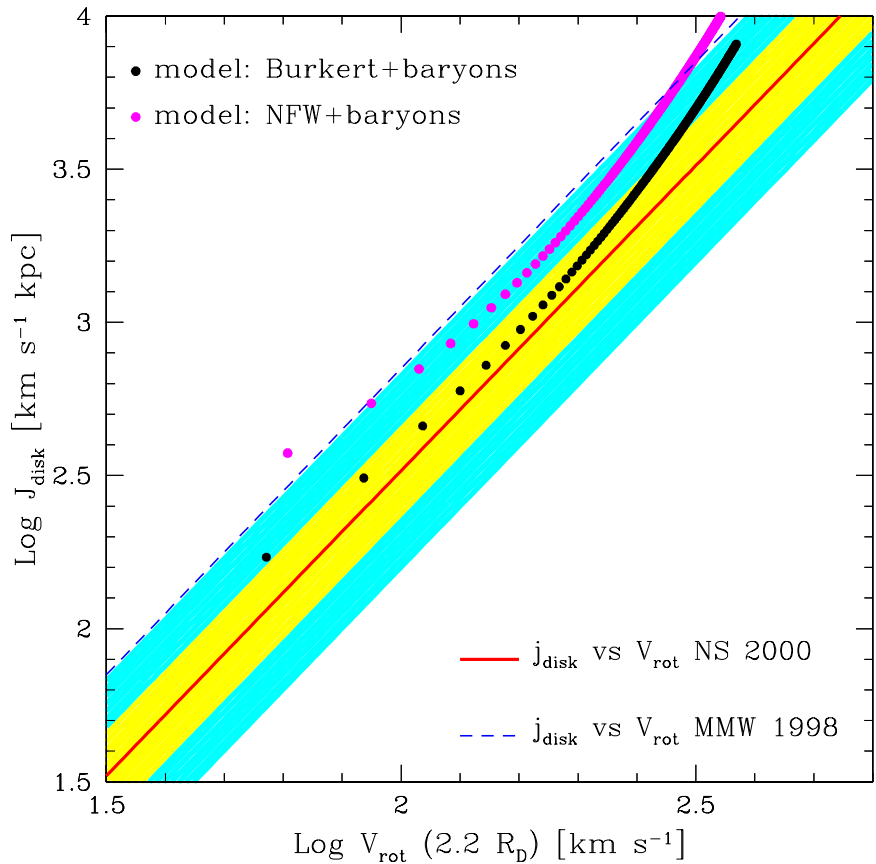,scale=0.6} 
\caption{The specific angular momentum
of the disk as a function of the rotation velocity at $2.2\,R_D$: Comparison
with the NFW $+$ Mo et al. model.
\textit{Black (lower) dots}: the result from this model, adopting the
Burkert profile; \textit{pink (upper) dots}: the result from this model, adopting the
NFW profile; \textit{blue dashed line} the Mo et al. model; \textit{red line}:
the best-fit relation by \cite{navarrosteinmetz}, with the 1-$\sigma$
(yellow) and 3-$\sigma$ (light blue) regions.}
\label{confronto}
\end{figure}

In Fig.~(\ref{tully}, \textit{left}) I show the specific angular momentum of
the disk, computed as $j_D=J_D/M_D$ from Eq.~(\ref{Jd}), as a function of
the total circular velocity at $r=2.2\, R_D$.
Plotted for comparison is also
the best-fit relation by \cite{navarrosteinmetz} from their
collection of data; note that these authors adopted a flat rotation
curve, so that $f_R=2$ and $j_D=2\,R_D\,V_H$. 

Fig.~(\ref{confronto}) shows a comparison between different models for the
halo and baryonic mass distribution. The \textit{black dots} represent again the
specific disk angular momentum as a function of the total circular velocity
in the current model, as in Fig.~(\ref{tully}); the \textit{pink dots} correspond to the same
model, but with a halo described by the NFW profile; the \textit{blue dashed line} is
the result by the Mo et al. model; the \textit{red line} is again the empirical
relation by \cite{navarrosteinmetz}, with the \textit{yellow} and \textit{light blue shaded
areas} representing its 1-$\sigma$ and 3-$\sigma$ regions respectively.   
Clearly, the model described here is the most successful in reproducing
the empirical data.

I derive the halo angular momentum by assuming the conservation of
the total specific angular momentum between the dark matter and the baryons:
\be
J_H=J_D \,\frac{M_{H}}{M_D}~, 
\label{haloangmom}
\ee
an \textit{ansatz} widely supported/adopted in the literature 
(\cite{momaowhite},\cite{vdb01b},\cite{vdb02},\cite{burkertdonghia},\cite{peirani}).
Notice that the specific angular momentum conservation 
yields the maximum disk angular momentum, in the absence of any dissipation
of $j_z$ during and after the baryonic collapse. In any other case, 
a more general expression should be of the form $J_H=J_D \, \alpha(t,M)
\,(M_{H}/M_D)$, where $\alpha(t,M) \geq 1$ takes into account dissipation of
$j_z$ and in principle depends on the halo mass and varies with time. Therefore the halo
angular momentum may in principle be larger than that yielded by
Eq.~(\ref{haloangmom}), even if unlikely (see the Discussion). 

Small variations of $J_D$ are
magnified by a factor $M_{H}/M_D$ in the value of $J_H$, i.e., the
latter is rather sensitive to the mass and radial extension of the baryons, 
as shown in Eq.~(\ref{Jd}). For this reason I included in the
computation, along with the stars, the gaseous component that
envelops the disk of spiral galaxies. 

The gas-to-baryon fraction in spiral galaxies is a decreasing function of disk
mass, with a maximum ratio of $\sim 0.5$ for small spirals;  
I derived the total mass of
the gas component from the disk luminosity (see above) through the
relation
\be
M_{\mathrm{gas}}= 2.13\times 10^6 \, M_{\odot}\,
\left(\frac{L_B}{10^6\, L_{\odot}}\right)^{0.81} \, \left[1-0.18 \,
\left(\frac{L_B}{10^8\, L_{\odot}}\right)^{-0.4}\right]~
\label{Mgas}
\ee
by \cite{ps99}, where I included a factor $1.33$
to account for the He abundance. 

The tiny contribution to the total mass leaves the rotation curve 
virtually unaltered (in fact, $V_{\mathrm{gas}} \sim
\sqrt{M_{\mathrm{gas}}/R_{\mathrm{gas}}}$). 
However, the gas is much more diffuse than the
stars, reaching out to several disk scale-lengths (\cite{corbelli},\cite{dame}),
and since most of the angular momentum comes from
material at large radial distances (\cite{vdb01b}, and Eq.~(\ref{Jd})),
I expect the gas to add a significant contribution to the total
angular momentum, especially in small spirals.

The detailed density profile of the gas in spirals is still under
debate in the literature. However, I am confident that the main
factors entering the computation of the gas angular momentum
$J_{\mathrm{gas}}$ are just the gas total mass $M_{\mathrm{gas}}$
and the radial extension of its distribution, in analogy with Eq.~(\ref{Jd}); 
in other words, I expect the details of the gas profile not to significantly
affect the results. 
In order to check this statement, I computed the total gas angular momentum 
for $3$ different gas models, \textit{i.e.} (i) a disk-like distribution
(DL), with scale length $\alpha R_D$; (ii) a uniform distribution
(U) out to a radius $\beta R_D$; and (iii) an $M33$-like gaussian
distribution (M33; \cite{corbelli}):
\bea
\nonumber\Sigma_{\mathrm{gas}}^{\mathrm{DL}}(r) &=&
{M_{\mathrm{gas}}\over 2 \pi\, \alpha^2 R_D^2}\, e^{-r/\alpha
R_D}\\
\nonumber\\
\Sigma_{\mathrm{gas}}^{\mathrm{U}}(r) &=& {M_{\mathrm{gas}}\over
\pi\, \beta^2 R_D^2}\, \theta\left(r-\beta
R_D\right)\\
\nonumber\\
\nonumber\Sigma_{\mathrm{gas}}^{\mathrm{M33}}(r) &=&
{M_{\mathrm{gas}}\over \pi\, (2 k_1^2+k_2^2)\, R_D^2}\, e^{-(r/k_1
R_D)-(r/k_2 R_D)^2}~, \label{gas}
\eea
where $\theta$ in the second equation is the Heaviside step
function. As fiducial values of the parameters, we adopt
$\alpha\approx 3$ in the first expression, $\beta\approx 6$ in the
second one \cite{dame}, and $k_1\approx 11.9$, $k_2\approx 5.87$ in
the last one \cite{corbelli}. Each profile has been
normalized to the total gas mass $M_{\mathrm{gas}}$ as computed from
Eq.~(\ref{Mgas}).

As in Eq.~(\ref{Jd}), the gas angular momentum will be
\be
J_{\mathrm{gas}}= 2\pi\,\int_0^{\infty}\, \Sigma_{\mathrm{gas}}(r)\,
\, r \, V_{c}(r)\, r\, dr = M_{\mathrm{gas}} \, R_D \, V_{H} \,
f_{\mathrm{gas}}~,
\label{Jgas}
\ee
where the shape factor $f_{\mathrm{gas}}$ encodes the specific
gas distribution. On comparing its values for the three models I
found differences of less than $15\%$, and so confidently choose the
gaussian profile as a baseline.

It is time to compute the halo angular momentum as a function of the total
baryonic one:
\be
J_H=(J_D+J_{\mathrm{gas}}) \ \frac{M_{H}}{M_D+M_{\mathrm{gas}}}~.
\label{tothaloangmom}
\ee
The gas is dynamically affecting the system mainly through its
different spatial distribution with respect to that of the stars,
adding an angular momentum component that is relevant at radii larger
than $R_D$. The final spin parameter turns out to be significantly different
if the gas is included; this is a conservative 
case, as will be clear in the next Section.

Note that I do not include a bulge component, since it would
contribute with a negligible angular momentum and a mass of $0.2\, M_D$
at most; in any case, this is again a conservative assumption, since the bulge
would slightly lower $J_H$ after
Eq.~(\ref{tothaloangmom}) and, as will be evident in the next
Section, would lower the spin parameter and strengthen my
conclusions.

\section{The spin parameter}

\noindent
The spin parameter is a powerful tool to investigate galaxy
formation, as it is strictly related to both the dynamics and the
geometry of the system. By studying its dependence on halo mass and its
distribution function across the galaxy population, I can gain some insight  
on the mechanisms of mass accretion and the history of halos. 

The spin parameter is defined as 
\be
\lambda=\frac{J_H\, |E_H|^{1/2}}{G \, M_{H}^{5/2}}~, 
\label{lambda}
\ee
where $G$ is the gravitational constant and $E_H$ is the total
energy of the halo. The latter is computed as
$|E_H|=2\pi\int{dr}\,r^2\,\rho_H(r)\,V^2_{c}$ after the virial
theorem, assuming that all the DM particles orbit on
circular tracks.

\begin{figure}
\epsfig{file=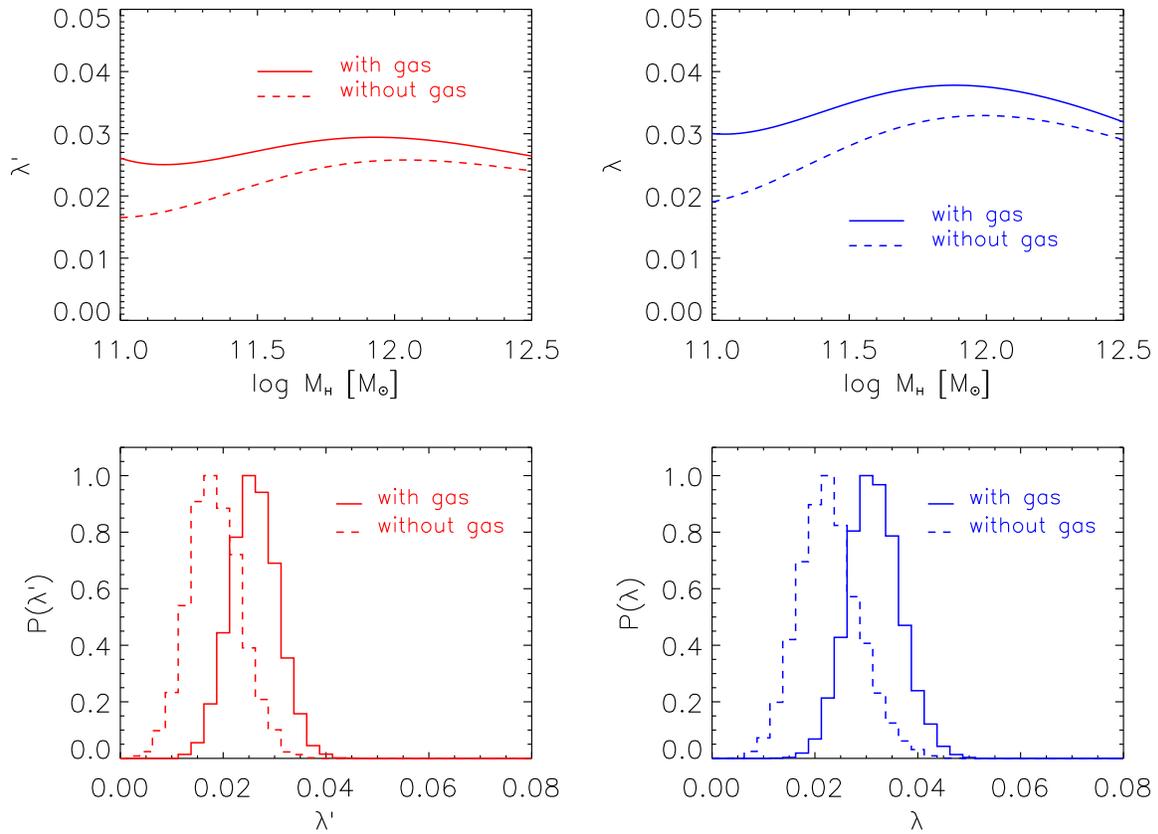,scale=0.5}
\caption{The spin parameter and its
distribution function. \textit{Top} panels: $\lambda'$
(\textit{left}) and $\lambda$ (\textit{right}) as a function of the
halo mass, when the gas component is included in the system
(\textit{solid line}) and when it is not (\textit{dashed line}).
\textit{Bottom} panels: the distribution function of $\lambda'$
(\textit{left}) and $\lambda$ (\textit{right}), again with gas and
without gas.}
\label{spinfig}
\end{figure}

An alternative definition used in simulations \cite{bullock} is given by the
first equality in the following:
\be
\lambda'=\frac{J_{\mathrm{H}}}{\sqrt{2}\, M_{H} R_{H}\,
V_{H}}=\frac{J_H+J_D+J_{\mathrm{gas}}}{\sqrt{2}
(M_{H}+M_D+M_{\mathrm{gas}})\, R_{H} V_{H}}~; 
\label{lambdap}
\ee
the definition is such that $\lambda=\lambda'$ for a NFW halo.
I found the second equality after Eq.~(\ref{tothaloangmom}), and
determined that for Burkert halos the ratio $\lambda/\lambda'$ is between
$1.1-1.3$ in the mass range $10^{11} - 3 \times 10^{12}\,
M_{\odot}$. In Fig.~(\ref{spinfig}) (\textit{top panels}) I plot both $\lambda$ and
$\lambda'$ as a function of the halo mass. I also highlight the
difference in the value of the spin parameter when the gas component
is included, especially in low mass halos.
As is clear in Fig.~(\ref{spinfig}), there is no strong evidence of a correlation between
the spin parameter and the halo mass; the halo angular momentum originates
from tidal torques during the episodes of mass accretion throughout its
history, and the similarity across the whole mass range suggests a 
common formation pattern for all halos, in agreement with the hierarchical scenario.   

To compute the probability distributions $\mathcal{P}(\lambda)$ and
$\mathcal{P}(\lambda')$ of the spin parameters, I exploited the
galactic halo mass function, i.e., the number density of halos with
mass $M_{H}$ containing a single baryonic core \cite{shankar}. A good fit is provided by the Schechter
function $\Psi(M_{H})=(\Psi_0/ \overline{M}) \, (M_{H}/
\overline{M})^{\alpha}\, \exp{(-M_{H}/\overline{M})}~,$ with parameters
$\alpha=-1.84$, $\overline{M}=1.12\times 10^{13}\, M_{\odot}$ and
$\Psi_0=3.1\times 10^{-4}\, $ Mpc$^{-3}$; note that within our range
of halo masses, this is mostly contributed by spirals. For the
computation of $\mathcal{P}(\lambda)$ or $\mathcal{P}(\lambda')$, I
randomly picked a large number of masses distributed according to
$\Psi(M_{H})$, computing then $\lambda$ and $\lambda'$ for each of them using
Eqs.~(\ref{lambda}) and (\ref{lambdap}), and I eventually built up the
statistical distributions. During this procedure I convolved
the relations (\ref{lambda}) and (\ref{lambdap}) with a gaussian
scatter of $0.15$ dex that takes into account the statistical
uncertainties in the empirical scaling laws adopted in this work; these are
mostly due to the determination of $R_D$ through Eq.~(\ref{mdrd}),
for which I obtained the scatter by using the disk mass
estimates of individual spirals \cite{ps90}.

As shown in Fig.~(\ref{spinfig}) (\textit{bottom panels}), I find a distribution peaked
around a value of about $0.03$ for $\lambda$ and about $0.025$ for
$\lambda'$, in the case when the gas is considered. 
This value of $\lambda'$ is close to the result of the simulations by \cite{donghia},
who on average find $\lambda'=0.023$ for spirals
quietly evolving (i.e., experiencing no major mergers) since
$z\approx 3$, see their Figure 4. In addition, \cite{burkertdonghia}
argue that this value of $\lambda'$ provides a very good fit
to the observed relation between the disk scale-length and the
maximum rotation velocity (see their Figure 1). 
Moreover, the peak value of $\lambda$ is in agreement with the
results by \cite{gardner}, \cite{vivitska} and \cite{peirani},
who find a distribution centered
around $0.03$ for halos that evolved mainly through smooth
accretion.

In addition, notice the effect of the gas contribution on the peak of the
distribution; if it is not taken into account, the values of the halos' angular momenta
are underestimated, and the offset with simulations is even more striking.

\section{Discussion and Conclusions}

\begin{figure}
\epsfig{file=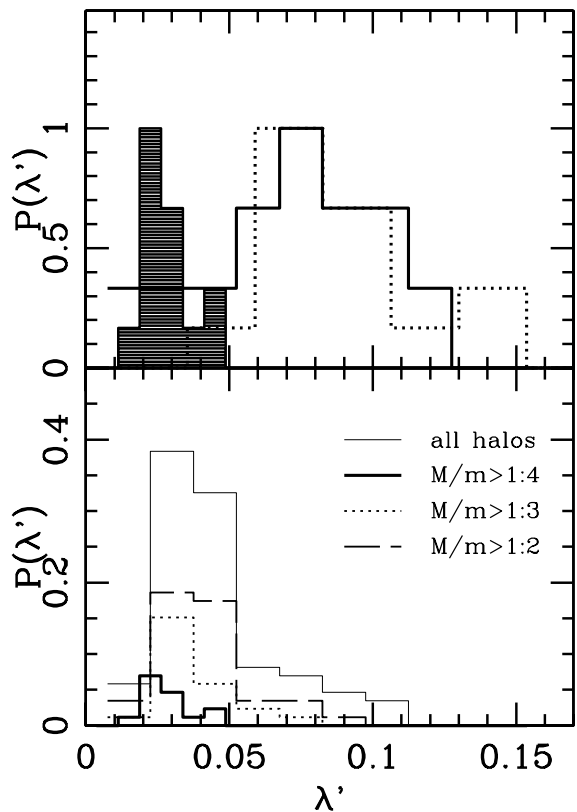,scale=0.7}
\caption{\textit{Top panel}: the spin parameter distribution function as found 
by \cite{donghia}, for halos that did not experience any major 
merger since z=3 (\textit{dashed region}). \textit{Bottom panel}: the spin
  parameter distribution 
function of the entire halo population (thin solid line).
Sub-populations are also plotted if one assumes that the definition of a major mergers requires
mass ratios of $>1:4$, $>1:3$, and  $>1:2$, respectively, since
z=3.} 
\label{donghiafig}
\end{figure}

\noindent
The spin parameter distribution functions obtained from numerical simulations
based on the $\Lambda$CDM framework performed by
various authors (\cite{bullock},\cite{donghia} and references therein),
show peak values of $\lambda' \geq 0.035$ for the whole halo catalogue,
significantly higher than our empirical value. 
However, \cite{donghia}
highlight the fact that, considering only halos that
have not experienced major mergers during the late
stages of their evolution ($z\la 3$), the average spin parameter
$\lambda'$ turns out to be around $0.023$, very close to our
observational result (see Fig.~\ref{donghiafig}). 
Moreover, \cite{gardner} and \cite{peirani}
showed that the spin parameter $\lambda$ undergoes different
evolutions in halos that have grown up mainly through major mergers
or smooth accretion: in the former case $\lambda$ takes on values
around $0.044$, while in the latter case $\lambda$ has lower values
around $0.03$.

The results presented in this Chapter have therefore a precise physical
meaning; the empirical determination of the spin parameter distribution
function, on the basis of the observed geometry and relative mass
distribution of the dark and luminous components,  
shows that spiral galaxies are hosted in halos characterized by a  
relatively poor history of major merging events, that grow mainly through
quiet accretion from $z \simeq 3$. 

Moreover, the distribution is quite tight, reflecting the very weak dependence
of the spin parameter on the halo mass, as pointed out in the previous
Section; this excludes significant frequent events of disk-disk collisions and
gravitational fly-bys in the majority of the spiral systems, 
because they would introduce a large scatter in the values of $\lambda$ and
$\lambda'$ due to the high number of different final states produced.  
 
This scenario is not inconsistent with the hierarchical picture of structure
formation, but challenges the proposed mechanisms for the formation of
spirals, based on major mergings up to recent times ($z \leq 1$) \cite{springel05}.

Another feature of these distribution functions strikes as impressive:  
the absolute value of $\lambda$ and $\lambda'$ is quite
small. On average, the degree of ordinate global rotational motions inside DM halos
is around $3\%$. 
By comparing these values with the disk's $\lambda \simeq 0.4$ (\cite{white2}; it
corresponds to $V_{rot}/V_c=1$), and
supposing that at the early stages of the baryonic cooling the luminous and
dark components share the same phase-space region (a case consistent with the specific
angular momentum conservation scenario), 
it is clear
that the formation of the galaxy inside the DM halo happens with a significant
transfer of angular momentum towards the central regions of the system, as  
the baryons carry their angular momentum with them during their collapse.

In a spiral galaxy, the collapse happens preferentially along one direction,
taken as the spin axis $z$, so that it is a good aproximation to suppose that
the specific $j_z$ along this direction is conserved. This is the reason why the disk angular
momentum is a good proxy (through a scaling) to the halo angular momentum. 

However, the total angular momentum $L^2$ of the baryonic component is not
constrained to be conserved during the collapse, and in fact, as will be
discussed in Chapter 5, it is plausible that a fraction of the baryonic 
total angular momentum, while carried towards the center of the halo,
is lost to the DM via processes like dynamical friction.

%% file: urc.tex
\chapter{Breaking the self-similarity}

\noindent
I present an observational counterpart to the NFW to model the mass
distribution of systems of disk galaxies embedded in dark matter halos.
The model is based on the observational scalings between the dark and luminous
components discussed in Chapter 2, and 
produces a one-parameter family of curves that can be used to fit the observed
rotation curves of spirals.
The validity of the model resides in its being observation-based and 
therefore reliable in reproducing real systems; 
it shows that dark matter halos hosting spirals are not
self-similar, and that the shape of their mass distribution features strong
trends with the galactic mass.

\section{Introduction}

\noindent
Numerical simulations of the hierarchical clustering predict the
equilibrium density profile of DM halos to be described by the one-parameter
NFW family of curves, that features self-similarity on the whole mass spectrum:
\be
\rho_H (r)= {M_{vir}\over 4\pi R_{vir}^3}\, {c^2\,g(c)\over x\,(1+cx)^2}~,
\ee
where $x\equiv r/R_{vir}$ is the radial coordinate, $M_{vir}$ and $R_{vir}$ are the virial
mass and radius, $c$ is the concentration parameter $c\approx 9.5
(M_{vir}/10^{12}M_{\odot})^{-0.13}$,
and $g(c)= [\ln(1+c)-c/(1+c)]^{-1}$ (see Chapter 1 for a more detailed description).
The corresponding velocity features the same self-similarity:
\be
V_{\rm NFW}^2(r)= V_{vir}^2 \frac{c}{g(c)} \frac {g(x)}{x}~,
\ee
with $V^2_{vir}=GM_{vir}/R_{vir}$. 
The total circular velocity of a system made of a disk galaxy embedded in a DM
halo, build through
the Mo et al. \cite{momaowhite} model described in Chapter 1, is predicted
to be a \textit{universal} function of radius.

On the other hand, observations of galaxy rotation curves, up to now still the
most reliable and abundant source of data on the mass distribution of the DM
in galaxies, are at odds with the theoretical predictions and show a shallower
cusp-less profile (among them, \cite{ps91},\cite{vdb01},\cite{swaters1},
\cite{weldrake},\cite{donato},\cite{gentile05},\cite{simon},\cite{gentile07}),
usually parameterised by the Burkert halo \cite{burkert}. 
A number of studies cast doubts on the reliability of the mass modelling
procedure and the data analysis of some galaxies (\cite{vdb01},\cite{swaters1},
\cite{valenzuela}), or propose
different explanations for the discrepancy, like triaxiality effects \cite{hayashi}.
The debate is still on, but the evidence for a cored, Burkert-like DM profile is getting stronger
(\cite{gentile04},\cite{gentile05},\cite{gentile07},\cite{deblok2},\cite{deblok3});
\cite{catinella} and \cite{swaters2} confirm the Burkert profile to be the one that most
successfully reproduces the observed profiles of spirals, with an analysis of
2200 curves and 60 extended curves respectively (see also \cite{willick},\cite{yegorova}).
Persic and collaborators \cite{pss96} presented an empirical model called the 
Universal Rotation Curve (from now on URC$_0$), a family of 
two-parameters velocity profiles that reproduce most of the systems, with the
exception of bulge-dominated spirals (Sa-type) and dwarves; the DM density profile
is described by the Burkert halo, and the disk luminosity and scale-length 
are the free parameters discriminating different systems. 
Independent analysis by \cite{rhee} and \cite{roscoe} of the same samples 
are in agreement with the URC$_0$ description;
the predictions of the model have been tested by \cite{corteau} and
\cite{verh} on additional different samples.

The discrepancy between observations and simulations regarding the structure
of DM halos may stem from the fact that we observe the luminous component to 
understand the dark one; if we combine this with the structural difficulties of Nbody-SPH codes to
penetrate the finest processes of galaxy formation and baryonic collapse, and
the limitations of observations due to the uncertainties and model-dependent
assumptions, the picture does appear a bit confused.

Yet, aknowledging the problem leads half-way to the solution. 
What this Thesis proposes is based on a simple consideration; if when
observing the baryons we don't recover what the theory predicts for DM halos,
then the cause of the discrepancy may well reside in the baryons themselves.
In Chapter 2 I showed how the structural parameters of disk
galaxies, as yielded by observations, lead to a different conclusion regarding
the spin of DM halos with respect to the theoretical predictions. The same
line of reasoning can be used to infer an observational counterpart to the 
NFW velocity profile, that features the same degree of universality even if it
entirely relies on empirical laws. And again, this will show how, by taking
the baryons into account, another feature of the theory fails to show up in real
systems, namely the self-similarity of dark matter structures.

\section{The rotation curves of spirals and the global properties of halos}

\noindent
To gain information about DM halos from the galactic observables, I need to 
overcome the difficulty of relating the baryonic structure, that dwells 
in the central regions of the system, to the whole equilibrium structure of the
halo. 

For such a purpose, rotation curves alone are not a reliable tool, since
the most recent and extended ones barely
reach radial distances of $\leq 30\%$ of the virial radius.
In the case of the URC$_0$, the model suffered from three main limitations; 
(i) it strictly held in a region extended less than $\sim 5\%$ of 
the halo size; (ii) the velocity profile of the halo component was fine-tuned to
reproduce the data in this small region, and was not suitable for
extrapolation to larger radii, in regions of cosmological interest; 
(iii) the free parameters of the
family of curves were the disk luminosity and scale-length, thus introducing
uncertainties due to the need of assuming a mass-to-light ratio, 
dependent on star formation rate, stellar evolution and extinction models.

Here I present a model for the velocity profiles 
of spirals that is based entirely on dynamical observations, and is able to make
predictions on the DM halo global structure. Such a model can be used as an observational
counterpart to the theoretical scenarios as the NFW, to predict the structure
of single halos as well as mass and spin distributions to be used in
statistical studies. It yields a one-parameter family of curves, so as to 
be as general as the NFW and equally straightforward to use and while it
reproduces the data coming from the
local matter distribution in the inner parts of the halo, 
at the same time it predicts the halo global properties, like
virial mass and spin, given the galactic mass alone.

The model described in Chapter 2, based on the observed scaling relations
between disk and halo, 
provides the tool to build a Universal Rotation Curve determined by
the whole equilibrium structure of
the DM halo, thus consistently extending it to the dynamical edge of the
system. 

Consider a Burkert halo hosting a spiral galaxy; the dark matter 
cumulative mass profile is described by
Eq.~(\ref{burkertmass}), and the corresponding halo circular velocity profile
is 
\be
V^2_{H}(r)= 6.4 \ G \ {\rho_0 R_0^3\over r} \Big\{ ln \Big( 1 +
\frac{r}{R_0} \Big) - \tan^{-1} \Big( \frac{r}{R_0} \Big) +{1\over
{2}} ln \Big[ 1 +\Big(\frac{r}{R_0} \Big)^2 \Big] \Big\}~, 
\label{burkertvelocity}
\ee
with $R_0$ and $\rho_0$ being the core radius and density
respectively. Provided that $R_0 \ll R_{vir}$, this converges to the NFW profile outside the
core. The disk velocity profile is given by Eq.~(\ref{vdisk}).

\begin{figure}
\centering\epsfig{file=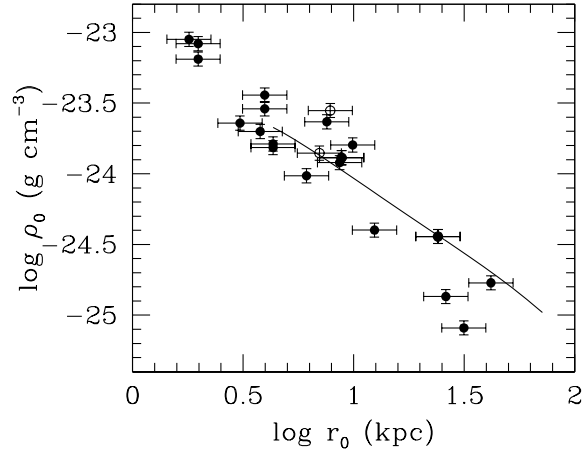,scale=0.4} 
\caption{The core radius \textit{vs} virial mass relations obtained in the
  present work (\textit{solid line}), compared with fitted values from single
curves (\textit{dots}, see text for references).} 
\label{rhor0fig}
\end{figure}

To completely determine the total velocity profile $V^2(r)=V^2_D(r)+V^2_H(r)$ 
as a function of one free
parameter, I use the empirical scalings described in 
Chapter 2, and
summarized in Fig.~(\ref{scalings}); namely, the halo mass - disk mass
relation of Eq.~(\ref{MvirMd}), obtained through a study of the halo
occupation statistics, 
the disk mass - disk scale-length relation of 
Eq.~(\ref{mdrd}), compared between a number of different authors (see 
Chapter 2), 
the halo central density - disk scale-length relation of
Eq.~(\ref{rhoMd}), determined by \cite{salucciburkert} from a sample of
extended rotation curves (up to $\sim 15\%$ of $R_{vir}$), and finally the halo mass - core radius
relation, derived by inserting the above equations into
\be
M_H(R_{vir})=M_{vir}~,
\label{mvirr0}
\ee
to extract $R_0$ for any given halo mass (see Eq. \ref{r0mh}).  
The present derivation of $R_0$ is very solid; in fact, errors 
up to a factor of $2$ in the mass determination
lead to an uncertainty in $R_0$ of less than $40 \%$, while errors in the
outer halo velocity slope in any case do not affect it by more that $\sim 10
\%$. In comparison, the pure determination of $R_0$ in the URC$_0$
obtained by fitting the central velocity profile  
($r \le 0.05 R_{vir}$), was subject to large uncertainties, namely $\delta
R_0/R_0 \sim 0.3-0.5$. 
In addition, the core radius obtained from global scalings works equally
well in reproducing the single inner rotation curves, than its counterpart
defined only by the inner kinematics, as is shown in Fig.~(\ref{rhor0fig}); 
the solid line represents the $\rho_0$-$r_0$ relation as obtained from the
current model, while the dots represent fitted values from single curves extending to a
maximum radius of $5-15 \%$ of the virial radius 
(\cite{donato},\cite{gentile04},\cite{salucciborriello}).

The velocity profile thus obtained is a one-parameter family of curves,
extended to the halo virial radius, and is determined by the whole halo
equilibrium structure (instead of being fitted to match the central $\sim
5\%$ of the profile). Being the free parameter the galactic mass, 
it does not suffer from the model-dependent uncertainties that affected
the URC$_0$, and makes the URC directly comparable with the results of numerical simulations.
Further improvement is brought about by the newer and more complete sample of
data for the determination of the scalings used in Chapter 2 (in particular,
curves reaching out to several disk scale-lengths, see \cite{salucciburkert}), 
with respect to \cite{pss96}.

Given the galactic mass (or the halo mass, or disk scale-length),
the halo structural parameters are determined, and 
the model yields the circular velocity at any radius,
with an error that is an order of magnitude smaller than the variations occurring
among different radii and different galactic masses (see Appendix A for a
complete description of the mass modelling uncertainties). Therefore, given
the observation of the inner rotation curve of a disk galaxy, the URC
can be practically used to determine the global disk and halo structural
parameters, with no degeneracy.

\section{Results and discussion}

\begin{figure}
\centering\epsfig{file=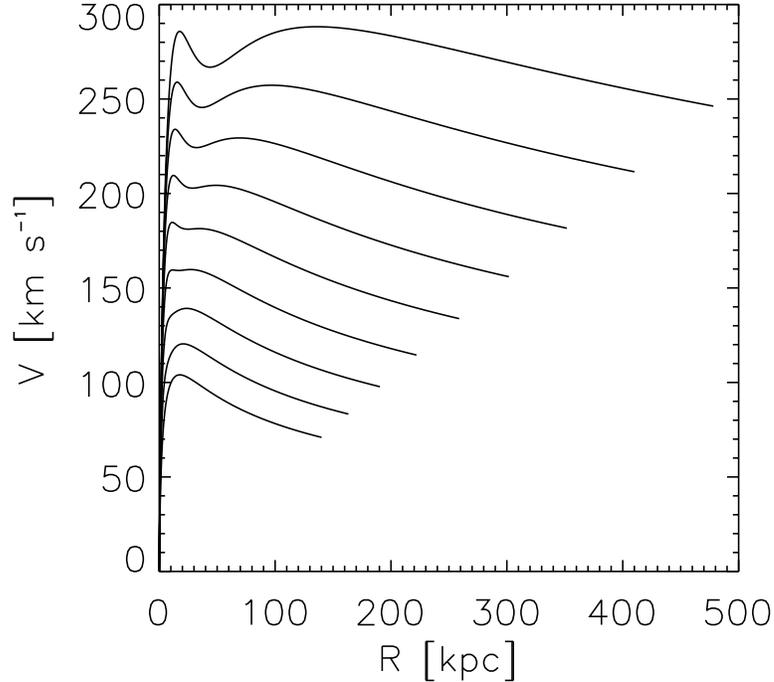,scale=0.7} 
\caption{The Universal Rotation Curve in physical units. Each curve
corresponds to $M_{vir}=10^{11} \ 10^{n/5} \rm M_\odot $, with $n =
1 \ldots 9$ from the lowest to the highest curve.} 
\label{urc1}
\end{figure}

\noindent
In Fig.~(\ref{urc1}) I show the URC in physical units, in the halo 
mass range $10^{11}\, \rm M_{\odot}\la M_{vir} \la 10^{13}\, \rm M_{\odot}$;  
the amplitude of the curve is obviously determined by the halo mass, but
the latter seems to affect the shape as well, contrary to any claim of
self-similarity across the mass spectrum. I'll be back on this issue with the
next plots. 

In the meantime, notice the contribution of
the baryonic component, negligible for small masses but increasingly
important in the larger structures, that mirrors the behavior of the
$M_{vir}-M_D$ relation. The baryonic peak becomes visible in the
curve for systems with $M_{vir} \sim 10^{12} M_{odot}$, while for lower masses
the inner curve is DM-dominated down to the center. For increasing
galactocentric distances, the halo eventually becomes
the dominant mass component in all systems, but it does so at different radii
according to the virial mass: from $\sim 10^{-2} R_{vir}$ for the
smallest objects, to $\sim 10^{-1} R_{vir}$ for the massive ones.
Remarkably, the maximum value of the circular
velocity occurs at about $15 \pm 3$~kpc, independent of the galaxy
mass, but due to different components: this seems to be a 
main dynamical imprint of the DM - luminous
mass interaction occurring in spirals. 

Furthermore, Fig.~(\ref{urc1}) shows
that the "Cosmic Conspiracy" paradigm has no observational support:
there is no fine-tuning between the dark and the stellar structural
parameters to produce the same particular velocity profile in all objects
(e.g. a flat one). Conversely, the scalings between the
parameters produce a variety of profiles.
Moreover, the peak velocity of the stellar component
$V_{disk}^{peak} = V_D(2.2 R_D) = G M_D/R_D\ k$, with $k =const$, is
not a constant fraction of the virial velocity as is found in
ellipticals, (i.e $\sigma \propto V_{vir}$), but it ranges between
the values $1$ and $2$ depending on the halo mass.

Notice how the URC profiles
are found to be (moderately) decreasing over most of the halo radial
extent; the available kinematical data (\cite{pss96},\cite{salucciburkert},
\cite{salucciborriello},\cite{donato},\cite{gentile04})
show that both at the last measured point (between $5$ and $15 \%$ of the virial radius)  
and at $r \sim 3R_D$, the velocity $V(r)$ is significantly higher than
$V_{vir}$ (of about $10 - 30
\%$). The same behaviour is found with the NFW and Burkert profiles, 
showing that the assumption
of flat rotation curves, often adopted to simplify calculations no matter the
density profile, is not
observationally supported, even as an asymptotic behavior at large radii. 

\begin{figure}
\centering\epsfig{file=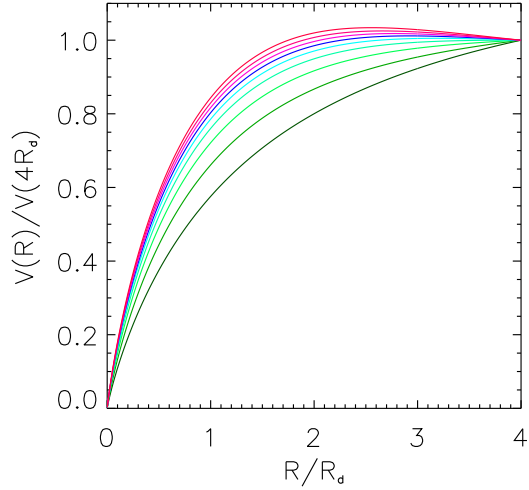,scale=0.5} 
\caption{The inner URC, normalized at
its value at $4R_D$, as a function of 
$r/R_D$. The single curves correspond to the same masses 
of Fig.~(\ref{urc1}), with the
lowest curve representing the lowest mass.}  
\label{innerurc}
\end{figure}

In Fig.~(\ref{innerurc}) I plotted the inner velocity profile, 
in the radial range including the luminous regions of spirals,
normalized at a radius $r/R_D=4$.
There is an inverse correlation between the average steepness of the profile 
slope and the halo mass, due manly to the $M_{vir}-M_D$ relation; this is
similar to the slope-luminosity relationship found by \cite{ps88}.

\begin{figure}
\centering\epsfig{file=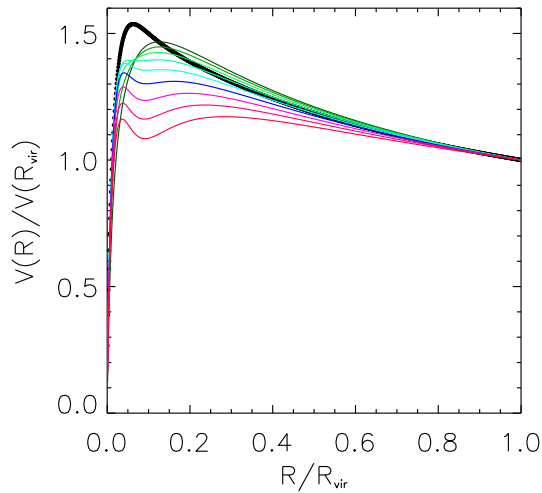,scale=0.5} 
\caption{The URC normalized at its
virial value $V_{vir}$, as a function of $ x = R/R_{vir}$. 
The single curves correspond to the same masses of Fig.~(\ref{urc1}), with the
lowest curve representing the highest mass. The solid black line is a pure NFW.}  
\label{urcDM}
\end{figure}

\begin{figure}
\centering\epsfig{file=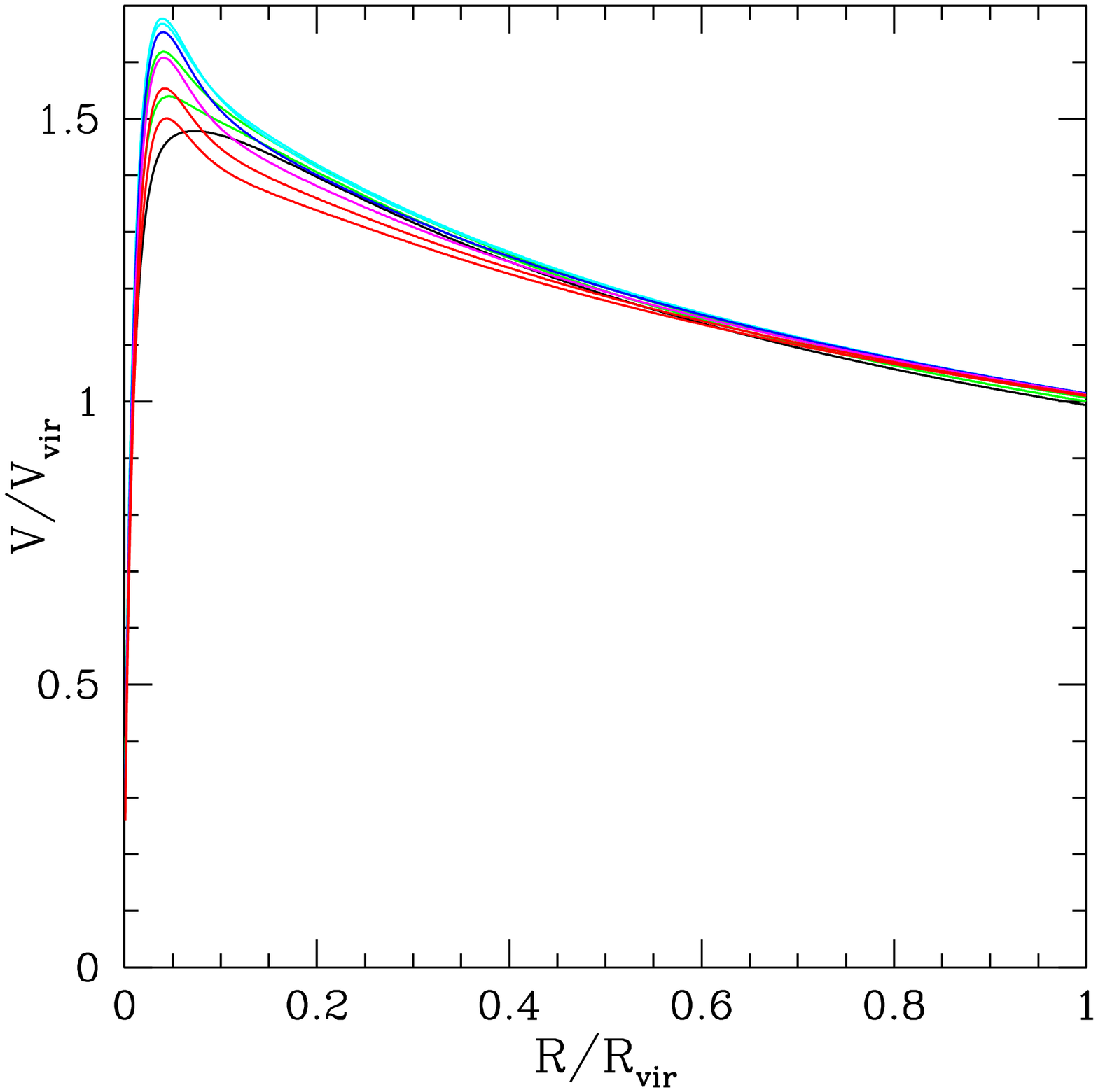,scale=0.33} 
\caption{The same set of curves of Fig.~(\ref{urcDM}), but with the NFW
  density profile substituting the Burkert; notice the amplitude of the ``baryonic break''
  of the self-similarity.}
\label{urcNFW}
\end{figure}

In Fig.~(\ref{urcDM}) I show the same curves from a DM perspective,
normalizing the velocities to $V_{vir}$ and the radii to $R_{vir}$, comparing
them with a pure NFW profile (\textit{black line}). 
Although the URC is a one-family of curves, there is a remarkable, strong 
mass-dependent systematics: clearly, this curves are not self-similar.

There is a duplice source for the non-similarity. On the one hand, the
scalings between the baryons and the DM adopted in this model are not linear
functions of the mass; the mass-to-light ratio is not constant on the mass
spectrum, nor is the disk scale length-to-virial radius ratio, and in general
terms the baryonic mass distribution strongly depends on the total mass,
leading to a ``baryonic break'' of the self-similarity. To quantify this effect, 
I plot in Figure~(\ref{urcNFW}) the rotation curves of galaxies of the same masses as in 
Fig.~(\ref{urcDM}), after
substituting their observed Burkert halos with the theoretical, self-similar NFWs 
(with the same normalization),
so that the offsets between
the curves depend only on the different baryonic distributions according to the
empirical scalings.

On the other hand, a comparison between Figs.~(\ref{urcDM}) and (\ref{urcNFW})
shows that in the observed systems, where the halo is better represented by the 
Burkert profile, the baryons alone cannot account for the differences in shape and amplitude
between curves of different masses; rather, this is an 
evidence of a ``DM break'' of the self-similarity, with
the halo mass distributon that changes 
depending on the galactic or virial mass, as shown also in Fig.~(\ref{urc1}).

Given the empirical base of the model, observations remarkably show that
galaxies of different masses reside in halos of different shapes, with the
luminous and dark components linked through smooth monothonic scalings. 
The amplitude of the ``DM break'' of the self similarity is significant but
small compared to the amplitude of the whole rotation curve, and it manifests itself
as a perturbation of the central halo equilibrium structure, as a function of the
disk mass; this strongly hints to a halo reaction to the baryonic
presence. 

An underlying physical mechanism that, most likely during galaxy
formation, may affect the equilibrium structure of both components, luminous and
dark, which react to each other and readjust their mass distribution (see
Chapter 5), could be a way out of an old debate.
The NFW halo is the final equilibrium state of DM structures in numerical simulations,
but is not the preferred profile yielded by observations; in Figure~(\ref{urcNFW}) it
looks quite static, equally unaltered by the presence of baryons, even in a wide range
of disk masses. The gap between Figs.~(\ref{urcDM}) and (\ref{urcNFW}) 
could be of an evolutionary nature; the primeval NFW halo, dominated
entirely by DM dynamics at early stages, is perturbed by galaxy formation,
to an amount depending on the size and mass of the galaxy itself, and changes its equilibrium 
structure into a cored configuration.

%% file: nfw.tex
\chapter{The dark matter distribution at the edge of spirals}

\noindent
I present some evidence that the discrepancy between observations and numerical
simulations regarding the shape of the mass distribution in dark matter halos
extends to the whole region where the luminous component is present, with the 
observed density profile not converging to the NFW even outside the halo core.
This indicates that such discrepancy
cannot be due to numerical effects or pure microscopic physics; the observed
excess of mass around $5-30\%$ of the virial radius
hints to a mechanism of mass redistribution, probably occurring during the disk
formation, that offsets the density profile in the whole central region of the
dark matter halo. 

\section{Introduction}

\noindent
The CDM halo structure, as predicted by numerical simulations, has not yet
found a physical explanation (\cite{taylor},\cite{white2}).
The phase-space structure of the NFW halo looks like an attractor for
hierarchical clustering in different cosmogonies \cite{white2}, and the self-similarity and the
intrinsic simplicity of its shape make for a very attractive picture of
structure formation, that is particularly successful in reproducing the
observed large-scale structure and the objects at cluster scales, and fails to
reproduce galaxy-size systems.

As discussed in Chapter 1, the invariance of the predicted halo 
structure at all scales, together with
the solidity of the final density profile against different initial conditions 
like the details of the power spectrum or the cosmological parameters,
points towards a universal mechanism of assembly characterised by 
processes like violent relaxation, 
that erase the memory of the initial state of
the system prior to each main accretion event, and makes the final structure
converge towards the NFW \cite{white2}. 

It is not clear to what amount the numerical treatment is responsible for this 
formation pattern. An example is the cusp feature, that may be a spurious
product of an intrinsic deficiency of simulations in treating small-scale
gravitational interactions. 
Mixing processes are not scale-invariant, and there are fundamental
differences in the behaviour of a system after coarse-graining;   
the numerical approach is not suitable to address such issues, almost by
definition, and the entity of the unavoidable approximation in representing 
such mechanisms is unknown.
These effects must manifest themselves at small scales, and it is not a
cohincidence that the claimed discrepancies between simulations and
observations mainly regard the inner halo structure and the amount of substructures.

However, in this Chapter I present recent observational evidence 
that the cusp-core discrepancy is just part of a more
serious and general offset betweent the CDM predicted profiles and the
observed ones, at least in some galaxies (as suggested by \cite{Mi6}).
The state-of-the-art
high quality rotation curves of spirals are extended out to several disk scale-lengths, and allow
to probe regions that previously were observationally off-limits;
for these systems, the density profile does not converge to the 
NFW even outside the halo core, but shows an
interesting excess of mass around $5-6 \ R_D$, corresponding to $5-30 \%$ of
the virial radius \cite{gentilechiara}.
It is fundamental to notice that these galactocentric distances represent 
scales large enough to be
unaffected by numerical effects, or by the smoothing of the cusp due to
self-interacting or warm DM (see \cite{hogan} for instance). 

As will be extensively discussed in Chapter 5, I don't believe 
these observations
are in conflict with the hierarchical paradigm, nor do they 
prove that the simulation
process is severely inaccurate, but rather, they highlight a lack
of a proper physical description of some of the mechanisms
accompanying galaxy formation inside the halos. In the specific case, the
baryonic collapse and the subsequent formation of the galaxy represent a
major perturbation in the halo structure; given the hierarchical clustering
scenario leading to the NFW halo, the new observations of galactic
dynamics are consistent with a halo response to galaxy formation causing
the evolution of its equilibrium structure. 

\section{The evidence of a discrepancy outside the core}

\noindent
In this Section I will analyze the data coming from two spirals' samples
spanning 3 orders of magnitude in disk mass, namely: 
i) the sample of high-quality
rotation curves selected by \cite{donato},
discarding the 4 galaxies with the 
smallest extension relative to the disk exponential
scale length $R_D$; ii) a sample selected from the literature with 
a criterium based on the extension of the rotation curve, \textit{i.e.} either
the last measured point is at a radius larger than a fixed distance 
(chosen as $r > \ 6R_D$ or $r>30 \ kpc$), or the velocity at the last measured point is
higher than a threshold (chosen to be $250 \ km \ s^{-1}$).  
These conditions ensure that each curve is extended enough to map regions
of the system where the baryons are markedly sub-dominant, so that the
velocity profile is regulated by the dark matter distribution;
moreover, these regions are
distant enough from the center of the halo not to be affected by the particular shape of the
inner profile. 
As an additional measure of precaution, only galaxies with curves 
regular out to the
last data point are considered. 
In addition to these two samples of galaxies, I will consider two particular
spirals, namely DDO 47 (\cite{salucciborriello},\cite{gentile05}) and
ESO 287-G13 \cite{gentile04}, due to the exceptional quality of their
observed curves ($H_\alpha$ and $HI$ respectively).
For all the 37 galaxies considered, the rotation curves reach maximum 
galactocentric distances between  
$\sim 5\%$ and $\sim 35\%$ (NGC 9133) of the virial radius, 
with an average
outermost radius of $24 \ kpc$. In Table (1) I list the
galaxies of sample (ii), marking with ``$*$'' those whose mass
decomposition was
provided in the reference. For those unmarked, the mass decomposition was
obtained following \cite{ps90b}.

\begin{deluxetable}{lc}
\tabletypesize{} \tablecaption{Sample (ii) with references.}
\tablewidth{0pt} \tablehead{\colhead{Galaxy} & \colhead{Reference}} 
\startdata
NGC $289^*$ & Walsh et al., 1997\\
NGC $1068$ & Sofue et al., 1999\\
NGC $1097$ & Sofue et al., 1999\\
NGC $1232^*$ & van Zee \& Bryant, 1999\\
NGC $3198^*$ & Blais-Ouellette et al., 2001 \\
NGC $3726$ & Verheijen \& Sancisi, 2001\\
NGC $4123^*$ & Weiner et al., 2001\\
NGC $5055$ & Sofue et al., 1999\\
NGC $5236$ & Sofue et al., 1999\\
UGC $5253$ & Noordermeer et al., 2004\\
NGC $5985$ & Blais-Ouellette et al., 2004 \\
NGC $6946^*$ & Carignan et al., 1990 \\
NGC $7331^*$ & Bottema, 1999\\
UGC $9133^*$ & Noordermeer et al., 2004\\
\enddata
\label{tablegalaxy}
\tablecomments{Selected galaxies with their references. The asterisk indicates that
  the original work provided also the dark-luminous decomposition of the rotation curve.}
\end{deluxetable}

The mass modelling of the rotation curves of these samples of galaxies
yields cored DM density profiles as best fits 
(\cite{salucciborriello},\cite{gentile05},\cite{gentile04}); 
however, I'm interested in
the outer regions of the disk, to check whether the discrepancy extends
to the whole DM profile. I found out these halos do not converge to NFW 
profiles anywhere; moreover, they show an interesting trend in 
their mass distribution.

In order to compare the observed density profile of these
galaxies, represented by the Burkert model, with an NFW halo,
there are different possible choices. Due to the poor performance of the
NFW profile in fitting these curves, instead of adopting the minimum $\chi^2$
criterium, I impose a more physical condition;   
the chosen NFW model must yield the observed total mass at the last
measured point, thus satisfying the condition    
\be
M_{\rm NFW}(r_{\rm f})=M(r_{\rm f})~.
\label{masseq}
\ee
Notice that, 
if the observed density profile indeed converged to an NFW, this equality would
hold for all radii outside the cusp/core, including the last measured point
(the contribution of the cusp to the total mass is negligible). 
On the contrary, if Eq.~(\ref{masseq}) is true $only$ for some radius
smaller than $r_f$, then the two profiles are significantly discrepant 
from that point outwards. In the case the equality holds for radii larger
than $r_f$, the two profiles are indeed completely different in all the radial
range considered.
\begin{figure*}
\centering\epsfig{file=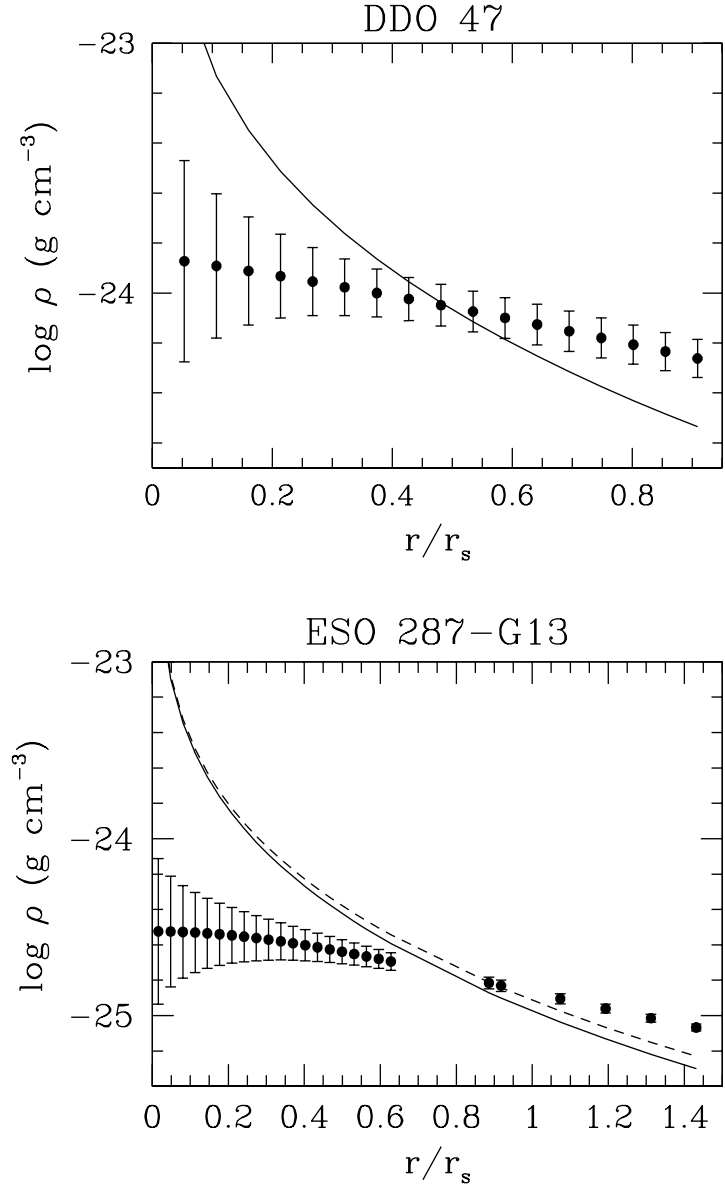,scale=1.5}
\caption{\textit{Dots}: DM density profiles for DDO 47 and ESO 287-G13, as yielded by the
  best fits (Burkert halo) in the original papers (\textit{dots}; \cite{salucciborriello},\cite{gentile04},\cite{gentile05}).
\textit{Solid lines}: NFW density
  profile such that $M_{NFW}(r_f)=M(r_f)$ (see text). \textit{Dashed line}:
  best-fit NFW for ESO 287-G13.} 
\label{densprof}
\end{figure*}

Fig.~(\ref{densprof}) shows the result of this comparison for
DDO 47 and ESO 287-G13, for which 
I find NFW halos of concentration and virial mass of
$c=18.4, \ M_{vir}=6 \times 10^{10} M_{\odot}$ and 
$c=13.3, \ M_{vir}=7 \times 10^{11} M_{\odot}$ respectively.
The dots represent the best-fit Burkert profile, and 
the errorbars mirror the uncertainties in the
fitting parameters 
(\cite{salucciborriello},\cite{gentile04},\cite{gentile05}), 
which are larger in the inner
parts, due to the difficulties of the mass decomposition.    
The solid lines represent the best NFW profiles, 
chosen according to Eq.~(\ref{masseq}). For ESO 287-G13,
the dashed line is the best NFW determined with a slightly different
mass modelling, with halo $+$ disk $+$ gas components \cite{gentile04}, and  
$M_{vir}$ and $M_D/M_{vir}$ as free parameters.
\begin{figure}
\centering\epsfig{file=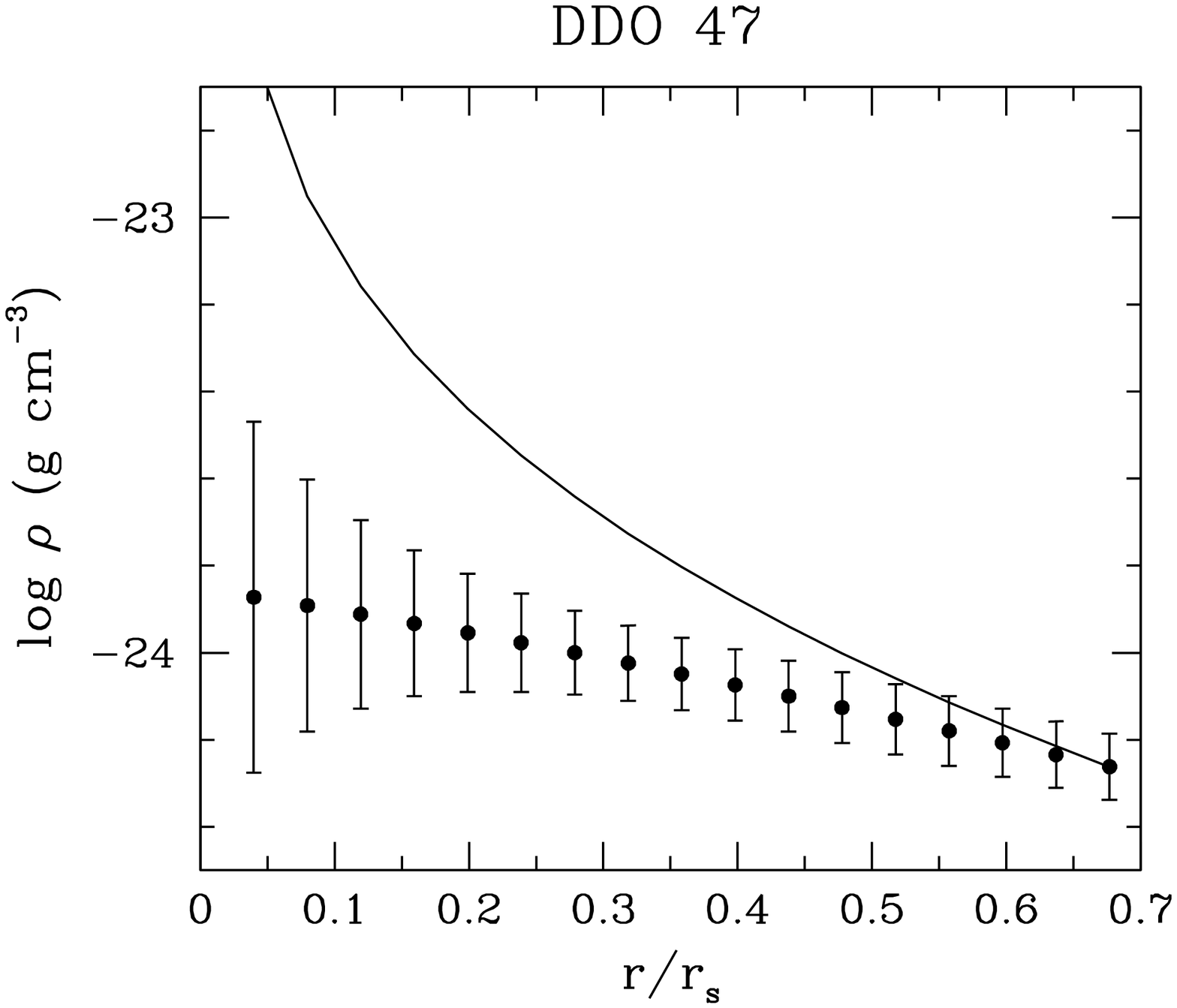,scale=0.5}
\caption{Density profile of DDO 47 assuming $\rho_{\rm NFW}(r_{\rm
    f})=\rho(r_{\rm f})$ (\textit{solid line}), compared to the data
    (\textit{dots, as in Fig.~(\ref{densprof})}).
}
\label{ddo47}
\end{figure}
Fig.~(\ref{densprof}) highlights a discrepancy between the NFW and the
observed profile that goes beyond the cusp/core issue: the observed profile
does not converge to the NFW anywhere. In the outer parts, the measured
density is higher than the NFW, and the slope of the profile is shallower. 
Notice that such a discrepancy was already present in some 
previous investigations 
(e.g. \cite{blais},\cite{borr},\cite{deblok1}),
although it was not claimed explicitly.

Another viable method to compare the observed dark matter mass distributions 
with the NFW profile
is to impose that, at the last measured point of the rotation curve, the densities coincide:   
\be
\rho_{\rm NFW}(r_f)=\rho(r_f)~,
\label{rhoeq}
\ee
as illustrated in Fig.~(\ref{ddo47}) for DDO 47. In this case, the NFW is well
above the observed profile anywhere inside the last point, showing an even wider discrepancy.

No matter the normalization, from these plots it is evident that the observed
density profiles of these two galaxies do not converge to an NFW; moreover, 
the NFW is steeper everywhere. 
The measured mass distributions show an inner density deficit, and an excess
of mass in the outer regions, when compared to the theoretical expectations; this may
happen for a number of reasons, the first being the possibility that 
these halos are not in equilibrium. 

To address this issue, I analyse the whole sample of galaxies, checking
whether their outer mass distribution converges to an NFW. In order to do this,
I take advantage of the fact that the NFW is a one-parameter family of curves,
and exploit the radial dependence of mass and density to build a curve
$M(\rho,r)$; as it turns out, an aproximation is given by
\be
M(\rho,r) \simeq \rho (r)^{3/4} \ r^{11/4}~.
\label{oddio}
\ee
Notice that this simply reflects a relation of the kind $M \sim \rho \ r^3$,
corrected by a shape factor accounting for the deviation of the NFW density
profile from a uniform sphere. As expected, for a given halo the plot is a
straight line with slope around $1$.  
However, the self-similarity of the NFW halo endows the nice property that the \textit{same}
straight line is followed by $all$ halos, regardless of the virial mass; 
this mirrors the phase-space stratification during the mass accretion of 
the halos in hierarchical clustering.
\begin{figure}
\centering\epsfig{file=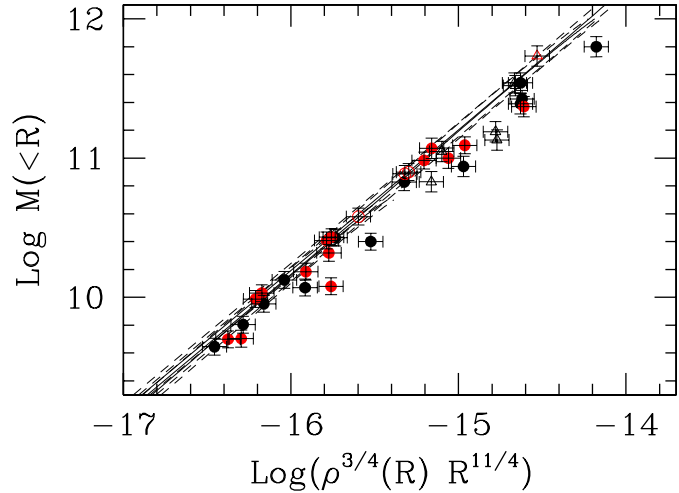,scale=0.5}
\caption{
The $M-R-\rho$ relation: the 
\textit{solid lines} represent NFW halos of virial masses 
$5 \times 10^{10} \ M_{\odot}$, 
$1 \times 10^{12} \ M_{\odot}$ and $1 \times 10^{13} \ M_{\odot}$.
The dashed lines enclose the $1 \sigma$ regions for the determination of $c$
(see \cite{wechsler}). The radial range for all the halos is between $1\%$
and $20\%$  of the virial radius. The \textit{dots} represent the galaxies of the
samples described in the text, at the last measured point. The three edge-on 
galaxies are denoted by \textit{empty circles}. 
\textit{Empty triangles} denote
the objects of sample (ii) for which we computed the mass decomposition.
Red symbols are isolated halos and black symbols are subhalos. 
}
\label{rela}
\end{figure}
This is shown in Fig.~(\ref{rela}), where I plotted 3 halos of masses $5 \times 10^{10}
\ M_{\odot}$, $1 \times 10^{12} \ M_{\odot}$ and $1 \times 10^{13} \
M_{\odot}$. The scatter among the lines is due to the uncertainties in the
determination of the concentration $c$ as a function of the virial mass.

I then compare this theoretical prediction with the observed properties of the
galaxies in the samples cited above.
For each galaxy, I compute the mass enclosed
inside the last measured point, and the density at that point. Notice that the
mass $M(<r)$ is mainly affected by the DM matter distribution near $r$, while the
presence of a cusp/core in the centre (as well as a baryonic component) is totally marginal. 

For these galaxies, the NFW is not the best-fitting
profile, so I do not expect them to sit on this relation. 
However, they do show an interesting pattern: all the galaxies are
sistematically offset on the same side of the curve, and in particular, 
at a given last radius and enclosed mass, the density is always higher than in the NFW
case, up to a factor of $\sim 3$. 
This is the equivalent of what was shown in Fig.~(\ref{densprof}), and
the conclusion is the same: in the majority of these galaxies, the DM density
profile around $5-6$ disk scale-lengths shows an excess of mass enclosed in
this region, compared to the NFW. 

Conversely, this plot can be read the other way around; 
if at a certain radius $r$ the density profile is described by the NFW, then
the mass inside that radius would be higher than observed, and this 
is the case represented in Fig.~(\ref{ddo47}), with an even larger overall 
discrepancy between the theoretical and observed profiles (see for 
comparison \cite{seigar}, with a case where, in order to fit two 
observed galaxies with an NFW profile and to match the outer mass
distribution, the inner rotation curve is significantly overestimated). 
Again in this case, all the galaxies
are offset on the same side of the theoretical curve.  
This result excludes the possibility that non-equilibrium effects are
responsible for this discrepancy. 

The mean offset of $\sim$0.1 dex is solid (but note that larger offsets
are also observed); in fact, 
the error propagation analysis shows that the 3\% error in the measure of
the circular velocity and the 0.05 error in the logarithmic gradient
$d{\rm log}~V(R)/d{\rm log}~R$ yield uncertainties of the order
of 0.025 dex in $M$ and 0.06 dex in $\rho$.   
The errors in the objects distances
are not included since they only
induce random uncertainties. 
 
It is interesting to see whether the environment has
an effect on our results: I distinguished between galaxies in
``isolated'' halos and galaxies in ``subhalos'' (for details, see \cite{saluccichiara}).
Qualitatively there
are no obvious trends between the two subsamples, 
although the result is consistent with the predictions of \cite{bullock}
regarding the difference
between the concentrations of halos and subhalos.

Three of the galaxies in the samples were edge-on ($i > 85^{\circ}$); the HI rotation curves in
these cases may suffer from unaccounted-for projection effects, while the H$\alpha$ curves may
be plagued by extinction \cite{bosma2}. However, they do not seem to
occupy any peculiar region of the plot, nor to be any more offset than the other 
galaxies.

Finally, the galaxies from sample (ii) for which the mass modelling was not
provided in the literature, and that were mass-decomposed following \cite{ps90b},
do not show any special trend with respect to the others,
indicating that the particular method for fitting the rotation curve
does not significanlty affect the result.

In the present analysis, spurious dynamical effects such as warps and
non-circular motions can bias the determination of the mass distribution. In fact, 
actual CDM halos are expected to be triaxial, which may induce
non-circular motions in the gas \cite{hayashi}; in addition,
gas moving along filaments 
\cite{dekel} may interact with the galaxies,
triggering the formation of warp-like features in the disks.
Nearly all the rotation curves collected in the present work
were derived using the tilted-ring fitting of the velocity field,
which can account for warps but not for non-circular motions.
The exception is DDO 47, which was studied in detail by \cite{gentile05}
using the harmonic decomposition of the velocity field \cite{wong}.
In summary, only the effects of warps are taken into account; however, 
non-circular motions are expected to produce only random scatter in the observations,
without any systematic effect.

\section{Discussion}

\begin{figure}
\centering\epsfig{file=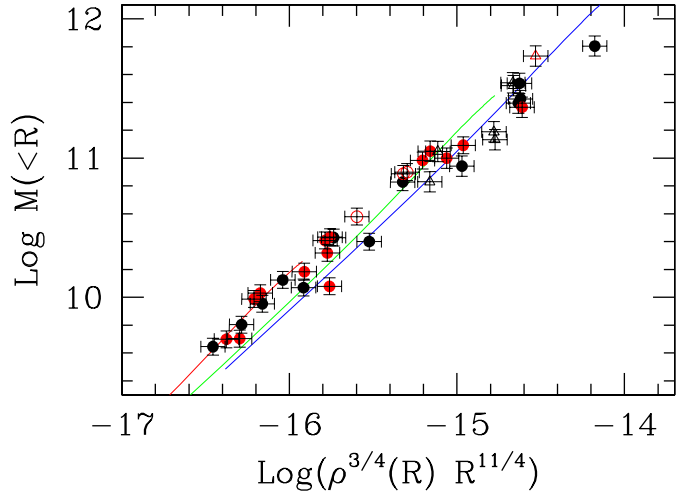,scale=0.5}
\caption{
The same as Fig \ref{rela}, 
but showing that Burkert halos 
predict a mass dependent $M vs (\rho^a~R^b)$ relation
in good agreement with observations \cite{salucciburkert}.
Symbols are the same as Fig \ref{rela}, and the 3 lines correspond to 
3 different virial masses ($5 \times 10^{10}$ M$_{\odot}$, 
$1 \times 10^{12}$ M$_{\odot}$ and $1 \times 10^{13}$
M$_{\odot}$.
}
\label{rela_bur}
\end{figure}

\noindent
The observed rotation curves were fitted with the Burkert profile, 
exploiting the empirical relation found in \cite{salucciburkert}, linking
the core radius and density (which is consistent with the URC model in the
halo's inner regions, as shown in Chapter 3). 
The virial masses were computed by integrating the density profile
until the mean density was $\Delta_{vir}$ times $\rho_{\rm c}$ (see Chapter 1).
In Fig.~(\ref{rela_bur}) I plot the observed galaxies against a relation
between the Burkert mass, density and radial distance that is equivalent to
the one described by Eq.~(\ref{oddio}), finding the expected random scatter around the curve.
This Figure highlights the main difference between the density profile 
inferred from observations and the NFW; the Burkert profile is not
self-similar, and in fact halos of different virial masses do not sit on the same
straight line, due to the marked mass-dependency of the Burkert shape factor.
In addition, the different position of the overall relation in the plot 
with respect to the NFW, highlights again that the two profiles are
significantly different at the edge of the disk, in the regions investigated by this dynamical analysis
($\sim 5 - 30$\% of the virial radius).

There are hints in the literature of an observed convergence of the DM 
density profile to the NFW in regions more external than the ones analyzed
here; \cite{prada} for instance investigated the DM profiles outside the optical
radii of isolated galaxies ($r > 0.2 - 0.3 \ R_{vir}$), by studying the 
kinematics of satellites. 
Similar conclusions were reached by (\cite{brainerda},\cite{brainerdb}) for scales $\geq
50~h^{-1}$ kpc, using weak galaxy lensing in addition to the dynamics of
satellites; \cite{vikh} and \cite{zappa} show the agreement between X-ray data and 
NFW profiles in galaxy clusters.

Although the statistical significance of the results of this Chapter
has to be improved with more data, so far in this Thesis the evidence 
presented indicates a trend for galactic halos to feature an inner mass distribution 
that is not universal, and does not resembre the one predicted by numerical simulations.  
In addition, this seems to apply to galaxies but not to more massive systems.

As already pointed out, the discrepancy is manifested at galactocentric
distances too high for numerical effects to be significant. 
In addition, for the same reason such a discrepancy cannot be explained by
effects due to self-interacting o annihilating DM, that would indeed erase the halo
cusp but would yield a profile convergent with the NFW already at small radii
(see for instance \cite{spergel},\cite{ahn} and references therein).

If one wants to speculate about the physics beneath this puzzle, she may find
herself wondering around the idea of baryons. 
The effects of baryons on DM halos, and in particular of baryonic collapse and galaxy
formation, are poorly understood. So far, adiabatic contraction
has been the only process extensively studied (\cite{blumenthal},\cite{gnedin},\cite{sellwood}),
and it is thought to increase the halo
concentration after the formation of the galaxy. However, it is difficult to
reconcile adiabatic contraction with the chaotic and random mass accretion process
that characterizes hierarchical clustering. 
On the other hand, a few have studied mechanisms with opposite effects,
like adiabatic expansion \cite{dutton06} or
dynamical friction \cite{chiaraangmom}, that transfers angular
momentum to the centre of the halo and mass to the regions outside the halo
core. In addition, \cite{seigar} show the failure of the NFW in
fitting the mass profile of a barred galaxy, the concentration parameter of
which has to be much smaller than the averaged 
predicted value; the authors discuss possible evidence of the absence of  
adiabatic contraction or alternatively of some dynamical effects that  
compensates for it.

Although baryons represent an almost negligible fraction of the halo mass, 
their collapse trasfers this mass and all the relative angular momentum to the
inner halo, as discussed at the end of Chapter 2.
From an initial state in which
they are distributed following the DM, the system ends up in a state where the
baryons completely dominate the center of the halo. Chapter 5 of this Thesis
will deal with the evolution of the halo structure following a perturbation in
its dynamical state, suggesting a kind of interaction, dynamical in nature, between the DM and
the luminous component, that re-shapes its equilibrium configuration.
Notice that, up to now, there is evidence of such a baryon-induced
evolution only in galaxy-size halos, where the ratio between the luminous and
dark mass is high enough, and the formation timescales short enough, to allow
the baryonic collapse to significantly affect the final equilibrium
structure of the system.

%% file: angmom.tex
\chapter{Dynamics of dark matter halo evolution and galaxy formation}

\noindent
I present a theoretical model of the dynamics of the halo reaction after a
perturbation of its equilibrium structure, highlighting the tight connection
between the spatial mass distribution and the shape of the velocity dispersion tensor. 
An unbalance in the components of the internal velocities, arising from
angular momentum injection into the halo, triggers a mass rearrangement
and the evolution into a new equilibrium configuration; if the tangential
motions dominate in the inner halo, the cusp is smoothed out into a corelike profile.
I also present a physical model describing such a halo evolution; during the 
baryonic collapse leading to the formation of the protogalaxy, the gas clouds 
infalling towards the center of the halo experience dynamical friction with
the background dark matter, locally transferring angular momentum to the
halo and thus enhancing its tangential velocity dispersions. The amplitude
of such a perturbation is big enough to unbalance the halo and cause
structural evolution. 

\section{Introduction}

\noindent
So far, the evidence gathered in this Thesis can be summed up in three main
points:

\textit{a-} the DM mass distribution in galactic halos is not self-similar, but
present strong trends with the galactic mass;

\textit{b-} the DM profile is not well fitted by the NFW in spirals, but
features instead shallower central slopes and excesses of mass in the
outskirts of the disk, that cannot be attributed to numerical effects or more
exotic variants than the CDM;

\textit{c-} the angular momentum measured in galactic halos strongly depends on the
mass and distribution of the baryonic component, and is not correctly
predicted by numerical simulations.

In brief, the observations of disk galaxies do not support the existence of a
universal, self-similar equilibrium DM halo, as naturally arising from the 
patterns of hierarchical clustering;  
rather, they suggest a more complex
scenario where the halo structure is determined by the global properties of 
both the dark matter and the baryonic components.

It is now the time to try and give some theoretical interpretation to these
facts. The equilibrium structure of halos is tightly linked to the mechanism
of halo formation, so one may think that doubting the NFW means
to doubt the whole hierarchical clustering theory; to the contrary, my take is
that the theory is correct, but there are some pieces of physics missing as
far as galactic halos are concerned.
The existence of determined scalings between the properties of the DM
and the baryons in disks, and the fact that the discrepancy between observations
and theoretical predictions regarding the structure of halos, manifest itself
precisely where the baryonic component is present (see \cite{donato}),
leads me to consider an evolutionary approach. 

Suppose that a pristine, unperturbed DM halo, as emerging from the non-linear growth
of perturbations in a hierarchical scenario, features a phase-space structure
and mass distribution as predicted by the CDM theory. The mass distribution
would be described by the NFW profile, the phase-space density would be a
power-law of slope $\sim 1.875$, 
and the velocity anysotropy profile would be null in the centre and slighlty
increasing in the outskirts (see Chapter 1 for reference). 

Some billion years later, we observe a disk galaxy inside the halo, we
recognize that its main features are probably related to the halo structure,
and yet we find that the same structure is non consistent with the picture of
a theoretical DM halo. This makes me wonder whether the baryons themselves 
could be at the origin of the disagreement. 

As the galaxy formation is 
undoubtedly determined by the halo dynamical properties, 
could the baryon collapse and galaxy formation induce a perturbation big enough to 
trigger an evolution of the halo, in response to the changing equilibrium conditions of the
whole system? And in particular, could such perturbation affect 
the dynamics and structure of the inner halo in such a way that the
original NFW profile is flattened into a corelike structure?

When the baryonic component in the Universe decouples from radiation, early dark
matter structures are already in place, hierarchically growing. 
The gravitational potential of the
overdense regions attracts the baryons, and in the simplest scenario they
settle into the wells in dynamical equilibrium with the dark matter. In other
words, the phase-space structure of the two components is similar in these
very early stages; the baryons share with the dark matter the mass and 
specific angular momentum distributions.
However, in the densest regions the baryons are subject to a series of
dissipative processes, namely the radiative cooling and the subsequent condensation, 
that lead to the formation of self-gravitating clouds, that may or may not
contain dark matter. Because of the high concentration of the baryons at this stage,
the clouds remain bound, and dynamically decouple from the background
halo, whose mass distribution in comparison is relatively smooth 
(see Chapter 1 for references). 

These baryonic substructures fall towards the centre of the halo under the
effect of gravity; as a consequence, the fraction of the total 
angular momentum of the system shared by the baryons is carried to the center. 
The cloud orbits inside the halo potential
well are determined by the cloud initial position and relative velocity (that
in turn may depend on the large-scale dynamical conditions, like the overall
velocity and density fields).

As the clouds fall into denser and denser regions, dynamical friction exerted
by the background dark matter slows them down, with a net transfer of
angular momentum locally from each cloud to the dark matter, thus enhancing 
the halo tangential random motions. The anisotropy profile, measuring the
balance between radial and tangential velocities, governs the equilibrium mass
distribution of the halo through the Jeans' equation; 
the predominance of tangential motions in the center of the halo
moves the particles on orbits of higher energy and increases the entropy, with
the result of partially unbounding the system and triggering a mass transfer 
out of the cusp, flattening the density profile (\cite{chiaraangmom}).

This Chapter is divided into two main parts. At first, I will give the
mathematical description of the halo evolution under perturbations of the
velocity anisotropy profile, and show in particular that an injection of random angular
momentum flattens the inner density profile. Secondly, I will produce a
physical model accounting for the dynamical coupling of the baryons to the
dark matter 
through dynamical friction, and show that random angular momentum is
transferred from the formers to the latter, with the right amplitude and
distribution to trigger the described halo evolution. 

\section{The NFW distribution function}

\noindent
As already mentioned in Chapter 1, 
the microscopic dynamical properties of a collisionless system of particles, like a dark
matter halo, are described in
phase-space by its distribution function (DF), a $7-$dimensional function
of coordinates, velocities and time $f(\textbf{x},\textbf{v}, t)$, that is the
solution of the collisionless Boltzmann equation;
as stated by the Jeans' theorem \cite{BT}, $f$ depends on the
phase-space coordinates only through the integral of motions in the halo potential,
thus defining the symmetries governing the system's evolution. 
It is not observable in itself, but its integrals yield the 
description of the macroscopic properties of the system;  
any macroscopic observable $O$ is obtained by means of
$f$ through the average 
\be
\langle O \rangle = \int O \, f \, \mathrm{d}^3 v / \int f \, \mathrm{d}^3 v.
\label{average}
\ee
Thus, in particular, the mass distribution of a DM halo is linked to its
microscopic dynamical properties through:
\be
\rho(r) = \int f(r,v) \mathrm{d}^3 v~, \label{def}
\label{frho}
\ee
at any particular time. In this Chapter I will assume that the halo evolves 
through a series of stationary states, thus discarding the explicit dependence of $f$
on time.   

In order to represent the NFW halo in phase-space, I need to find its
distribution function. This is almost never a trivial task, even in the simplest of
scenarios, and for the NFW the additional complication is that the process is not
analytical. 

I remind the reader of the characteristics of my adopted 
unperturbed equilibrium density profile:
\be
\rho = {M_H\over 4\pi R_H^3}\, {c^2\, g(c)\over x\, (1+cx)^2}~,
\label{rhoNFW}
\ee
where $c$ is the concentration parameter, and
$g(c)\equiv [\ln{(1+c)}-c/(1+c)]^{-1}$. From now on, all the physical 
quantities will be expressed in units of the virial mass $M_H$ and radius $R_H$.  
The gravitational potential is obtained through the Poisson's equation 
\be
\nabla^2 \Phi (x)= 4 \pi G \rho(x)~,
\label{Poisson}
\ee
and is expressed in terms of the virial velocity $V_H\equiv \sqrt{G\, M_H/R_H}$:
\be
\Phi(x) = -V_H^2\, g(c)\, \frac{\ln{(1+cx)}}{x}~.
\label{phiNFW}
\ee 

In the simple case of a totally isotropic halo, 
$f$ is explicitly a function of
energy alone (both positions and velocities of the particles are defined by
their energy), and its determination from the density profile is unique. By
conventionally defining the relative potential and binding energy as
$\Psi = -\phi$ and $\varepsilon = -E = \Psi-\frac{1}{2}v^2$
\cite{lokasmamon}, the DF describing the equilibrium is obtained from the
potential-density pair through the Eddington's inversion formula 
\cite{BT}
\be
f(\varepsilon) = \frac{1}{\sqrt{8}\pi^2} \frac{d}{d \varepsilon}
\int_0^\varepsilon \frac{d \rho}{d \Psi} \frac{d
\Psi}{\sqrt{\varepsilon - \Psi}}~. 
\label{eddie}
\ee

However, the simulated halos show a nontrivial anisotropy profile;
the degree of anisotropy is commonly expressed through the parameter
\be
\beta(r)=1-\frac{\sigma^2_t}{\sigma^2_r},
\label{beta}
\ee
where $\sigma^2_t$ and $\sigma^2_r$ are the $1D$ tangential and radial
velocity dispersion profiles respectively. In simulations the halos turn out
to be isotropic in the center ($\beta=0$), and radially anisotropic
($\beta > 0$) outwards (\cite{colin},\cite{fuku2}).

For anisotropic, spherically symmetric systems, $f$ is an explicit function of 
two integrals of motions, commonly taken as the energy and the total
angular momentum $L^2$ \cite{BT}. The quantity
$\vec{L}(r)=\vec{r}\times\vec{v_T}$ is defined in terms of the
tangential velocity $\vec{v_T}(r)$; this is the $2D$ vectorial sum,
on spheres of radius $r$, of all the 
velocity dispersions orthogonal to
the radial direction. 
For this reason, $\vec{L}$ is not an 
angular momentum in the strict sense, as it does not define any
preferential direction of rotation; rather,
the quantity $L^2(r)$ is a measure of the tangential component
of the internal, randomly-oriented motions of the halo, and for this reason
I will from now on 
refer to it as the halo's \textit{random angular momentum}. 

With two explicit variables, there are infinite allowed DFs
that satisfy Eq.~(\ref{def}). 
However, only some of the possible DFs correspond to equilibrium
configurations of the system, or in other words, satisfy the Jeans' equation.
A generalization of the Eddington's solution, for systems with generic anisotropy
profiles, yields a DF of the form \cite{cudderford}:
\be
f(Q, L^2) = f_0\left(\varepsilon-{L^2\over 2 r_a^2}\right) \,
(L^2)^{\alpha}~. 
\label{fnostra}
\ee
$f_0(Q)$ is an equilibrium
DF describing systems characterized either by isotropy or by radial
anisotropy, with $r_a$ being the
anisotropy radius, at which $\beta=0$ (\cite{osipkov},\cite{merritt},\cite{BT});
the orbital energy $L^2/2r_a^2$ associated with $L$ lowers the particle binding
energy, that becomes $Q=\varepsilon - L^2/2r_a^2$.
To account for tangential anisotropy, a pure
angular momentum component is given to the DF, taking the simple functional
form of a power-law of index $\alpha$.

This DF represents a very general family of equilibrium solutions of the 
Jeans' equation \cite{cudderford},
for spherically symmetric systems with mass distribution and anisotropy
profile depending on $f_0$, $r_a$ and $\alpha$. Systems described by this DF
include the NFW (both in the isotropic and radially anisotropic realizations),
and the cored profiles, as I will show in the next Section. 

The spherically averaged density profile of a system described by a DF 
of the family of Eq.~(\ref{fnostra}) can be recovered by  
transforming the coordinate system from $(v_r,v_T)$ to $(Q,L^2)$
in Eq.~(\ref{def}), yielding
\be
\rho(r)=\frac{2 \pi}{r^2} \int_0^\Psi f_0(Q) \mathrm{d}Q \
\int_0^{2r^2 (\Psi-Q)/(1+r^2/r_a^2)} \frac{(L^2)^\alpha \ \mathrm{d}
L^2}{\sqrt{2(\Psi-Q)-(L^2/r^2) ( 1+r^2/r^2_a)}}~ 
\label{rhoNFWric}
\ee
(for comparison see \cite{BT}, sections 4.4 - 4.5). In
spherical symmetry, the averaged $1D$-components of the velocity are
null, and the first non-zero moments are the radial and tangential
velocity dispersions:
\be
\sigma_r^2 (r) = \frac{2 \pi}{\rho r^2}\, \int_0^\Psi f_0(Q)
\mathrm{d}Q \, \int_0^{2r^2 (\Psi-Q)/(1+r^2/r_a^2)} (L^2)^\alpha \
\sqrt{2(\Psi-Q)-\frac{L^2}{r^2} \left(1+\frac{r^2}{r^2_a} \right)} \
\mathrm{d}L^2 ~, \label{sigmar}
\ee
\be
\sigma_t^2 (r) = \frac{\pi}{\rho r^4}\, \int_0^\Psi f_0(Q)
\mathrm{d}Q \, \int_0^{2r^2 (\Psi-Q)/(1+r^2/r_a^2)}
\frac{L^2\,(L^2)^{\alpha}}{\sqrt{2(\Psi-Q)-(L^2/r^2)(1+r^2/r^2_a)}}
\ \mathrm{d}L^2 ~, \label{sigmat}
\ee
with the total velocity dispersion being
$\sigma^2(r)=\sigma^2_r(r)+2\sigma^2_t(r)$. 

By substituting $\sigma^2_r$ and $\sigma_t^2$ in Eq.~(\ref{beta}) as in \cite{cudderford},
the anisotropy profile now reads
\be
\beta(r) = \frac{r^2-\alpha r^2_a}{r^2+r^2_a}~. 
\label{betar}
\ee
Notice that, for positive $\alpha$, the anisotropy is tangential in
the inner regions where $r^2 < \alpha r^2_a$, zero at $r^2 = \alpha
r^2_a$ and radial in the outer regions where $r^2 > \alpha r^2_a$;
on the other hand, if $\alpha$ is negative the model is radially
anisotropic everywhere and for all values of $r_a$.

\begin{figure}
\centering\epsfig{file=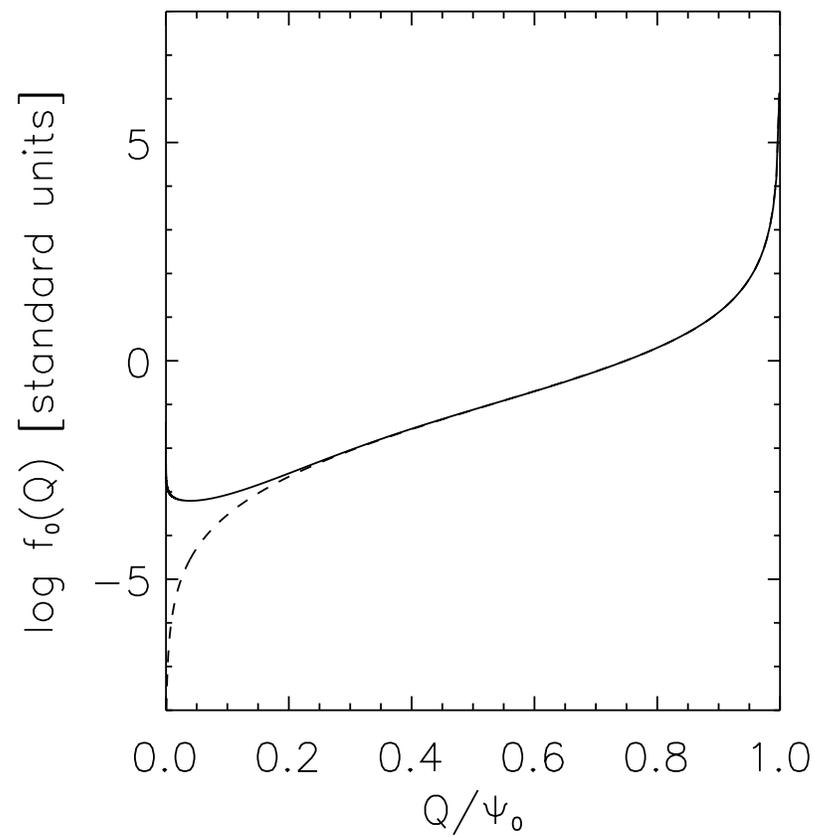,scale=0.8}
\caption{The phase-space
distribution function for a NFW halo, in standard units of $G = M_H
= R_H/2 = 1$, see \cite{lokasmamon}. \textit{Solid line}: halo as simulated,
isotropic in the center and radially anisotropic at the outskirts;
\textit{dashed line}: totally isotropic.} 
\label{figDF}
\end{figure}

I remind the reader that the simulated halos are centrally isotropic and radially
anisotropic in the outskirts so that, with the DF chosen above, the correct
value is $\alpha=0$ for the NFW. 
Moreover, I set the value of the anisotropy radius $r_a
\simeq 1$ \cite{lokasmamon}. With this choice of $\alpha$ and
$r_a$ I obtain a precise representation of the NFW halo.
Under these conditions, the relation between the DF and the density
profile reads 
\be
f_0(Q)= \frac{1}{2^{5/2} \pi^2} \frac{d^2}{dQ^2} \int_0^Q \left( 1 +
\frac{r^2}{r^2_a} \right) \rho(\Psi) \mathrm{d}\Psi~,
\label{f0q}
\ee
after integrating in $dL^2$ \cite{cudderford}.
In Fig.~\ref{figDF} I plot the ``energy part'' $f_0$ of the DF 
defined by Eq.~(\ref{fnostra}, \ref{f0q}) that represents the NFW, with $\alpha=0$ 
and $r_a \simeq 1$ (\textit{solid line}),
compared with that of a totally isotropic halo (\textit{dashed
line}).

\begin{figure}
\epsfig{file=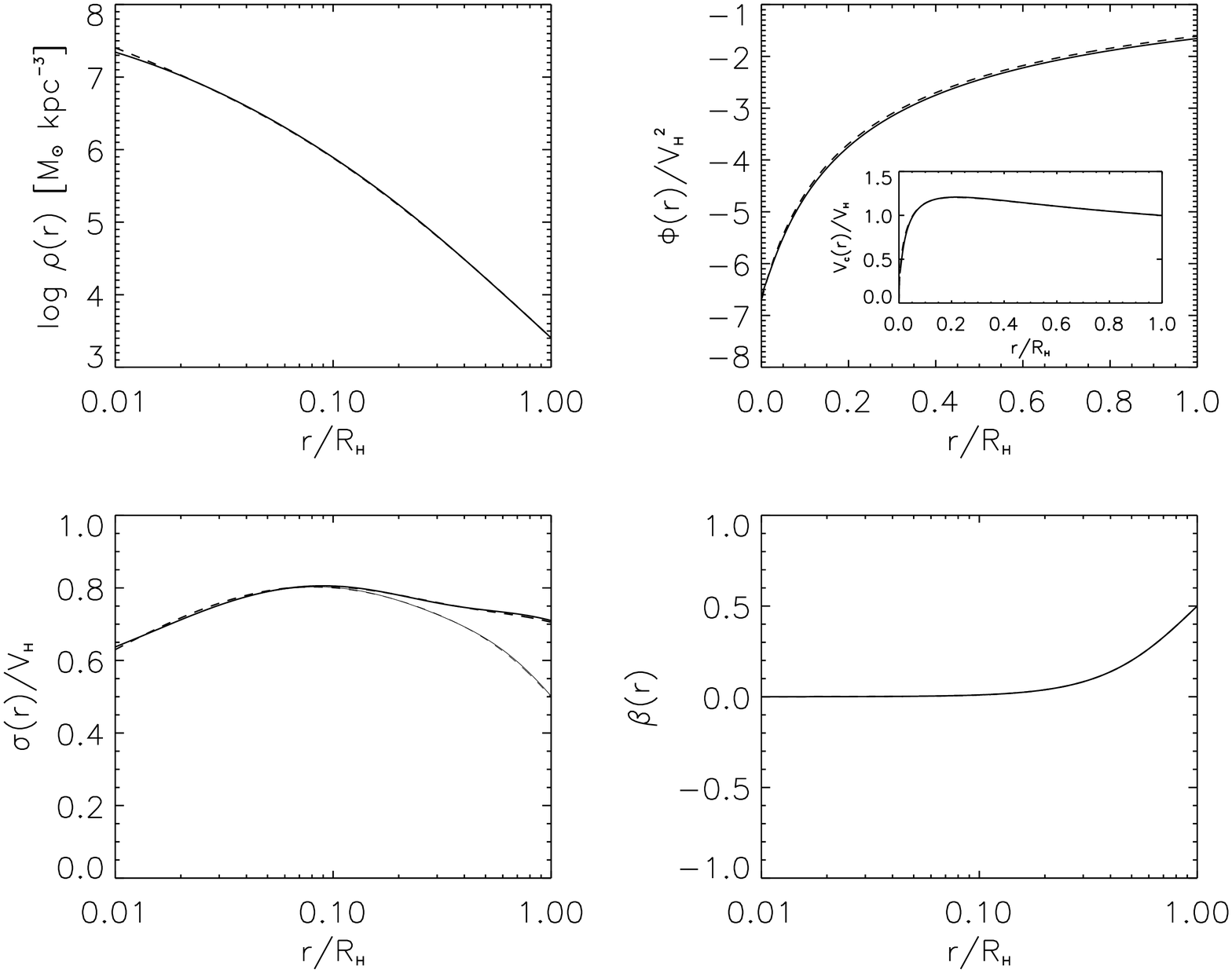,scale=0.5}
\caption{The NFW halo (\textit{solid lines}) as reconstructed from the 
DF of Eq.~(\ref{fnostra}), compared to the original one from simulations
(\textit{dashed lines}, \cite{colin},\cite{fuku2}). \textit{Upper left panel}: logarithmic
density profile; \textit{upper right}: gravitational potential and
rotation curve (\textit{inset}); \textit{lower left}: velocity
dispersions profiles, radial (\textit{thick}) and tangential
(\textit{thin}); \textit{lower right}: anisotropy profile.}
\label{figNFW}
\end{figure}

In Fig.~\ref{figNFW} I show the halo as reconstructed through the
DF defined by Eqs.~(\ref{fnostra}, \ref{f0q}) (\textit{solid
lines}), compared with the original NFW (\textit{dashed lines}) taken from
simulations (\cite{colin},\cite{fuku2});
in the upper panels I show the logarithmic density profile
(\textit{left}), and the gravitational potential (\textit{right}),
as well as the rotation curve (\textit{inset}). In the lower left
panel, I show the velocity dispersion profiles for the radial
(\textit{thick}) and the tangential (\textit{thin}) components; this
halo is isotropic in the inner $10\%$ of the virial radius, and
becomes radially anisotropic in the outer regions, as mirrored by
the anisotropy parameter profile (\textit{right}).

\begin{figure}
\centering\epsfig{file=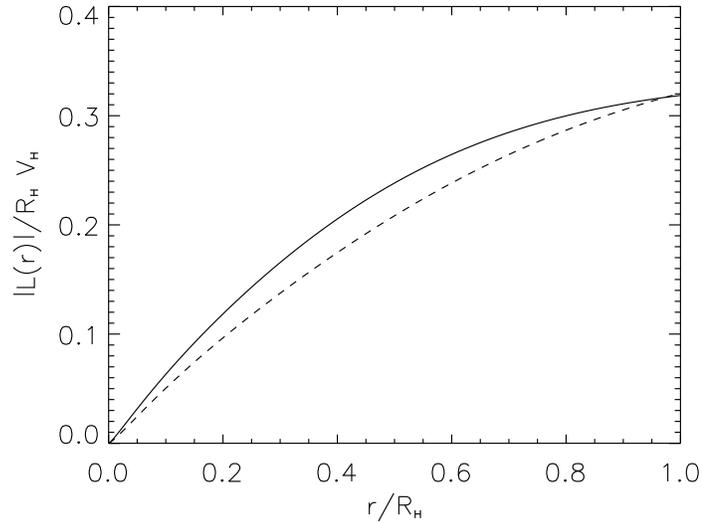,scale=0.5}
\caption{Angular momentum
profile yielded by the new phase-space DF (\textit{solid line}),
compared to the unperturbed NFW one (\textit{dashed line}), under
the same unevolved gravitational potential.} \label{figAM}
\end{figure}

Once I know a suitable DF, I can investigate the specific
angular momentum profile yielded by the dark matter 
tangential random motions, that is defined as follows:
\be
\langle L(r) \rangle = \frac{2 \pi}{\rho r^2}\, \int_0^\Psi f_0(Q)
\mathrm{d}Q \, \int_0^{2r^2 (\Psi-Q)/(1+r^2/r_a^2)} \frac{L \
(L^2)^{\alpha}}{\sqrt{2(\Psi-Q)-(L^2/r^2)(1+r^2/r^2_a)}} \
\mathrm{d}L^2~. \label{jmedio}
\ee
Notice that, even if the averaged $1D$ velocities are null, the
averaged angular momentum is nonzero, due to the symmetry of the DF,
as discussed after Eq.~(\ref{fnostra}). This is plotted as the
\textit{dashed line} of Fig.~(\ref{figAM}).

\section{Perturbing the halo: the angular momentum transfer}

\noindent
From the above description of the halo, one can argue that the mass
distribution of the system is strictly linked to its dynamics. The
question that now arises is the following: is the halo stable
against perturbations in its dynamical state? In other words,
suppose that the halo becomes involved in a process that causes a
variation in its velocity dispersion tensor, such as an increase of energy and
angular momentum $L$; will the macroscopic observables, like the density
profile, the gravitational potential and the anisotropy profile, be
affected?

I defer the reader to the next section for a toy model of angular
momentum transfer between baryons and DM during the first stages of
galaxy formation, and I proceed now to analyze its effects on the
equilibrium state of the halo.

Consider a system described by the DF of Eqs.~(\ref{fnostra},
\ref{f0q}), and suppose to inject random angular momentum, of the kind $L$ described
above, into it; the measured value of $L$, given by an equation
like (\ref{average}), increases, and this is achieved by assigning a positive
value to $\alpha$ (and a suitable value to $r_a$). From Eq.~(\ref{fnostra}) one
can see that the particle orbital energy increases; more importantly, from
Eq.~(\ref{betar}) it is clear that the anisotropy of the halo decreases and
becomes negative, meaning that the tangential motions dominate. 
 
The DF is function of both
energy and angular momentum, and is a solution of the Jeans' equation;
the halo is bound to conserve $E$ and
$L^2$ before the perturbation, and $E+\Delta E$ and $L^2+\Delta L^2$
afterwise, redistributing the excess and rearranging the DM
particles in the $6D$-space of coordinates and velocities. This in
turn implies an evolution of the gravitational potential; hence, the
system moves towards a new equilibrium configuration of density and
velocity.
This process is governed by the Poisson's equation
\be
\frac{\mathrm{d}^2\Psi}{\mathrm{d}r^2} +
\frac{2}{r}\frac{\mathrm{d}\Psi}{\mathrm{d}r} = 4\pi\,G\,\int_0^\Psi
\int_0^{2r^2 (\Psi-Q)/(1+r^2/r_a^2)}\, f(Q,L^2)\,\mathrm{d}L^2\,
\mathrm{d}Q~, 
\label{poisson}
\ee
where $\alpha$ and $r_a$ are now to be intended as the new,
perturbed parameters. This integro-differential equation has to be
solved for $\Psi$; the consistent density and anisotropy profiles are linked
to the evolved potential through the Jeans' equation, or alternatively
are yielded by the DF through 
Eqs.~(\ref{rhoNFWric}, \ref{sigmar}, \ref{sigmat}, \ref{betar}); they 
are the observables of the new equilibrium state of the
halo. While the complete integration has to be done iteratively, the
solution for small radii is analytical, and gives an interesting
insight on the behavior of the halo.

Before proceeding, note that the density profile of
Eq.~(\ref{rhoNFWric}) can be written as
\be
\rho(r)=\frac{(2\pi)^{3/2} \ 2^\alpha \
r^{2\alpha}}{(1+r^2/r_a^2)^{\alpha+1}}
\frac{\Gamma(\alpha+1)}{\Gamma(\alpha+3/2)} \ \int_0^\Psi f_0(Q)
(\Psi-Q)^{\alpha+1/2} \mathrm{d}Q~, \label{rhomezza}
\ee
after the integration in $L^2$ is performed explicitly. For small
radii, \textit{i.e.} when $Q \rightarrow \varepsilon \rightarrow
\Psi_0$, with $\Psi_0$ the central value of the potential, it is
easy to see that $\rho(r) \propto 1/r \propto 1/[\Psi_0-\Psi(r)]$;
by changing variable in Eq.~(\ref{f0q}) from $\varepsilon$ to
$(\varepsilon-\Psi)/(\Psi_0-\varepsilon)$, the energy part of the DF
behaves like
\be
f_0(\varepsilon) \propto (\Psi_0-\varepsilon)^{-5/2}~.
\label{f0aprox}
\ee
I then put this expression into Eq.~(\ref{rhomezza}), and pass from
$\varepsilon$ to $(\Psi-\varepsilon)/(\Psi_0-\Psi)$; I find the density
profile at small radii to behave as
\be
\rho(r) \propto [\Psi_0-\Psi(r)]^{\alpha-1} r^{2\alpha}~.
\label{rhosmall}
\ee
I now insert this expression into the Poisson equation
(\ref{poisson}), to obtain the self-consistent solution for the new
potential $\Psi(r)$, which reads
\be
\Psi_0-\Psi(r) \propto r^{2\,(\alpha+1)/(2-\alpha)}~.
\label{psismall}
\ee
Finally, the new density profile $\rho(r)$ from Eq.~(\ref{rhosmall})
reads
\be
\rho(r) \propto r^{-2\,(1-2\alpha)/(2-\alpha)}~. 
\label{rhosmall2}
\ee
Thus I find that for $\alpha \rightarrow 0$ the inner profile behaves like
$r^{-1}$ (NFW), while for $\alpha \rightarrow 1/2$, \textit{i.e.} when the DF
is linear in $L$, I obtain $\rho(r) \rightarrow \ constant$, which is a core.
For intermediate values of $\alpha$ I obtain anything between a
cusp and a core; for values of $\alpha$ larger than $1/2$ the density profile
is not realistic, featuring a ``hole'' in the centre, 
while for negative values of $\alpha$, corresponding to enhancing the radial
motions over the tangential (or, alternatively, to subtracting angular
momentum from the halo) the cusp's slope steepens.

Notice that, from Eqs.~(\ref{betar}) and~(\ref{rhomezza}), I conclude that
$\rho(r) = \rho(r,\beta)$, and for $r \rightarrow 0$, 
\be
\rho(r) \propto r^{-2\,(1+2\beta)/(2+\beta)}~,
\label{rhobeta}
\ee
\textit{i.e.} the density profile is a function of the anisotropy
parameter, that represents the balance between the radial and tangential motions
inside the halo; it is then clear that the shape of the velocity dispersion
tensor determines the halo mass distribution. In particular, for
$r \rightarrow 0$ the anisotropy parameter behaves like $\beta(r) \rightarrow
-\alpha$, therefore for $\alpha=1/2$
the halo features a constant tangential anisotropy in the inner
regions. The corelike feature in a halo is always accompanied by tangential
anisotropy, and it is possible to verify that radially-dominated halos cannot
develop a core.   

The value of $\alpha$ determines the inner specific angular momentum 
profile as well, which is obtained as
\be
L(r) \propto r \ [\Psi_0-\Psi(r)]^{1/2} \propto r^{3/(2-\alpha)}~;
\label{jsmall}
\ee
notice that for the unperturbed halo $L(r) \rightarrow r^{3/2}$ and for
$\alpha=1/2$ I get $L(r) \rightarrow r^2$.

Along with these analytical results, I performed the full numerical
integration of Eq.~(\ref{poisson}). As a boundary condition
throughout this computation I adopted the halo mass conservation;
I further normalized the evolved potential in order to obtain the
same behavior of the outer rotation curve as before.

It is interesting to analyze this process in two steps: since the
density and anisotropy profiles and the potential well evolve
together, it is impossible to evaluate the variation in the angular
momentum of the system after the new equilibrium state is reached from Eq.~(\ref{jmedio}),
because the system has then lost memory of its initial conditions.
Instead, suppose to picture the halo at the moment when it
receives its input in energy and angular momentum, but the potential
well has not yet evolved; then I can evaluate the amount of angular
momentum that has been injected into the halo. In Fig.~\ref{figAM}
I show the specific averaged angular momentum profiles as yielded
by the old, NFW-like (\textit{dashed}), and new (\textit{solid})
DFs, in the same potential. Keep in mind that this is not a stable
state, the system is unbalanced and is going to evolve into a new configuration that
satisfies the Jeans' equation.

\begin{figure}
\epsfig{file=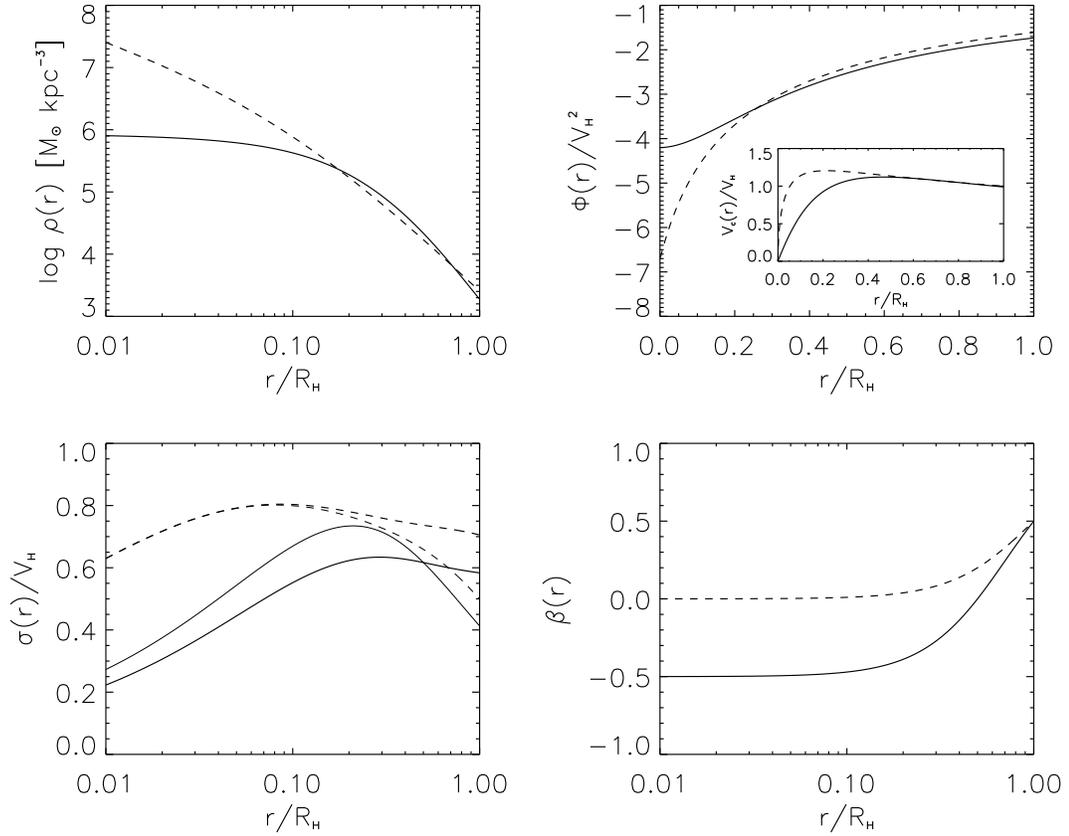,scale=0.5}
\caption{New equilibrium
configuration of the perturbed halo (\textit{solid lines}), compared
to the initial NFW (\textit{dashed lines}). \textit{Upper left panel}:
logarithmic density profile; \textit{upper right}: gravitational
potential and rotation curve (\textit{inset}); \textit{lower left}:
velocity dispersions profiles, radial (\textit{thick}) and
tangential (\textit{thin}); \textit{lower right}: anisotropy
profile.} \label{figNEW}
\end{figure}

The final configuration of the system, corresponding to its new
equilibrium state (with $\alpha=1/2$),
is showed in Fig.~(\ref{figNEW}). The density
profile (\textit{upper left}) has been smoothed and the cusp erased;
this mirrors the mass loss experienced by the inner regions of the halo, since
the DM particles have been moved to more energetic orbits. Notice
the corresponding flattening of the potential well (\textit{upper
right}). Accordingly, the rotation curve (\textit{inset}) is more
gently rising in the inner regions and is nearly unchanged beyond
half of the virial radius.

As for the velocity dispersion, the symmetry between the radial and
tangential motions has been broken in favor of the latter
(\textit{lower left}); correspondingly, the anisotropy parameter profile
(\textit{lower right}) is now negative in the inner halo and changes
sign at a radius corresponding to the new $r_a=1/\sqrt{2}$, that is
set by requiring the same outermost value of $\beta(r)$ as before,
although it scarcely affects the results shown above.

In sum, given the DF of equation (\ref{fnostra}), the couple ($\alpha=0, \
r_a \sim 1$) describes the known NFW halo, cuspy and almost entirely
isotropic, while the couple ($\alpha=1/2, \ r_a \sim 1/\sqrt{2}$)
produces a corelike halo, with a constant inner
density profile and a tangentially-dominated inner anisotropy
profile.

\begin{figure}
\centering\epsfig{file=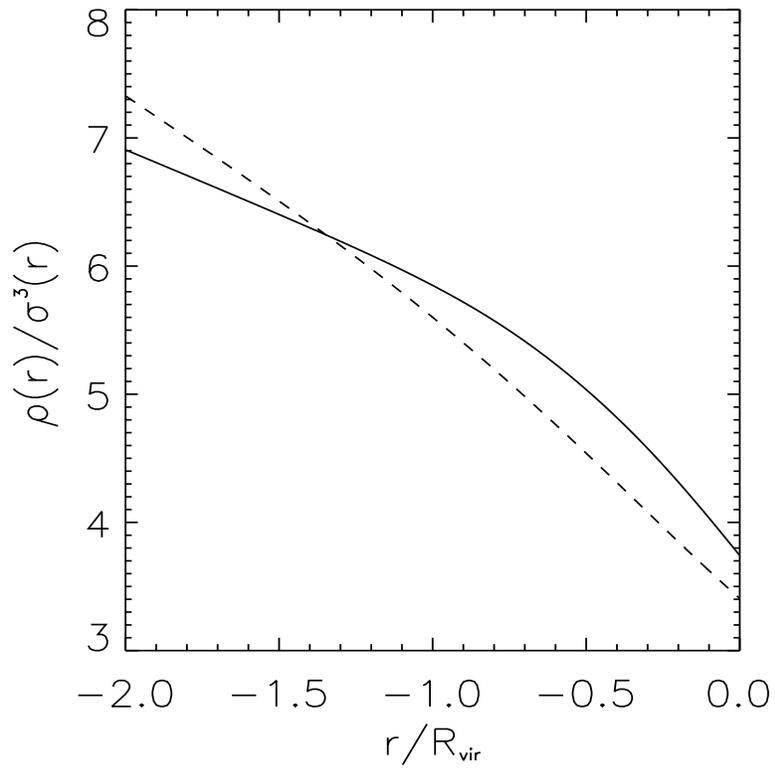,height=11cm, width=12cm}
\caption{The phase-space density $\rho/\sigma^3$ of the unperturbed NFW (\textit{dashed}) and
  of the perturbed halo (\textit{solid}) as a function of radius, to be compared with
  Fig.~(\ref{tay1}). The injection of tangential motions increases the entropy
  in the center of the halo and flattens out the power-law.}
\label{rhosigma}
\end{figure}

It is also interesting to see the behaviour of the evolved halo in phase space;
in Fig.~(\ref{rhosigma}) I show the phase-space density $\rho/\sigma^3$ for
the original NFW (\textit{dashed line}) and for the perturbed halo
(\textit{solid line}), to be compared with Fig.~(\ref{tay1}) in Chapter 1. Notice that the
DF described by Eq.~(\ref{fnostra}) yields the power-law dependence on
radius expected for the NFW, with the correct slope. The injection of 
tangential motions into the halo, however, perturbs the phase-space structure
by increasing the entropy in the center of the halo, thus flattening out the 
power-law.

I point out that the injection of angular momentum of the kind described here
generates additional random tangential motions, that do not produce an
\textit{ordinate} rotation;
in fact, the symmetry of the halo and its dynamics are determined
by the DF, that does not define a preferential direction of rotation 
(\textit{i.e.} this halo has no spin). 
The DM particles move on orbits
around the center of mass with random orientations, so that the
resulting angular momentum $L^2$ is nonzero, but the average value
of each velocity component is null.

Notice also that the assumption of injecting angular momentum into the halo is 
quite realistic, since halos interact only gravitationally; 
any tidal encounter, accretion of satellite or merging is accompanied by
angular momentum exchange. A pure energy transfer on the contrary, is quite
difficult to achieve, and given the DF described by Eq.~(\ref{fnostra}), it is
clear that it would have no effect on the equilibrium state of the halo; in
fact, injecting or subtracting energy would enhance or diminish the velocity
dispersions isotropically, without unbalancing the anisotropy profile. With
the same dispersion tensor, the mass distribution would conserve its shape,
and only expand or contract.  

\section{Dynamical friction as an angular momentum engine}

\noindent
Is there a physical process that can account
for the evolution described above? In this Section I 
present a toy model that provides the halo with such an amount of
random angular momentum to allow it to evolve from a cuspy to a cored
configuration. The natural framework is galaxy formation, when the
baryons collapse inside the halo potential well and exchange
angular momentum with the dark matter through dynamical friction.

Specifically, in the very early stages of galaxy formation, the
baryons trapped inside the potential wells of the halos undergo
radiative dissipation processes that cause them to lose kinetic
energy and to condense in clumps inside the relatively smooth dark halo. If
radiative cooling is effective, the gas will organize into
self-gravitating clouds before it collapses to the halo center and
fragments into star-forming units; moreover, 
the clouds are likely to survive the tidal
stripping due to the DM, because of their relatively high binding
energy \cite{momao}. The infalling, clumpy gas component decouples from
virial equilibrium, and while
the clouds spiralling down get closer and closer to the halo center,
increasing their tangential velocity along their orbits and reaching
regions with higher and higher density, they dissipate their orbital energy.

In these regions in fact, a gravitational effect becomes relatively
efficient in slowing down the clouds, namely the dynamical friction
exerted by the background DM particles, that causes part of the
cloud tangential velocity to be transferred from the baryons to the DM itself. 
Notice that, for clouds spiralling down the potential well, the istantaneous
orbit is elliptical, and the radial component of the velocity is on average
much smaller than the tangential; thus, the radial velocity transfer is
negligible. 
A complementary analysis of this process, focussing on the energy transfer 
from the baryons to the dark matter, that causes halo expansion, 
can be found in (\cite{elzant1},\cite{elzant2}). Conversely, I specifically focus
on angular momentum transfer, the reason being that energy
transfer alone is not viable to unbalance the halo anisotropy profile; in this
case, any expansion could be contrasted by the deepening of the potential well caused
by the baryonic mass accumulating in the halo center.

As a result of the dynamical friction, the inner part of the halo 
is therefore granted with a surplus of
angular momentum $L$ (as described above) and energy, depending on the number, mass and
initial velocity of the clouds.

Consider a cloud of mass $M_c$ that at time $t=0$ is at a certain
distance from the center of the halo, with initial velocity
$v^2=v^2_r+(L/r)^2$ and angular momentum $L$; I define the initial
pericenter of its orbit as $r_+(0)$, the eccentricity as $e(0)$, and
the apocenter as $r_-(0)= r_+\, (1-e)/(1+e)$. The cloud is
self-gravitating and hence I consider it as a point mass, immersed
in the halo potential well; in the orbit-averaged approximation
\cite{laceycole}, the equations of motions for the cloud energy
and angular momentum are given by
\be
{\mathrm{d} E\over \mathrm{d} t}  = - {\int_{r_-}^{r_+}{(1/v_r)}~~
v\, |F_{\mathrm{frc}}| / M_c \ \mathrm{d}r \over
\int_{r_-}^{r_+}{(1/v_r)} \ \mathrm{d}r}~, \label{derLdyn}
\ee
\be
{\mathrm{d} L\over \mathrm{d} t}  = - {\int_{r_-}^{r_+}{(1/v_r)}~~
L\, |F_{\mathrm{frc}}| / (M_c v) \ \mathrm{d}r \over
\int_{r_-}^{r_+}{(1/v_r)} \ \mathrm{d}r}~, \label{derEdyn}
\ee
with initial conditions set by
\be
L(0)=\sqrt{2[\Psi(r_+)-\Psi(r_-)]\over
1/r_+^2-1/r_-^2}~,~~~~~~~~~~~E(0)=\Psi(r_+)+\frac{v^2(r_+)}{2}~.
\label{LE}
\ee
At each instant, the force exerted by the background DM particles on
the cloud is \cite{BT}
\be
|F_{\mathrm{frc}}| =-4\pi G^2\, M_c^2\, \ln{\Lambda}\ {\int_0^v
f(v')~{\mathrm{d}^3v'}\over v^2}~, \label{Fdyn}
\ee
in terms of the NFW phase-space distribution function $f$ (see
Section 2), of the cloud speed $v=\sqrt{2[\Psi(r)-E(t)]}$ and of the
Coulomb logarithm $\ln{\Lambda} = \ln{(M_H/M_c)}$ \cite{BT}. At each timestep,
$r_\pm(t)$ are given by the condition $v_r^2=\sqrt{v^2-L^2/ r^2}=0$.
Due to the dynamical friction, the orbit shape and the velocity of
the cloud evolve in time, so that this set of equations has to be
solved iteratively.

For a halo of virial mass $M_H$ I performed a series of Montecarlo
simulations for different realizations of the baryonic component,
organizing it in ensembles of clouds, characterized by a
mass function scaling as $M_c^{-\delta}$, with index $\delta$
ranging from $0$ to $2$; both these extremes are to be considered unrealistic, 
$0$ corresponding to a constant mass function, and $2$ yielding an excessive clumping
factor. In each realization, I allowed the cloud
masses to range from $10^{-5}$ to $10^{-2}\, M_H$ \cite{elzant2}.
The number of clouds is actually constrained by the total
amount of baryons, set to equal the cosmological fraction $0.16\,
M_H$. The initial spatial distribution of the clouds is uniform
between $r=0$ and $r=R_H$; at time $t=0$ the clouds are in
statistical equilibrium with the background halo, therefore I 
randomly sampled their initial velocities from a Maxwellian
distribution, with mean $0$ and variance $\langle \sigma^2_{t, r}
\rangle/2$.

For each $\delta$ I performed $100$ runs, and computed the average
specific angular momentum transferred by the clouds to the halo
after $2$ Gyr; I expect that after this time the inner part of the
halo becomes so crowded with clouds that they start to collide and
disrupt, and the star formation effects dominate over the dynamics
of the gas \cite{elzant2}. However, in the outer regions the
process continues with longer timescales, so that I also followed
the evolution of the system for about $6$ Gyr.

\begin{deluxetable}{lccccc}
\tabletypesize{} \tablecaption{Dynamical friction results}
\tablewidth{0pt} \tablehead{\colhead{$\delta^a$} & \colhead{$\langle
N_i \rangle^b$} & \colhead{$\langle M_c \rangle^c $} &
\colhead{$\langle \Delta L_{DM} \rangle^d$} & \colhead{$M_g$ ($2$
Gyr)$^e$} & \colhead{$M_g$ (6 Gyr)$^f$}} \startdata
0 & 31.45 & 5.03e-3 & 1.04 & 0.039 & 0.074  \\
1 & 109.66 & 1.46e-3 & 1.86 & 0.032 & 0.064 \\
2 & 2276.98 & 7.05e-5 & 7.54 & 0.019 & 0.043 \\
\enddata
\label{clouds} \tablecomments{Column label: (\textit{a}) cloud
power-law mass function index; (\textit{b}) average number of
clouds; (\textit{c}) average cloud mass (units of $M_H$);
(\textit{d}) average of the exchanged specific angular momentum
(units of $R_H\ V_H$) integrated over the profile; (\textit{e})
average baryonic mass that falls inside $0.1\, R_H$ after $2$ Gyr
and, (\textit{f}) after $6$ Gyr (units of $M_H$).}
\end{deluxetable}

In Table~\ref{clouds}, for each $\delta$ I give the average number
and mass of the clouds, the total angular momentum gained by the
halo, and the mass accumulated in the center of the halo after $2$
and $6$ Gyr. Note that with my power-law mass functions, the
massive clouds constitute a small fraction of the total; on the
other hand, the dynamical friction is more effective on them, and
therefore they have a large probability to lose all their angular
momentum quickly, and to collapse in the center of the halo soon.
The small clouds instead take more time to spiral down in the halo
potential well, and retain a larger fraction of their initial
angular momentum. 
Notice that this is 
consistent with a scenario of rapid initial baryonic collapse followed by
smooth subsequent infall, and that depending on the initial cloud mass
function, the morphology of the galaxy could be affected \cite{chiaraangmom}. 
In the end, a steeper mass function, that selects
a high number of small clouds, results in a more effective transfer
of angular momentum $L$ to the halo.

\begin{figure}
\centering\epsfig{file=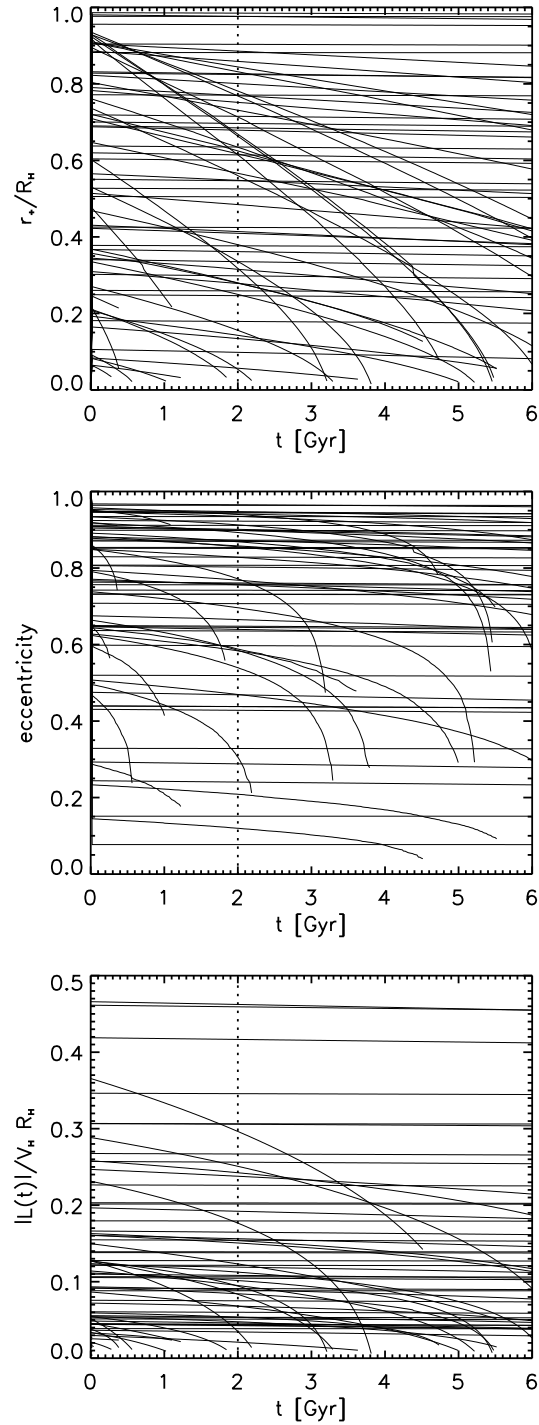,scale=0.8}
\caption{Time evolution of
a set of clouds sampled with power-law index $\delta=1$.
\textit{Upper panel}: apocenter of the orbit; \textit{middle panel}:
orbit eccentricity; \textit{bottom panel}: angular momentum retained
by the clouds.} \label{figclouds}
\end{figure}

In Fig.~\ref{figclouds} I illustrate an example of the time
evolution of a set of clouds sampled with power-law index
$\delta=1$. In the time interval from $t=0$ to $t=6$ Gyr a fraction
of the clouds reaches the inner $10\%$ of the virial radius (\textit{upper
panel}), building up the mass that is likely to end up in the
spheroidal component of the forming galaxy. In the middle panel I
show the evolution of the orbit eccentricity, and in the lower panel
the angular momentum lost by the single clouds and transferred to
the halo. Notice that the fraction of clouds that effectively
transfer angular momentum to the halo is relatively small, and that
the majority of the clouds remains on almost unperturbed high
orbits, meaning that the timescales of dynamical friction are long.
This effect is enhanced for increasing $\delta$, as shown in the
last two columns of Table~\ref{clouds}; a steeper mass function
results in a slower accumulation of baryons in the center. In any
case, the final amount of baryons inside the inner $10\%$ of the
virial radius after $2$ and $6$ Gyr is overabundant with respect to
the known galactic masses (spheroidal and/or disk components; see
\cite{shankar}), due to the fact that the total baryonic
component initially matches the cosmological fraction. However,
while baryons accumulate in the center of the potential well and
organize themselves into the protogalactic structure, star formation
and AGN activity start, along with the ensuing feedback processes
that eventually regulate the actual amount of baryons.
Notice that these feedback mechanisms can in no way affect the
transfer of angular momentum between the visible and dark
components, as they take place after the cloud collapse; as will be
discussed in the next section, the mass of the growing baryonic component at the
center of the halo can affect the final dynamical state of the
system only marginally.

\begin{figure}
\epsfig{file=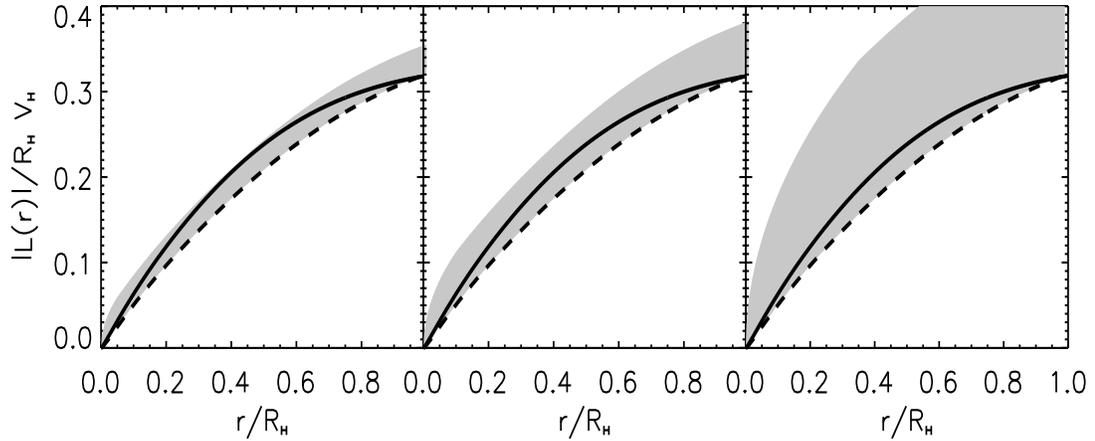,height=6.5cm, width=15cm}
\caption{Comparison between
the angular momentum profiles yielded by the phase-space DFs and by
dynamical friction. \textit{Dashed line}: unperturbed NFW halo
(see Section 2); \textit{solid line}: perturbed halo, new DF (see
Section 3); \textit{shaded regions}: NFW angular momentum profile
plus $\Delta L$ from dynamical friction, for $\delta=0$
(\textit{left}), $\delta=1$ (\textit{center}) and $\delta=2$
(\textit{right}).} \label{figAM3}
\end{figure}

Notice also that the angular momentum transfer from the clouds to the dark
matter happens locally, and with random orientations; this corresponds to the 
kind of random angular momentum $L$ described in the previous Sections. 

In Fig.~\ref{figAM3} I present the final angular momentum profile
of the halo, for $\delta=0, 1, 2$. From each run of my simulation
I extract a transferred momentum profile $\Delta L$ after $2$ Gyr and then add
it to the original NFW (\textit{dashed lines}, see also
Fig.~\ref{figAM}); the \textit{shaded areas} represents the overlap
of all the resulting new profiles. For comparison, I plot the
profile obtained in Section 3 through the transformation of the DF
(\textit{solid lines}, see also Fig.~\ref{figAM}).

\smallskip

I find that the angular momentum profile produced through the dynamical friction
mechanism is clearly compatible with that resulting from the
perturbation of the halo DF described in Section 3. In fact, the
effect of the dynamical friction on the halo is of enhancing
the random tangential motions with respect to the radial; 
this produces an unbalance in the velocity dispersion tensor, with an
anisotropy profile that becomes more tangential in the center of the halo,
where the dynamical friction is more effective.

\section{Discussion and conclusions}

\noindent
In the previous Chapters I collected some observational evidence about 
an incompleteness in the theoretical description of galaxy formation in the
CDM hierarchical scenario, resulting in a discrepancy between the predicted
and observed mass distribution of DM halos, and the predicted and observed
spin parameter distribution function. The limited mass range of halos affected
by this issue (galactic halos), along with the scale where the discrepancy
manifests itself (galactic scale), on the one hand excludes that this is a
fundamental problem of the whole hierarchical scenario, that still works on
groups and clusters scales, and on the other rules out that this is merely a
numerical problem originating from the failure of N-body codes to represent
very small scales. 

I approached this topic from the point of view of the coevolution of 
the baryons and their host DM halo, and investigated whether the mechanism of
galaxy formation involves perturbations in the halo structure, big enough to 
rearrange its mass and velocity distribution, 
flattening the inner density profile with a mass transfer to outer regions, 
and thus reconciling the observational 
evidences with the standard theory of hierarchical clustering.

The NFW halo is rather sensitive to variations of its anisotropy profile; 
if the tangential components of its velocity dispersion are enhanced, for
instance by means
of a transfer of random angular momentum as shown in this Chapter, its
phase-space structure enters a new equilibrium configuration, and a flat
density profile can be attained.
 
From the macroscopic side, I looked for a mechanism of random angular
momentum transfer that directly involved the baryons, that was
especially efficient during galaxy formation, and that directly
affected the microscopic state of the system; thus I focused on
dynamical friction (\cite{elzant1},\cite{elzant2}).
Angular momentum can be transferred to the halo also by tidal interactions and
mergers; however, such events are less and less frequent with
decreasing redshift both for spheroidal galaxies (\cite{koopmans},\cite{dom})
and for spirals (\cite{donghia},\cite{chiaraspin}, and Chapter 2). 
Moreover, such interactions between the
halo and its neighbours occur on scales comparable to that of the halo
itself, thus giving rise to perturbations of its dynamical state that produce
global ordinate motions; these events are thought to produce the halo spin.

When modelling the phase-space DF, I had to face the degeneracy
intrinsic to this kind of problem, that makes its determination not unique
(contrary to the case of isotropic halos).
With the method described in this Chapter, I ensure that the adopted DF describes
an equilibrium phase-space configuration \cite{cudderford}, and that it reproduces the
observables of a given halo (density, anisotropy and velocity dispersion
profiles, rotation curve and potential). Moreover, regardless of the particular
shape of the function, the symmetry of the system in phase-space
is set by the DF explicit dependence  
on the integrals of motion, thus the halo dynamics is set with no ambiguities. 

For an isotropic NFW,
$f(\textbf{x},\textbf{v})=f(\epsilon)$, \textit{i.e.} 
the phase-space coordinates are completely determined by the particle's energy. 
A symmetry breaking between the radial and tangential components of the
velocity dispersions causes an evolution of the system into $f(\epsilon,L^2)$,
thus the shape of the mass distribution depends explicitly on the shape of the
anisotropy profile: $\rho(r)=F(\beta(r))$. In physical terms, the excess or
deficit of tangential motions determines a mass rearrangement to 
attain equilibrium. 

Notice that this kind of DF cannot account for the halo spin $L_z$, that is
found to be $\lambda \sim 0.03-0.06$ from observations and simulations
(see \cite{chiaraspin} and Chapter 2). This
value of $\lambda$ is small, nonetheless it yields information on
the true shape of the halo in phase-space. 
Following from the Strong Jeans theorem (\cite{BT}; see Chapter 1),
a real collisionless system such a DM halo 
would be described by a DF function of 3 integrals of motion, 
and following the approach of this Chapter, it would be
of the form $f(\varepsilon,L^2, L_z)$; the explicit dependence on $L_z$,
setting a preferential direction of rotation, would manifest itself at the
breaking of the symmetry between the two tangential
velocity components.

A generalized DF could be written as 
$f(Q,L^2,L_z)=f_0(Q) \, f_1(L^2, L_z)$, where
$f_1$ may be a power-law or a more complex function, odd in $L_z$ to
produce net bulk rotation around the symmetry axis. With this DF the average 
$1D$ tangential velocity $\vec{v_t}$ parallel to the equatorial plane is not null,
and the halo features a macroscopic total angular momentum in the
$z$-direction, $\vec{L_z} = \vec{r} \times \vec{v_t}$, that is
aligned with the spin, $\lambda \simeq v_t/\sigma$.

In the computation of the perturbed gravitational potential I did
not include the baryonic mass; although the baryons piling up in
the center of the halo tend to deepen the well, this process does not
interfere with the flattening of the density profile. In fact, three
considerations are needed here: 1) the final amount of baryons that
is actually observed in the center of halos is deeply connected with
feedback processes, that are responsible for removing at least half
of the initial baryonic mass \cite{shankar}; 2) as hinted in
Section 1, the feedback processes themselves can transfer energy to
the DM and cause halo expansion; 3) most importantly, 
the dominant effect
in changing the equilibrium configuration of the halo is the
symmetry breaking of the velocity dispersion tensor, that determines the 
final density profile, regardless of the depth of the potential well. 
For this reason, the excess of mass carried by the baryons simply produces an isotropic
enhancement of all the $\sigma$ components, that may
cause contraction or expansion without preventing the flattening of the density
profile. Notice that, for the same line of reasoning, mechanisms to erase the cusp that
are based on feedback alone (like \cite{eke} where the baryonic mass loss produces a core)
are implausible because they cause temporary expansion out of equilibrium, without 
transforming the halo into a stable new configuration.  

Up to now, there is no definite knowledge of the details of the
baryon collapse into the protogalactic structure. Nevertheless, if
dynamical friction indeed plays a major role in the collapse, it
could affect the morphology of the galaxy that is to form, possibly
through its timescale that in turn depends on the cloud mass
function, and in conjunction with feedback processes. 

In this scenario, spheroids may be formed by the massive clouds, 
that lose all their angular momentum and collapse early and quickly; 
rapid and intense star formation would arise at the center of the halo, where the
clouds crash into one another and settle down. 
On the other hand, small clouds are slower in losing their momentum and
tend to stay on higher orbits, thus being more likely to end up in a rotating
disk; feedback from the forming galaxy is likely to be more efficient on small clouds,
thus preventing star formation in the more external regions until later times. 
The slow, gradual accretion of small clouds and the consequent propagation of star formation
can be assimilated to the \textit{inside-out} formation of disks \cite{chiappini}. 
As a possible evidence of this
process, I can point out the molecular clouds that today travel
across the Galactic disk, trapped inside the potential well of the
DM halo; they may be the low-mass tail of the cloud mass function.
So small as to have been only marginally affected by dynamical
friction, they are still on high orbits and are  
not yet actively participating to the growth 
of the disk.
In this picture, the initial conditions of the baryonic collapse play
an important role.

From the dark point of view, a similar argument holds; depending on the
details of the baryon collapse and dynamical friction, the equilibrium
structure of the halo evolves into different final states. This hints to a
very complex picture, where the dark matter density profile is not universal, but
strongly depends on the baryonic structure that forms within the halo. In my
opinion, it would be very interesting to look for patterns in the co-evolution
of the luminous and dark structures; I find this approach most promising to
understand the mechanisms of galaxy formation.


%% file: simulazioni.tex
\chapter{Simulating the infalling of substructures}

\noindent
I present my work in progress, on the topic of dark matter substructures and
halo evolution. I am currently engaged in devicing a simulation
able to correctly reproduce the dynamical friction inside halos, for a double purpose;
(i) to test the semi-analytical model presented in Chapter 5, and (ii) to  
address the issue of the disruption of satellites infalling into a dark matter halo, 
in an attempt to find a solution to the problem of the overabundance of substructures
predicted in galactic systems. 
I describe the method for writing initial conditions
for a simulation featuring stable halos with different phase-space
equilibrium structures, and eventually report my plans for the suite of
simulations. 

\section{Introduction}

\noindent
The hierarchical evolution of a generic dark matter halo can be roughly divided into
two phases; the mass accretion phase, during which smaller subunits merge and
build up its mass, shaping its equilibrium structure, and the satellite phase,
during which the halo itself is accreted onto a larger object. 

The universal structure of pure DM halos that grow hierarchically arises from physical
mechanisms that smooth out the clumpy satellite component infalling onto the
parent halo (see Chapter 1); on the other hand, the inner regions of early 
virialized objects often survive accretion on to a larger system, 
thus giving rise to a population of subhaloes. This substructure 
evolves as it is subjected to the forces that try to dissolve it: 
dynamical friction, tidal forces and impulsive collisions. 
Depending on their orbits and their masses, these subhaloes therefore 
either merge, are disrupted or survive to the present day (\cite{vdb2005}).
From Chapter 5, it should be clear that the processes that determine the 
satellite disruption are the same that shape the halo equilibrium structure
in reaction to the accretion, whether of luminous or dark components. 

A longstanding prediction of the theory of hierarchical clustering 
is that the subhalo population is self-similar, with low-mass
systems such as galaxies being simply scaled-down versions of larger systems
like clusters (\cite{donghia3},\cite{vdb2005} and references therein). This
would imply that the subhalo mass function is independent of the parent halo mass.
The Milky Way for instance 
is predicted to have nearly the same distribution
of substructures (scaled down in mass) as the Virgo cluster
(\cite{kly},\cite{moo}). These expectations are supported by numerical
simulations, that predict about 500 satellites in Milky Way-like objects.  
Observations of galactic systems (see for instance the SLOAN data, 
\cite{wilma},\cite{kleyna}) find $\sim 30$ satellites in the Local Group, 
highlighting an offset in the theoretical predictions of more than one order
of magnitude. Moreover, \cite{donghia2} showed that fossil groups with
intermediate mass between the Local Group and the Virgo cluster feature the
same lack of substructures. Some authors (see for instance 
\cite{stoehr}) suggest that the discrepancy in general 
arises because a large number of satellites are dark, \textit{i.e.} they
contain no stars; the problem with this approach is that disk galaxies 
would be perturbed by the dark substructure nevertheless, with the consequent
heating of the thin disk and visible, and disruptive, effects. 

The offset between the predicted and observed substructure mass function and
distribution points towards what has been called the
``missing satellite problem''.
Obviously, this is rather an ``overabundant substructure problem'' of numerical
simulations and of the hierarchical clustering picture. 
Numerical simulations have been on the scene for more than 20 years \cite{whitefrenk};
in time, the coding techniques have been constantly refined, and the
resolution and dynamical range of the runs has improved enormously. 
Still, simulations are not always the best of tools in studying dark matter
structures, and despite their terrific performance on large scales, they
cannot address many questions regarding the detailed physical 
processes taking place inside galaxies. Among these in particular, the effects
of tidal stripping and dynamical friction on the orbits of satellites are not
satisfactorily reproduced 
(\cite{vela},\cite{vdb99},\cite{colpi},\cite{mayer},\cite{chiaraangmom}),
along with the aforementioned survival and evolution of 
substructures (\cite{moore},\cite{taffoni},\cite{hay});
as for the baryons, the heating of galactic disks due to
substructure remnants, 
(\cite{quinn},\cite{vela2},\cite{babul},\cite{font}),
the susceptibility of
disks to bar instabilities \cite{mihos},
and the effects of these bars on the halo
central density cusps (\cite{deba},\cite{atha},\cite{valez})
are not well traced \cite{kaz}.

In this Chapter, I address the issue of simulating the infalling of
substructures in a parent halo, focussing on the correct implementation of the 
physics of dynamical friction, in order to test it as a viable
mechanism to alleviate the satellite discrepancy; in fact, the dynamical
coupling between the parent halo and an infalling object affects the
structure of both, with the result that the subhalo can be unbound and
destroyed. 
There have been recent attempts at specifically simulating dynamical friction on dark
matter satellites accreted into larger systems, with
positive results for cluster and galactic halos (\cite{nipoti},\cite{ma}).
In particular \cite{ma} find that the same mechanism is able, under certain
conditions, to flatten the parent halo's inner density profile.

The numerical representation of dynamical friction, as well as tidal stripping and all the
small-scale processes, is indeed a delicate procedure, with many caveheats and
problems.
Let alone the implementation of the baryonic physics, the numerical modelling
of the gravity and dark matter dynamics encounters serious difficulties at small
scales.
In fact, the scale-invariance of the NFW mass distribution resulting from simulations
has induced many into thinking that 
the accuracy and robustness of the numerical treatment of the halos
is not affected by
scale either, and that the method can be safely extrapolated 
to the small scales. Actually, this operation is subject to a number of risks,
arising from problems intrinsic to the numerical method. For instance, the
extreme coarse-graning of a simulated system (the halo is modelled
as an ensemble of particles of given phase-space dimension) affects the mixing
mechanisms in an unknown way (see \cite{BT},\cite{dehnen}). Moreover, the
limited mass range accessible with the numerical treatment generates two main 
problems. First, a simulation of dynamical friction needs a very high
resolution, because the process efficiency is extremely sensitive
to the ratio between the test particle (the cloud for instance) and the 
background particle (the DM); but even with the state-of-the-art computational
facilities, it is virtually impossible to correctly   
model gravitational effects that stem from the behaviour of the DM as a 
smooth fluid of microscopic particles 
and from \textit{local} variations of the smooth background potential, while 
considering objects of the size of dark matter halos.
Second, and unfortunately opposed to the
previous point, the higher the ratio between the test particle and the
background particle, the more the 2-body scatterings between the two components
become important, making the infalling cloud behave like a bowling ball among
the pins. If the softening length applied to the potential in order to avoid these
spurious interactions is set on the small background particle's dimension, it would
have no effect whatsoever in preventing the scattering;
on the contrary, if set on the
cloud's dimension, it would be so large that the 
background particles would not feel the cloud at all, and they would let it pass
through the halo undisturbed (C. Frenk, private communication).

Still, I am currently making an attempt at numerically modelling the dynamical
friction inside halos. A way around the main difficulties
with the method involves the use of dark matter
substructures instead of solid bodies like clouds; in this case, the infalling
object is made of particles itself, of size comparable to the
background ones, and this should in part smooth out the unwanted 2-body
interactions. On the other hand, the object is not totally self-gravitating
once it enters the parent halo, but it is subject to tidal disruption; this
works against dynamical friction, in that the mass of the object is decreasing
as it falls towards the centre of the halo.

On the bright side, having dark matter substructures fall into the parent
halo allows me to learn something about the fate
of the satellites at the same time, studying their disruption and the mass accretion mechanisms,
as depending on a number of boundary conditions.
In particular, I am interested in investigating the fate of satellites with
different phase-space structures, like for instance cored \textit{versus} 
cusped, or with isotropic \textit{versus} anisotropic velocity dispersion
tensors; obviously, I need a very high number of
particles, in order to model the satellites with an accuracy high enough to 
properly represent different inner structures. 
In addition, other parameters like the satellite's initial velocity or orbit 
eccentricity are likely to affect its evolution.  

My simulations are performed with the use of the facilities of the Institute
for Computational Cosmology in Durham. The simulation code is Gadget-2
(\cite{springel01},\cite{springel05}).

\section{Setting the initial conditions for equilibrium DM halos}

\noindent
Since I need to build isolated halos with different and well defined 
phase-space structures,
and sort them into parent and satellites, the latters organized following
some distribution in mass and orbital parameters, I cannot use halos taken
from cosmological simulations, but rather I need to write controlled 
initial conditions (ICs).
If the halos are anywhere near realistic, building numerically their
equilibrium structure is not an easy task. For brevity I will
describe the procedure for an isotropic NFW, while
a generalization to a halo of any generic phase-space
structure follows from Chapter 5.

There are two steps in constructing a numerical halo:
(1) finding the phase-space distribution function (DF) producing the desired
equilibrium structure;
(2) using Monte Carlo
samplings of this DF to generate the N-body realization. 
It should be clear, from Chapter 5, that
determining an equilibrium DF for a given halo structure is the actual main 
difficulty of this process. Simple, analytical DFs are known only for
a handful of models, such as Plummer spheres \cite{plummer}, 
lowered isothermal models (e.g. King models \cite{king}), lowered power-law
models (\cite{evans3},\cite{ku}), and a few special cases
(e.g. the Hernquist model \cite{herny}, and \cite{jaffe},\cite{dehnen93}).
There is no universal procedure to achieve the determination of the DF, but
according to the single cases, different methods can apply.
For the NFW, one way is to follow Chapter 5, and use the
``theoretical'' DF described there. However, the N-body NFW is only an
approximation of the theoretical one, being truncated at large radii to yield
a finite mass, and being softened at small scales; not sure about the effects
of these features on the DF, I chose to find a steady-state DF numerically, 
that reproduces the desired density and velocity anisotropy profiles.

Following the suggestion of \cite{kaz}, I avoided a shortcut 
often used in the literature (see for example \cite{spri},\cite{boily}),
\textit{i.e.} of finding the velocity distribution by means of the 
\textit{local maxwellian approximation}; with the latter, the velocity at each
point in space consistent with a given potential is approximated by a 
multivariate Gaussian, whose \textit{mean} velocity and velocity dispersion tensor 
satisfy the Jeans' equation at this point (see \cite{hern}).
In fact, there is a dangerous
shortcoming of this approximation when it is used to generate initial
conditions for high-resolution numerical simulations; 
most of the models of interest have local self-consistent velocity profiles
that become strongly non-Gaussian, especially near the center so that,
if one uses the local Maxwellian approximation 
to construct an N-body realization of such a model, the center of 
the resulting N-body system will be far from equilibrium. When the halo 
is evolved in isolation,
it rapidly relaxes to a steady state whose density and 
velocity profiles differ significantly from the initial, intended ones. 
In particular, satellites constructed
using the local Maxwellian approximation can undergo rapid artificial 
tidal disruption \cite{kaz}.

With no other help, the numerical determination of the DF for a given halo
follows the theoretical treatment, with some computational issues to consider.

The first step is to set the mass distribution of the halo. The NFW
\textit{per se} has an infinite mass, so I need to cut it at some point, that
is taken far outside the virial radius in order to avoid boundary effects on
the halo particles; the cutting function must be smooth, since a sharp truncation would
lead to unphysical
models with $f<0$. Of course, this results in a waste of particles used to
represent uninteresting regions of the system.
The density profile I adopt features an exponential cutoff \cite{spri}; I will
refer to it as the ``modified NFW'':
\be 
\rho(r)=\frac{c^2 \ g(c)}{4 \pi} \ \frac{1}{r \ (1+cr)^2} \ \ \ for \  r \ <1
\label{kaznfw}
\ee
\be
\rho(r)=\frac{c^2 \ g(c)}{4 \pi} \frac{1}{(1+c)^2} \ r^{\gamma} \ e^{(1-r)/R_{decay}} \ \ \ for \ r \ >1
\label{kaz}
\ee
here all the distances are in units of the virial radius $R_{vir}$; 
$r$ is the radial coordinate defined on a grid uniformly spaced in logarithm,
$R_{decay}$ is a parameter setting the sharpness of the
profile decay that starts after the virial radius, and
$\gamma=(-1-3c)/(1+c)+R_{vir}/R_{decay}$ is set by the condition of a continuous
logarithmic slope, that insures the smoothness of the transition at $R_{vir}$.
The minimum radius of the system is chosen in order to be sufficiently close
to the centre of mass (for example, $r_{min}=1e-3$); the maximum
radius $R_{cut}$ is chosen as the distance from the centre where the density
falls below a given threshold value ($\rho < 1e-10 \ \rho_{central}$ for instance).
As an example, for a halo of concentration $c=10$ and $R_{decay}=5$ 
the cut radius is $R_{cut}\simeq 17$. 
This density profile yields a finite total mass, that in this example is
$M_{cut} \sim 2 M_{vir}$. So, only about half of the particles used to simulate
this halo in a stable configuration end up inside the virial radius.
 
The second step is to initialize the particles' positions. These are
obtained from the cumulative mass profile, that in the case of the 
modified NFW lacks an analytical expression.
Given $M(r)$, I randomly pick radial distances out of the inverse relation $r(M)$,
and in each position I assign a particle of mass $m_{part}=M_{cut}/N_{part}$,
where $N_{part}$ is the total number of particles. To obtain a spherically
symmetric system, I pick two random angles and split the radius $r$ in the
($x,y,z$) components.
In this way, I obtain a spherically symmetric system of particles, whose mass
distribution is the NFW inside the virial radius, and decays exponentially 
outside. 


The third step is to determine the gravitational potential generated by this
particle distribution. This is the solution of the Poisson's equation, but
numerically it is more straightforward to evaluate the force
acting on each particle, as the sum of the contributions by all the others; the
complication is that the forces need to be softened at small scales.
For this purpose, I define a \textit{softening length} 
\be
S=\frac{4 \ R_{vir}}{N_{vir}}
\label{soft}
\ee
following from \cite{power}, with
$N_{vir}$ being the number of particles inside the virial radius. A correct
softening length should be of the order of $<1 \%$ of the virial radius. 

There are several prescriptions for the softened potential, and at the
beginning I was using the one by Springel
(\cite{springel01},\cite{springel05}). 
In the end, I
found it more consistent to simply feed my particle distribution to Gadget-2,
and have it evaluate the potential directly.

Once the potential is set as a function of the particles' positions, the
allowed range of velocities for each particle is determined by the
prescription that its binding energy must be positive or null:
\be
\epsilon(r)=\Psi(r)-1/2 \ v(r)^2 \ge 0~;
\label{binding}
\ee
here $\Psi(r)=-\Phi(r)$, with $\Phi$ being the gravitational potential.
It follows that the escape velocity is defined as
$v_{esc}(r)=\sqrt{2 \Psi(r)}$ at each radius, where the allowed velocities 
span from $0$ to $v_{esc}$.
Notice that, under the condition $\epsilon \ge 0$, a physically 
meaningful constraint on the DF is to
set $f=0$ for unbound particles.

In order to assign a velocity to each particle, I need now to evaluate the
single particle's DF. In the case of the isotropic halo, I obtain it from the density
profile and the potential through the Eddington's inversion, that I re-write
here in a version of more practical use (for halos of any phase-space
structure, the generalization of this formula follows from Chapter 5):
\be
f(\epsilon)=\frac{1}{\sqrt{8}\pi^2} \left[ \int_0^\epsilon
  \frac{d^2\rho}{d\Psi^2} \ \frac{d\Psi}{\sqrt{\epsilon-\Psi}} \ +
  \frac{1}{\sqrt{\epsilon}} \left( \frac{d\rho}{d\Psi} \right)_{\Psi=0} \right]~.
\label{ed}
\ee  
The second term on the r.h.s. is null for any sensible potential \cite{BT}.
The first term contains the second derivative of the density
profile with respect to the potential, and for both the NFW and the modified NFW
the relation $\rho(\Psi)$ is not analytical. The attempt to evaluate it
numerically, with different methods, failed already for the first derivative, 
due to the very high noise. At the end, I found that the best alternative 
is to fit the $\Psi-\rho$
relation and find the second derivative of the fit analytically.

For each particle at a position $r$, I need the values $f(\epsilon)$ on a grid
$0 \le \epsilon \le \Psi(r)$, where $\Psi(r)$ is the maximum value of the
binding energy for a particle at $r$; 
$f(\epsilon)$ is then given by an intregral in the potential from $0$ to $\epsilon$.
Notice that, at a fixed particle's position, $f(\epsilon)=f(v^2)$, thus the
energy DF evaluated at $r$ directly yields the probability density for the particle's total velocity
to have a value in the volume $d^3v$.  

It follows that the particle's velocity probability function is obtained as
\be
f(v)dv=4 \pi v^2 f(v)dv~,
\label{vDF}
\ee 
where at each radius the values of $v$ are given by $v=\sqrt{2(\Psi(r)-\epsilon)}$
and span the interval $0 \le v \le \sqrt{2\Psi(r)}$. The probability function correctly
goes to $0$ at the extremes of the allowed velocity range, with a peak that
moves towards higher velocities as the particles are closer to the center of
the halo. 

Notice that this method is computationally expensive, since it involves the 
evaluation of a set of integrals on a different grid for each particle; instead of
doing this, I evaluate $f(\epsilon)$ on a grid of
values $\epsilon$ equispaced in $log$, that spans the whole range defined by 
the halo potential; then for each particle I perform a series of
interpolations in $f$ and $\epsilon$ to build its velocity DF. The accuracy of
the two methods are comparable \cite{kaz}.

Finally, I have everything I need to assign the velocities to the particles.
Given the relation $v - f(v)$, I use the acceptance-rejection technique to set
$v(r)$ (\cite{press},\cite{ku}); 
it consists in randomly sampling a velocity in the range $0 -
v_{esc}(r)$, and a value of $f$ in the range $0 - f_{peak}$, and confronting the point
I obtain with the plot $v - f(v)$: if it falls under the curve, the random
velocity is taken and assigned to the particle.
The single $(v_x,v_y,v_z)$ components of the velocity vector are then assigned
in analogy with the coordinates, with two random angles, thus obtaining an
isotropic velocity distribution. For different shapes of the velocity
dispersion tensor, the components are set through the anisotropy profile
parameter $\beta(r)$ (see Chapter 5).  

\begin{figure}
\epsfig{file=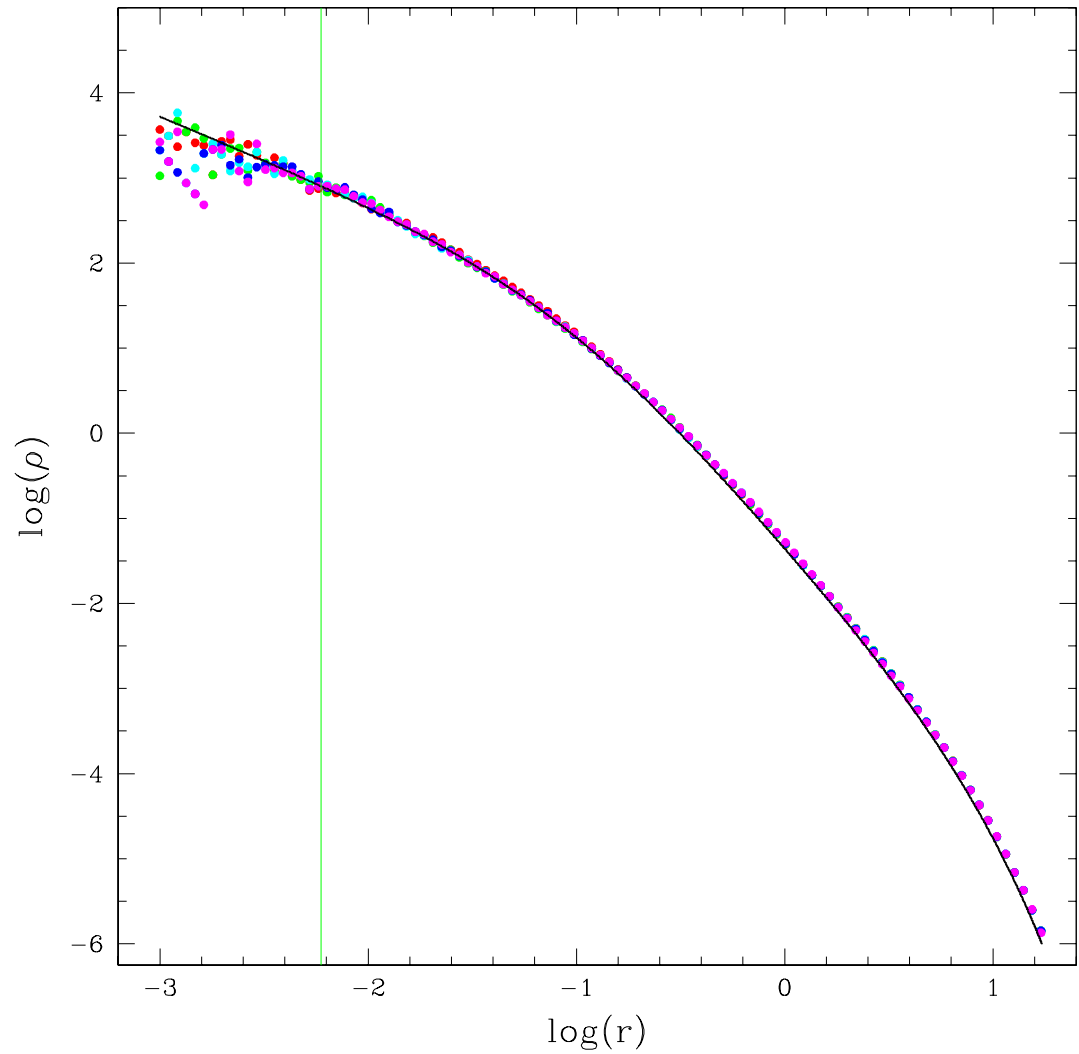,scale=0.75}
\caption{Stability test for the initial conditions of a simulated NFW of
  $10^6$ particles: density
  profile at different snapshots, up to 3 dynamical times (\textit{different
  colors}). \textit{Solid line}: theoretical modified NFW. \textit{Vertical
    green line}: softening length limit; inside this radius any result is
  dominated by numerical artefacts.}    
\label{densteo}
\end{figure}

\begin{center}
\begin{figure}
\epsfig{file=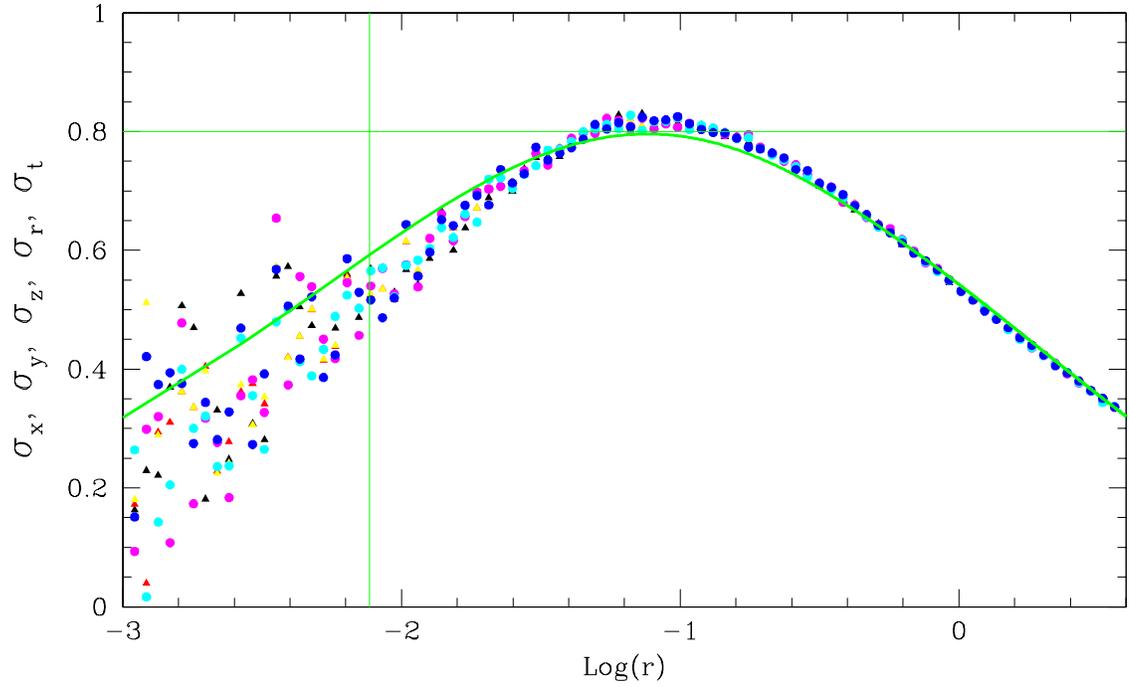,scale=0.8}
\caption{The velocity dispersions' profiles for an isotropic $10^6$-particle
  modified NFW (different colors for
  $\sigma_x,\sigma_y,\sigma_x,\sigma_r,\sigma_t$); 
  the \textit{green line} represent the theoretical solution of the
  Jeans' equation for the modified NFW density profile. The softening length
  is marked by the \textit{vertical green line}.}
\label{sigmateo}
\end{figure}
\end{center}

In Figure (\ref{densteo}) I show the result of a stability test on a simulated
modified NFW with isotropic velocity dispersions, made of $10^6$ particles.
The density profile is recovered from different simulation snapshots, up to 3
dynamical times (\textit{different colors}). 
The halo looks stable from the very beginning of the simulation, with no
oscillations and no deviations from its correct mass distribution. As already
discussed here and in the previous Chapters, the stability condition is
verified by the use of the Eddington's inversion (or its generalizations).
Of course, this does not apply to radii smaller than the softening length,
where the N-body approximation fails and numerical artefacts dominate the
dynamics. In Figure (\ref{sigmateo}) I show the velocity dispersion profiles
$\sigma_x,\sigma_y,\sigma_x$ for the same halo (in \textit{different colors}), and
confirm that the halo feature the correct isotropy; 
in addition, I show the 
velocity dispersions profiles $\sigma_r,\sigma_t$
obtained by projecting the total velocity
on the radial and tangential directions; notice
that there is no directional dependence whatsoever of the velocity
distribution profiles.
The \textit{green line} represents the theoretical
one-dimensional $\sigma$ obtained from the Jeans' equation for the 
isotropic modified NFW; notice a bit
of noise for small radii, due to the numerical implementation and the
randomness of the velocity selection process. As it is immediately apparent,
anything at radii smaller than the softening length (\textit{vertical green line}) is
out of equilibrium.    

\section{Work in progress}

\noindent
I intend to start with a suite of simulations specifically studied to model the
satellite disruption. To this purpose, 
I am going to make an ensemble of satellites infall into a parent, trying
different equilibium configurations for the halos.
In particular, the parent will be a modified NFW, with
anisotropy profile as resulting from cosmological simulations.
For the satellites, I will try various solutions, for instance (i)
the same as the parent, (ii) cored configurations with tangential
anisotropy.
In addition, I am going to set the satellites' orbits following \cite{benson}.
In order to test the parent halo response to the substructure infall, I plan
to try different subhalos mass functions, as discussed in Chapter 5. 

The first issue to consider is the simulation resolution; the inner regions of the
satellites must be resolved with an accuracy high enough to appreciate
the differences between cores and cusps, and radial and tangential
anisotropies. In order to reach this level of precision, I initially plan to use 
$5 \times 10^5$ particles for the biggest substructures. This makes the total
number of particles for the parent halo $ \sim 5 \times 10^7$ at least.

The tidal stripping of the satellites is going to offset the dynamical
friction efficiency in the central regions of the halo;
I expect this effect to depend on the orbital parameters, and on both the parent
and the satellite structure. Since 
I cannot evaluate the entity of the offset a-priori, I plan to
run a number of tests before the actual simulations.

The main problem is that the dynamical friction efficiency 
strongly depends on the mass of the satellite
as compared to the mass of the background particle; 2-body effects
will also depend on the same parameter. If I use dark matter particles of the same mass for
the satellite and the halo, the softening length is unique for all particles, and will have
to be tuned in order to minimise the spurious gravitational scattering and at
the same time maximise the dynamical friction efficiency.
Again, a number of preliminary tests
are required to settle this question. 

Regarding the evolution of the parent halo, there are some interesting side
issues that can be investigated. 
The first is linked to the lack of self-similarity
at small scales. In galaxies we do not observe a large number of
substructures, while on cluster scales their mass function as 
predicted by simulations is much more in accord with observations. This
suggests a mass-dependent mechanism of disruption, or alternatively an
environment dependency, and I would like to examine this matter further.

The second regards the mass accretion of the parent halo due to
substructure infall. As explained in Chapter 1, the phase-space density of
NFW halos is described by a power law of critical slope, that maximizes the
phase mixing. I am interested in testing this result against different boundary
conditions for the mass accretion, including different phase-space structures
of the satellites, and with the inclusion of an accurate treatment of
dynamical friction, to see whether the same relation still holds, or is replaced 
by some other functional form, or even whether a universal relation continues to exist.

%% file: conclusioni.tex
\chapter{Conclusions}

\noindent
In this Thesis I investigated the structure of galactic dark matter
halos, encouraged by the lack of a consensus about their mass profile 
and angular momentum distribution. In fact, while the hierarchical clustering
scenario leads naturally to the NFW halo profile, the
latter is challenged by observations of the inner structure of galaxies.

My analysis started from the  
study of the properties of halos in relation to the host galaxies,
and the determination of scaling relations between the dark and luminous
components, that allowed the construction of an observation-based 
spin parameter distribution function. During this analysis, it became 
apparent that the halo mass distribution is not a universal feature
deriving from hierarchical clustering, but is instead heavily dependent
on the baryonic mass hosted in the halo; I quantified this with an 
extension of the Universal Rotation Curve to include the whole virialized
halo. One interesting feature emerging from this study was that, 
when compared to the density profile predicted by numerical simulations, 
the dark matter mass distribution appears not only to be flat 
in the centre of halos (the core-cusp problem), but also 
to be overdense around the galactic optical radius, thus not 
converging to the NFW profile even at these scales.

The overall picture offered by these dynamical evidences suggested me to  
investigate the interaction between the baryonic component and the
dark matter during galaxy formation, and the consequent evolution of the 
halo. In order to do this, I firstly made a detailed analysis of the 
phase-space dynamical properties of dark matter halos as collisionless
systems of particles whose structure is governed by the evolution in 
a gravitational potential. I quantified the connection between the 
mass density profile and the velocity dispersion tensor, and showed how a 
perturbation in the velocity structure, triggered by angular momentum
injection, causes the cuspy halos predicted by the CDM N-body simulations
to evolve into cored configurations. I then proposed a physical mechanism
to account for such a halo evolution, that describes the 
dynamical coupling between the dark and luminous components; 
dynamical friction of the dark matter
on the collapsing baryonic component at the time of protogalaxy formation
is powerful enough to alter the halo inner motions and transform the halo
profile, thus making it possible to reconcile theory and observations.  

\bigskip
\bigskip

\noindent
In the first part of this Thesis I studied the structure of
dark matter halos as inferred from observations of galaxy rotation curves.
I built a series of scaling relations that link
the mass distribution and geometry of the luminous components in the disk 
to the dynamical parameters of the host dark
matter halo, i.e. mass and angular momentum. 
The results are at odds with the predictions of numerical simulations. 
In detail:

$\bullet$  the spin distribution function of galactic halos inferred from
  observations peaks at significantly lower values than predicted by numerical
  simulations, indicating a mass accretion history devoid of major mergers
  since $z\sim 3$;

$\bullet$ the shape and amplitude of the dark matter mass distribution 
depend strongly and non-linearly on the baryonic mass; 

$\bullet$ in single halos, the mass distribution not only is not cuspy in the 
  central regions, but also it does not converge to the NFW even outside 
  the halo core. Rather, it features an excess of mass at the optical radius 
  not attributable to numerical artefacts or exotic DM behaviour. 

I conclude that there is no evidence of a universal, 
self-similar halo hosting spiral galaxies, in contrast with the prediction
by hierarchical clustering. The dependence of the halo density profile 
and angular momentum on the baryonic mass and geometry suggests a dynamical 
coupling between the dark matter and the baryons, originating at the epoch of 
galaxy formation.

\bigskip
\bigskip

\noindent
In order to understand the evolution of dark matter halos under 
external perturbations, such as the assembly and evolution of the 
baryonic component, I modeled the dynamical properties of halos in 
phase-space, and studied the effects of perturbations on their equilibrium
structure. 

Specifically, I assumed a ``pristine'' pure dark matter halo
to be well described by the NFW, and modeled its evolution during 
galaxy formation into a cored configuration. 
The results I obtained can be summarized as follows:

$\bullet$ the phase-space dynamical structure of dark matter halos, 
  including the NFW and the Burkert (cored) halos, is well represented 
  by the same family of solutions of the Jeans
  equation, \textit{i.e.} a distribution function depending on the binding 
  energy and total random angular momentum of the halo; 

$\bullet$ the halo mass distribution is governed by the inner dynamics; 
  the symmetry of the distribution function determines the shape of the 
  velocity dispersion tensor, which in turns determines the density profile.
  The halo mass distribution is very sensitive to the balance
  between the tangential and radial motions. In particular, I 
  found the following relation between the anisotropy parameter and the 
  slope of the inner density profile: 

\be
\rho(r) \propto r^{-2\,(1+2\beta)/(2+\beta)}~;
\label{rhobeta}
\ee

$\bullet$ a perturbation in the anisotropy profile triggers a 
  rearrangement of the halo mass into a new
  equilibrium configuration; in particular, angular momentum transfer to the 
  halo induces an increase of the tangential motions
  in the inner halo, leading to the formation of a corelike feature;

$\bullet$ the baryonic collapse inside the halo leading to the formation of the
  protogalaxy represents a perturbation of the phase-space halo structure that
  enhances the dark matter tangential motions; the self-bound clouds
  infalling into the halo potential well are subject to dynamical friction by
  the background dark matter, and exchange angular momentum locally with the
  halo, with the result of unbalancing the halo anisotropy profile. 
  The amplitude and distribution of the angular momentum
  exchanged are compatible with triggering the halo evolution into an
  equilibrium configuration compatible with the observations.   

\medskip

Galactic dark matter halos, grown from primordial perturbations through
hierarchical clustering, and having acquired a characteristic structure through
processes like violent relaxation and entropy stratification, are then
perturbed by galaxy formation, when a non-negligible amount of
baryonic mass and angular momentum are transferred to the center of the halo. 
The dark matter halo does not behave as a static entity during this process; 
as the
baryonic component is shaped depending on the halo dynamics, conversely the
halo reacts to the formation of the galaxy by rearranging its structure and
adjusting to the changed equilibrium conditions. 
The dark and luminous component are 
dynamically coupled, and jointly evolve into new equilibrium states.

%% file: appendix.tex
\chapter{The resolution of mass modelling from rotation curves}

\noindent
In this Appendix I present a method for the mass modelling from rotation
curves (RCs) of spirals that, provided the curves are of high quality,
 ensures a unique decomposition of the luminous and
dark matter components, for a general halo profile.  

Consider a two-components system, made of an exponential disk of scale-length
 $R_D$ embedded into a dark matter halo, of mass distribution taken as 
the pseudo-isothermal sphere (PI). 
The reason for this choice is the following: the PI model has two free
parameters determining the mass distribution, a fact that increases the
degeneracy of the mass modelling with respect to one-parameter profiles, 
like the NFW or the URC. For this reason, the conclusions in this Appendix
 safely apply to the NFW and the URC as well, less problematic from this point of view.

The total rotation curve is given by 
\be
V^2(r) = V^2_d(r) + V^2_h(r)~, 
\ee
where $r$ is the radial distance normalised to $R_D$; in these units, 
the disk component is 
$V^2_D(r)=GM_D/R_D \ \nu (r)$, with $\nu(r) r/2 \ (I_0K_0 - I_1K_1)|_{r/2}$
\cite{freeman} and the halo one is $V^2_H(r)=GM_H/(r \ R_D)$.

In the most general case, the contribution of the luminous matter 
to the total rotational velocity is 
determined by two parameters: the disk scale-length $R_D$
and the total disk mass $M_D$. 
 The dark component is determined by other two parameters, usually taken as the central halo 
density $\rho_0$ and the halo characteristic scale length $R_C$.

\section{The mass modelling for PI halos}

\noindent
The DM density profile of the PI sphere is given by
\be
\rho(r) = \frac{ \rho_0 }{ 1 + (r/ R_C )^2 }~; 
\ee
for convenience I define $r \equiv r/R_D$ and $R_C  \equiv  R_C/R_D$.
The mass profile of the halo is
\be
M_H(r)=4\pi G \rho_0 \ R_C^3 R_D^3
\left[\frac{r}{R_C}- tan^{-1}\left(\frac{r}{R_C}\right)\right]~;
\ee
after setting
\be
\lambda (r) \equiv \frac{R_C^3}{r} \ 
\left[\frac{r}{R_C} -  tan^{-1}\left(\frac{r}{R_C}\right)\right]~,
\ee
the total velocity is written as
\be
V^2(r) = V^2_D(r) + \frac{G 4\pi \rho_0 \ R_D^3}{R_D} \ \lambda(r)~.
\ee
Now consider the following transformation:
\[\alpha \equiv \frac{G M_D}{R_D \ V^2_1}, \hspace{0.7cm} \beta \equiv 
\frac{4 \pi G \rho_0 \ R_D^3}{R_D \ V^2_1}, \hspace{0.7cm} \lambda_1 \equiv 
R_C^3 \ \left[ \frac{1}{R_C} -
  tan^{-1}\left(\frac{1}{R_C} 
\right)\right]\]
where $V^2_1 \equiv V^2(r=1))$, and the parameters $\alpha$ and $\beta$ 
are directly proportional to the fraction
of the luminous and dark mass respectively, evaluated at $R_D$.
The normalized rotation curve is then
\be
\frac{V^2(r)}{V^2_1} = \alpha \ \nu(r) + \beta \ \lambda(r)~;
\ee
$\alpha$, $\beta$ and $\lambda$ determine the shape of the curve, while
 $V^2_1$ determines its amplitude, as a normalisation parameter.
Notice that 
\be
\alpha \ \nu_1 + \beta \ \lambda_1 = 1~,
\ee
where $\nu_1=\nu (r=1)$; the quantities $\alpha \nu_1$ and
 $\beta \lambda_1$ represent the fractional contributions to the total 
velocity at $R_D$ due to the luminous and dark matter 
respectively; I define $f_{DM} \equiv \beta \lambda_1$. 

Finally, the rotation curve can be written as
\be
\frac{V^2(r)}{V^2_1} = (1-f_{DM})\ \frac{\nu(r)}{\nu_1} +f_{DM} \ \frac{\lambda(r)}{\lambda_1}~,
\ee
with the transformed free parameters governing the shape of the curve defined 
in the ranges 
\[0 \leq f_{DM} \leq 1  \hspace{1cm} 0 \leq R_C \leq 3~.\]
The range for $R_C$ is conservative; high-quality rotation curves described in
Chapter 3 and 4 reach out to several disk scale-lengths. The last parameter
$V_1$ is a normalisation that can be neglected in the present argument, since
it does not affect the precision of the curve determination.

\begin{figure*}
\begin{center}
\epsfig{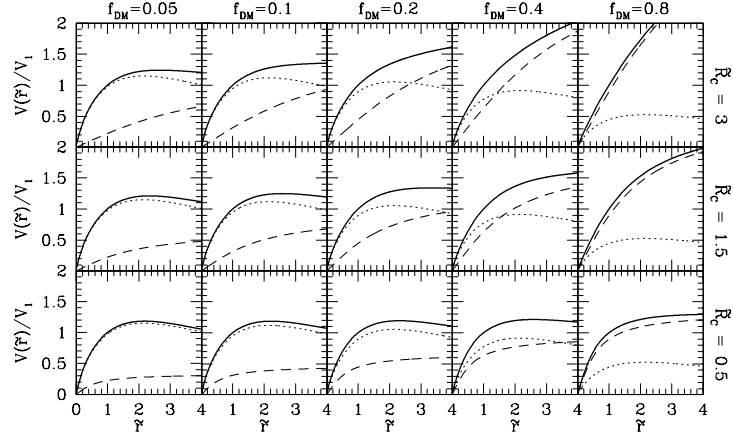}
\caption{PI rotation curves, for different $f_{DM}$ and $R_C$.
\textit{Solid lines}: total rotational velocity; \textit{dashed lines}: dark
matter component; \textit{dotted lines}: baryonic component.} 
\label{iso15}
\end{center}
\end{figure*}
\begin{figure*}
\begin{center}
\epsfig{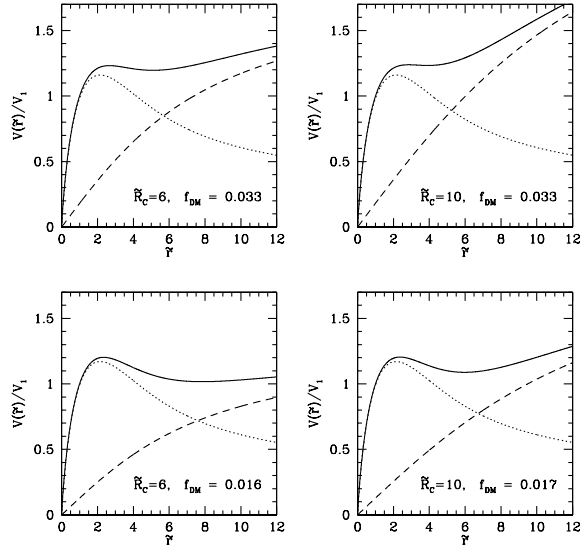}
\caption{PI rotation curves with a second-order anomaly, indicating a 
sharp transition between a disk-dominated to a halo dominated region.}
\label{isof}
\end{center}
\end{figure*}
In Fig.~(\ref{iso15}) I show some examples of rotation curves obtained for
different values of the parameters, to illustrate the visual degeneracy of the problem.
Apparently, very simular curves can be obtained with different combinations of the
parameters (\textit{dashed lines} represent the halo component, \textit{dotted
  lines} the baryonic one), when the curves are smooth like in the majority of
the observations. 
In Fig.~(\ref{isof}) I show some easy examples instead, of curves where the
baryonic component produces a visible feature in the velocity profile, thus facilitating the
mass-modelling process.

\section{Disentangling a high-quality RC}

\noindent
From a smooth, featureless rotation curve, it is possible to obtain the
correct mass modelling and uniquely disentangle the luminous and dark components of the
mass distribution, provided that the quality of the curve is acceptable, that
is, that the observational errors at each point of the curve are not too large. 

Here I illustrate a virtual experiment: I simulate the obervation of a number 
of reference curves, and apply the mass modelling described above to recover
the parameters characterising the luminous and dark components.

In detail, consider 25 reference curves, each one consisting of 25 data points
between $0 \le r \le 4$ in units of $R_D$; the error in the value of the 
velocity is on average $\epsilon_V = 0.02$ and that in the slope of the curve
is $\epsilon_D = 0.05$ (\cite{ps95}; these errors are of the order
of those in high-quality RCs measured today).
The parameter space is defined as 
$f_{DM} =\{0.1,0.3,0.5,0.7,0.9\}$ $\otimes$ $R_C =\{0.1,1.0,2.0,3.0,4.0\}$.
A semplification coming from the observational technique is that $V_1$
can be set direclty from the data:
\be
V_1 \equiv V(1) \pm \mathcal{O} (10^{-2})~, 
\ee
if and only if the curve is accurate enough. Actually, I would define a
high-quality RC as a curve allowing for this measurement. 

\begin{figure}
\begin{center}
\epsfig{file=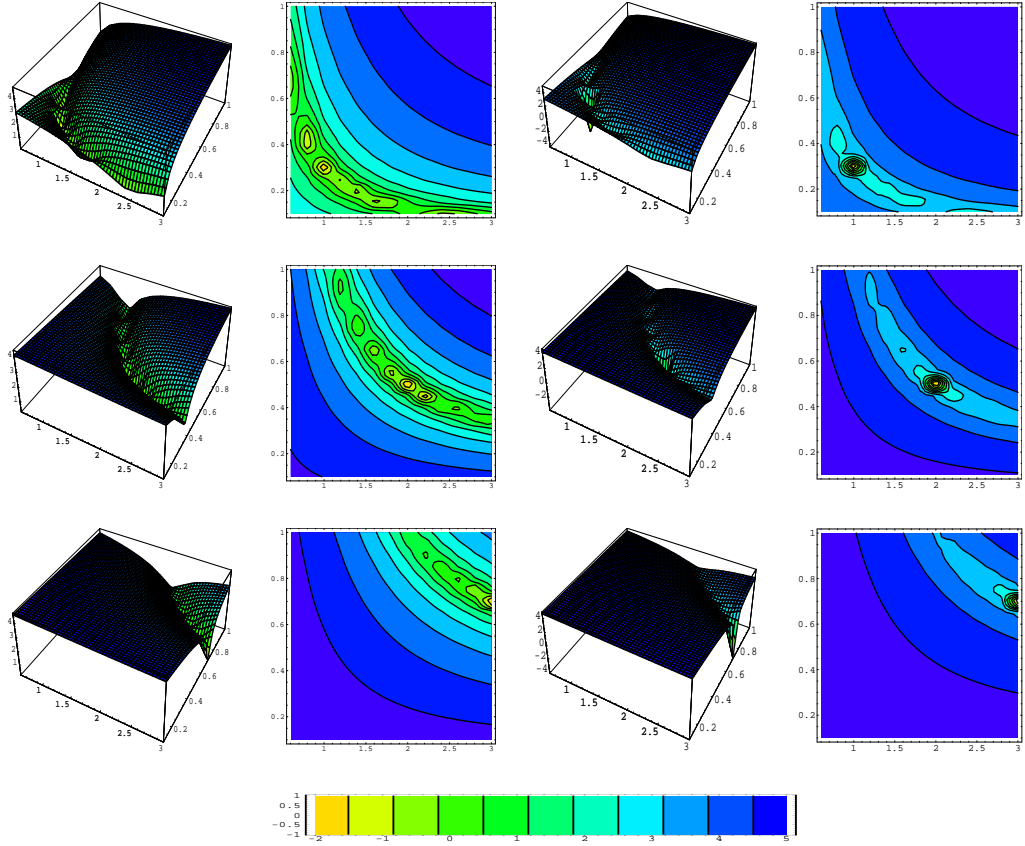,scale=0.5}
\caption{$\chi^2(V)$ (\textit{left}) compared to $\chi^2_{TOT}$ (\textit{right})
  for 3 reference curves, as a function of the model 
parameters: $R_C \in [0.2,4]$ (\textit{x-axis}) and $f_{DM} \in [0.05,1]$
(\textit{y-axis}). The ``true'' parameters are: ($R_C=1.0,$$f_{DM}=0.3$) \textit{top panels}, 
($R_C=2,f_{DM}=0.5$) \textit{middle panels}, ($R_C=3,f_{DM}=0.7$)
 \textit{bottom panels}. The colour scale is in $log(\chi^2)$.} 
\label{quadretti}
\end{center}
\end{figure}

The distance in parameter space between a model curve $V_{mod}$ (reconstructued thorugh
mass modelling) and a reference curve $V_{obs}$ can be defined by a  
likelihood-type parameter, taken as the sum of the $\chi^2$ 
computed on the velocity and on the RC slope: 
\be
\chi^2_{TOT}=\chi^2_V+\chi^2_D~.
\ee
The two contributions are given by:
\be
\chi^2_V([f_{DM},R_C]_{mod})= \frac{1}{\epsilon_V^2} \ 
\sum_{n=1}^{25} \left(V_{mod}(r_n;[f_{DM}, R_C)]_{mod})-V_{obs}
(\tilde{r}_n)\right)^2~,
\ee
\be
\chi^2_D([f_{DM},R_C]_{mod})=\frac{1}{\epsilon_D^2} \ \sum_{n=1}^{25}
 \left(\frac{r_n}{V_{mod}} \ \frac{dV_{mod}(r_n;[f_{DM},R_C)]_{mod})}{dr}
-\frac{r_n}{V_{obs}} \ \frac{dV_{obs}(r_n)}{dr}\right)^2~.
\ee
The novelty of the approach is in considering the slope of the curve as an aid
in minimising the smallest volume in parameter space that is resolved by the
mass modelling. In Fig.~(\ref{quadretti}) I illustrate the difference in the
mass-modelling resolution, if the $\chi^2_V$ is considered alone (\textit{left
panels}), and if the $\chi^2_{TOT}$ is considered instead (\textit{right panels}).
It is clear that in the second case the resolution increases dramatically. 
In fact, the $\chi^2_{TOT}$ rises sharply to a value of $\sim 10^2$ in an interval
 of $\Delta f_{DM} \sim 0.05$ and of $\Delta R_C < 0.25$; these values
 set the resolution scale of the mass modelling.

\begin{figure}
\begin{center}
\epsfig{file=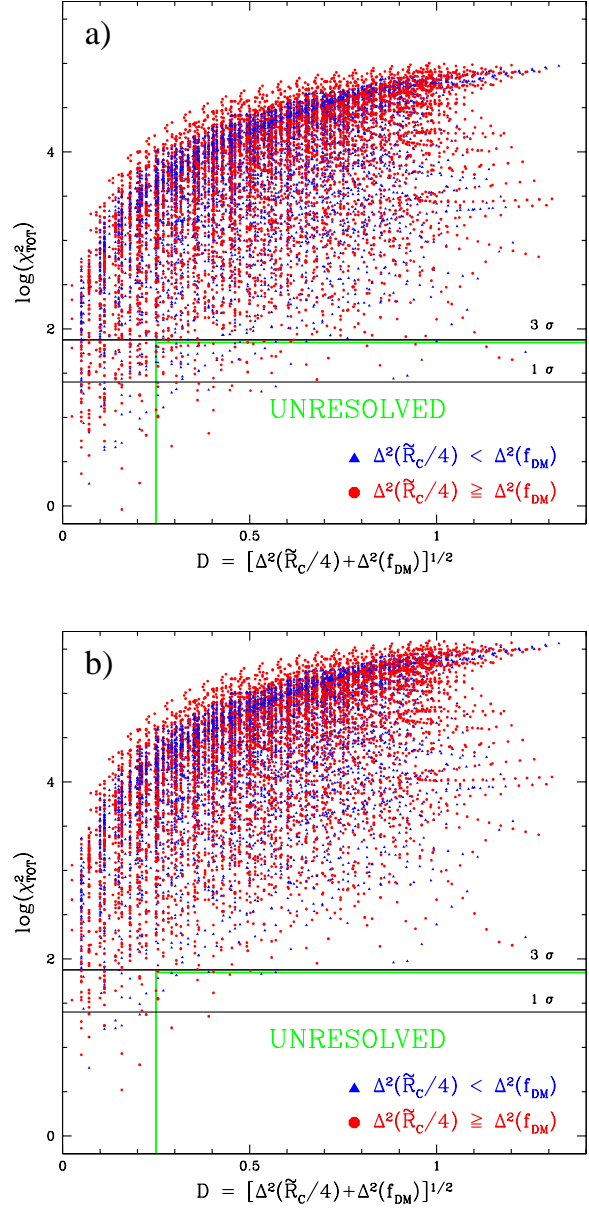,scale=1.2}
\caption{$\chi^2_{TOT}$ of the 25 reference ``observed'' curves mapped with 
400 model curves, as a function of the distance in the parameter space. Red
 points (circles) are for distances dominated by the variation of
 $\tilde{R}_C$,
 blue points (triangles) are for distances dominated by the variation of
 $f_{DM}$.
 The straight lines represent the $1\sigma$ and $3\sigma$ limits. Panel
 \textit{a)} $\epsilon_V=0.02$ and $\epsilon_D=0.05$. Panel \textit{b)} 
$\epsilon=0.01$ and $\epsilon_D=0.03$. }
\label{distanze}
\end{center}
\end{figure}

Fig.~(\ref{distanze}) illustrate te behaviour of $\chi^2_{TOT}$, for an
ensemble of 400 model curves used to fit the 25 reference curves: each of the
10.000 points in the plot represents the $\chi^2_{TOT}$ for a couple
$V_{obs}-V_{mod}$, mapped as a function of their actual distance in parameter
space, defines as 
\be
D \equiv (\Delta^2(R_C/4)+\Delta^2(f_{DM}))^{1/2}~,
\ee
with $\Delta(x)=x_{mod}-x_{obs}$; since the range of variation 
of $R_C$ is four times that of $f_{DM}$, the former is normalised by a 
factor 4 to make the two contributions to $D$ comparable. 
The result is shown in the \textit{upper panel}; 
the \textit{red circles} represent cases in which the distance is dominated by 
variations of the core radius ( $\Delta^2(R_C/4) >
 \Delta^2(f_{DM})$), 
the \textit{blue triangles} are cases in which the distance is mainly due 
a different amplitude of the dark matter contribution. 
The straight lines represent the values of $\chi^2_{TOT}$ corresponding to the
$3\sigma$ and $1\sigma$ limits, thus selecting out the couples
$V_{obs}-V_{mod}$ unresolved by the model. Notice that most of the points
lay over the $3\sigma$ limit, showing that the model can uniquely resolve the
mass distribution parameters.   

In addition, the vertical line defines a distance in the parameter space of
$D=0.25$; couples lying in this range are characterised by parameters so
similar that the offset between them is quite negligible for the purpose of
determining the mass distribution of the components. Hence,  
the points in the range $D\leq0.25$ and laying below the $3\sigma$ limit are
not to be considered as system's failures.
Summing up, the mass modelling described above is successfull in the 
$\sim 99 \%$ of cases, while about $\sim 1 \%$ of the couples $V_{obs}-V_{mod}$ are unresolved. 

The observational errors adopted are typical of about 
the top $50\%$ of the observed RCs available today. 
There are cases however, where the errors on the RC are even smaller; 
if one considers errors of $\epsilon_V=0.01$ and $\epsilon_D=0.03$ (curves
with these errors are available today, in increasing number), the same
analysis leads to the \textit{bottom panel} of Fig.~(\ref{distanze}).
In this case, about $0.17\%$ of the couples are unresolved; if the lower limit
of the resolution was set by the worst performance, the uncertainty in the mass
modelling would still be reasonable: $(\Delta(f_{DM}))_{MAX}=0.25$ 
and $(\Delta R_C)_{MAX}=1.2R_D$. However, considering the negligible amount of failures
however, I can safely state that the average resolution of this mass-modelling
procedure is set by the density of points in the parameter space,
\textit{i.e.} $(\Delta(f_{DM}))=0.05$ and $(\Delta R_C)=0.2R_D$.